\documentclass[prd,preprint,tightenlines,floatfix,preprintnumbers,showpacs,nofootinbib,eqsecnum]{revtex4}
 \usepackage[dvips,final]{graphicx}
  \usepackage{amssymb,amsmath,bm,pifont}
   \graphicspath{{./Figs/}}

\usepackage[cp1251]{inputenc}
 \usepackage[english,russian]{babel}

\begin{document}

\title{Глобальная дробно-аналитическая теория возмущений в КХД и ее некоторые приложения}
\author{А.~П.~Бакулев}
\email{bakulev@theor.jinr.ru}
\affiliation{Лаборатория теоретической физики им.~Н.~Н.~Боголюбова, ОИЯИ,\\
    141980 г.~Дубна, Россия}

\begin{abstract}
Представлено глобальное обобщение
Дробно-Аналитической Теории Возмущений (ДАТВ),
учитывающее пороги тяжелых кварков и позволяющее работать
с КХД-амплитудами как в евклидовой,
так и во временноподобной областях значений
квадрата передачи импульса $q^2$.
Кратко изложен аналитический подход в теории возмущений КХД,
инициированный работами Джонса, Соловцова и Ширкова.
Также кратко изложены основные положения ДАТВ при фиксированном
числе кварковых ароматов.
Более подробно обсуждается проблема порогов тяжелых кварков
и построение глобальной версии ДАТВ,
в том числе вопрос об аналитических константах связи
во временноподобной области значений
$q^2=s>0$ для описания $e^+e^-$-аннигиляции и формфактора пиона.
Достаточно подробно описаны приложения развитой глобальной версии
ДАТВ к феноменологически интересным процессам
(формфактор пиона и ширина распада хиггсовского бозона $H\to b\bar{b}$),
а также к суммированию пертурбативных рядов.
\end{abstract}

\maketitle

 \begin{flushright}
 \textit{Посвящается памяти Игоря Соловцова,\\
         одного из создателей Аналитической Теории Возмущений.}
 \end{flushright}

\renewcommand{\thefootnote}{\arabic{footnote}}

\section{Введение}
 \label{sec:intro}

Теория возмущений в КХД в области пространственно-подобных
передач 4-импульса
($Q^{2}=-q^2>0$ --- в дальнейшем будем называть эту область евклидовой)
основана на разложениях в ряды по степеням
эффективного заряда
(или эффективной константы связи)
$\alpha_\text{s}(Q^2)$,
который в однопетлевом приближении имеет вид
\begin{eqnarray}
  \alpha_\text{s}(Q^2)
=
  \frac{4\pi}{b_0}~a[L]=\frac{4\pi}{b_0}~\frac{1}{L}~
\label{eq:1}
\end{eqnarray}
с $\displaystyle ~b_0=11-\frac{2}{3}N_f$, ~$L=\ln(Q^2/\Lambda^2)$,
где $\Lambda^2\equiv \Lambda_\text{QCD}^2$,
а ``нормированный''
заряд $a[L]$
удовлетворяет уравнению ренормгруппы (РГ)~\cite{SP53,BS55-rus,BS56}
\begin{eqnarray}
 \label{eq:RG.alpha.s}
  \frac{da[L]}{dL}
   = - a^{2}\left(1
                + c_{1} a^{}
                + c_{2}a^{2}
                + \ldots
            \right) \, .
\end{eqnarray}
Здесь $c_1=b_1/b_0^2$ и $c_2=b_2/b_0^3$ --- вспомогательные параметры
разложения
(см. приложение \ref{App:RG-solution-2L}).
Однопетлевое (при $c_1=c_2=0$) решение этого уравнения есть просто $1/L$
и оно, как видно,
имеет сингулярность в виде полюса в точке $L=0$,
называемую полюсом Ландау.
По этой причине применение теории возмущений
в области малых передач импульса
$Q^2\sim\Lambda^2$ или $L\ll1$
затруднено,
так что характеристики адронных процессов,
рассчитываемые в КХД на партонном уровне в виде
разложений в ряды по степеням эффективного заряда,
не являются всюду хорошо определенными величинами.

В свое время появление такой сингулярности в КЭД~\cite{LAH56,DG05rus},
названной ``призрачной'' из-за отрицательности вычета
в появляющемся полюсе пропагатора,
было проинтерпретировано как указание
на противоречивость квантовой теории поля.
Однако, как было показано в работе~\cite{BS59,BSvvtkp-rus},
это всего лишь свидетельствует о неприменимости теории возмущений в области,
где параметр разложения не мал.
Указание на то, что эта сингулярность не настоящая
следует также из изучения характера этой сингулярности
в высших приближениях.
Рассмотрим, например, двухпетлевое уравнение~(\ref{eq:RG.alpha.s})
c $c_1\neq0$ и $c_2=0$.
Вблизи сингулярности $a[L]\gg1$ и поэтому уравнение
можно переписать в виде $da[L]/dL \approx - c_{1} a^{3}$,
что немедленно определяет поведение эффективного заряда
вблизи сингулярности при $L\ll1$:
\begin{eqnarray}
 \label{eq:alpha.sing}
  a^\text{2-loop}[L] \approx \frac{1}{\sqrt{2\,c_{1}L}}\,.
\end{eqnarray}
Мы видим, что тип сингулярности изменился:
вместо полюса появляется точка ветвления типа $L^{-1/2}$,
а при учете вкладов высших петель степень сингулярности
очевидным образом уменьшается все сильнее
(на самом деле, она просто равна $1/l$,
 где $l$ --- число петель).

Появление таких ``призрачных'' сингулярностей
с теоретической точки зрения противоречит
принципу причинности квантовой теории поля~\cite{BSvvtkp-rus,BLT69-rus},
поскольку делает спектральное представление Челлена--Лемана невозможным.
С другой стороны, оно также осложняет определение
эффективного заряда КХД во временноподобной области ($q^{2}>0$).
С момента появления КХД многие исследователи
пытались определить соответствующий эффективный заряд
в области временноподобных передач 4-импульса
(в дальнейшем будем говорить ``область Минковского''),
который был бы пригоден для описания процессов
типа $e^{+}e^{-}$-аннигиляции в адроны, распадов
кваркониев и $\tau$-лептонов в адроны, и т.~п.
Многие такие попытки, см., например,~\cite{PR81,PRR83,Mar88},
использовали аналитическое продолжение эффективного заряда
из глубоко евклидовой области,
где хорошо работала теория возмущений КХД,
в область Минковского,
где проводились реальные эксперименты:
$\alpha_\text{s}(Q^2)\to\alpha_\text{s}(s=-Q^2)$.

С течением времени стало ясно, что в инфракрасной (ИК) области
малых значений $Q^2$ эффективный заряд
$\alpha_\text{s}(Q^2)$
может иметь устойчивую неподвижную точку и перестать возрастать.
Такое поведение означало бы, что при очень малых импульсах
в окрестности этой точки
цветовые силы насыщаются  и глюоны перестают видеть кварки
как  цветные объекты,
поскольку воспринимают их как целое квазибесцветное состояние.
Так, Корнвел~\cite{Cor82} в модели с конденсацией вихрей
изучил формирование эффективной массы у глюона,
которое приводит к такому насыщению эффективного заряда и
делает его значение конечным в полюсе Ландау.
Похожие попытки предпринимались
при изучении других моделей~\cite{PP79,MS92,GHN93,Sim93},
причем всюду глюон приобретал эффективную массу,
которая играла роль ИК регулятора сингулярного поведения
эффективного заряда в области малых импульсов.
В работе~\cite{Ste99} было показано,
что такая ИК защищенная модель эффективного заряда
может быть связана с формфактором Судакова,
подавляющим излучение мягких глюонов,
так что глюоны с длиной волны,
превышающей некоторую характерную (непертурбативную) длину волны,
не могут различать индивидуальные кварки внутри бесцветного
адронного состояния.

Параллельно с этими работами Радюшкин~\cite{Rad82}\footnote{
Эта работа была опубликована 26 февраля 1982 г.
в виде препринта ОИЯИ, № E2-82-159,
затем послана для публикации в журнал \textit{Physics Letters B},
но была отвергнута редакцией.} и
Красников и Пивоваров~\cite{KP82},
используя интегральное преобразование функции Адлера $D(Q^2)$
в отношение сечений
$R(s)=\sigma\left(e^{+}e^{-}\to~\text{адроны}\right)/
      \sigma\left(e^{+}e^{-}\to\mu^{+}\mu^{-}\right)$,
получили аналитические выражения
для однопетлевого эффективного заряда
(и его первых степеней) сразу в области Минковского
(за подробностями отсылаем заинтересованного читателя
 к работам~\cite{Shi00,BRS00,Shi01}).
Этот тип аналитизации эффективного заряда в КХД во временноподобной области
был переоткрыт позже в подходе пересуммирования фермионных пузырей
Бенеке и Брауном~\cite{BB95},
а также Баль, Бенеке и Брауном~\cite{BBB95},
причем в последней работе в связи приложениями к расчету ширины распада
$\tau\to\nu_{\tau}+$адроны.

Систематический подход, называемый Аналитической Теорией Возмущений (АТВ),
появился и утвердился в последнее десятилетие благодаря работам,
инициированным
Джонсом и Соловцовым, Ширковым и Соловцовым,
Милтоном и Соловцовым~\cite{JS95-349,JS95-357,SS96,MS96,SS97,SS98}.
Главными объектами в этом подходе являются спектральные плотности,
с помощью которых аналитический эффективный заряд
и его целые степени определяются в евклидовой области в виде
дисперсионных интегралов.
Те же спектральные плотности используются для построения
эффективного заряда
и его степеней в области Минковского с помощью дисперсионного
соотношения, связывающего $D$-функцию Адлера
и $R$-отношение~\cite{MS96,MiSol97}.
Эти интегральные преобразования,
названные Ширковым $\hat{R}$- и $\hat{D}$-операциями\footnote{%
Д.~В.~Ширков объясняет такие обозначения двойственно:
преобразование $\hat{D}$ связано с $D$-функцией Адлера
и с работой Докшицера с соавторами~\cite{DMW96},
а преобразование $\hat{R}$ является обратным (reverse) по отношению к
преобразованию $\hat{D}$, но также связано с именами
Радюшкина~\cite{Rad82}, Красникова и Пивоварова~\cite{KP82}.
\label{p.footnote.R.D}}
(см.~следующий раздел),
дают возможность определить одновременно аналитический эффективный заряд
как в евклидовой области значений квадратов передач импульса,
так и в области Минковского.

Вскоре были развиты аналитические и численные методы,
необходимые для расчетов в двух- и трехпетлевых
приближениях~\cite{Mag99,Mag00,KM01,Mag03u,KM03,Mag05}.
Этот подход был применен для расчета характеристик ряда адронных процессов,
в том числе ширины инклюзивного распада
$\tau$-лептона в адроны~\cite{JS95-349,MSS97,MSSY00},
зависимости от выбора схемы и масштаба перенормировок
в правилах сумм Бьоркена~\cite{MSS98} и Гросса--Левеллин~Смита~\cite{MSS98GLS},
ширины распада $\Upsilon$-мезона в адроны~\cite{SZ05}, и т.~п.
Более того, АТВ была применена также для анализа процессов,
в которых имеется не один, а два масштаба, а именно:
переходного $\gamma^*\gamma\to\pi$-формфактора~\cite{SSK99,SSK00}
и электромагнитного формфактора пиона
в $O(\alpha_\text{s}^2)$-порядке~\cite{SSK99,SSK00,BPSS04}.

Суммируя, можно сказать, что этот аналитический подход
(см. обзоры~\cite{SS99,Shi00,SS06})
дает достаточно разумное описание адронных величин в КХД,
хотя имеются как скептические мнения на этот счет~\cite{GIZ01},
так и альтернативные подходы к трактовке сингулярности
эффективного заряда в КХД
\cite{Piv91a,LeDiPi92,DMW96,Dok98,CMNT96,Gru97,GGK98,Piv01,Nes03,Ale05,CV06,Rac06} ---
в частности,
в отношении глубоко инфракрасной области $Q^2\leq \Lambda^2$,
где, в конечном счете,
может стать важным появление ненулевых адронных масс~\cite{DPTar89,NP04}.
Кроме того, подход АТВ имеет строгие ограничения
в приложении к описанию КХД-процессов,
поскольку неявно подразумевает, что единственными величинами,
подлежащими аналитизации,
являются эффективный заряд $\alpha_\text{s}(Q^2)$
и его целые степени $[\alpha_\text{s}(Q^2)]^n$.

Но, как было показано в~\cite{DR81,BT87,Mul98,MNP99a,MNP01a,MMP02},
трехточечные функции, такие, например,
как используемые при описании электромагнитного формфактора пиона или
переходного $\gamma^*\gamma\to\pi$-формфактора,
в следующем за ведущим порядке КХД-теории возмущений
содержат логарифмические вклады,
связанные с дополнительным масштабом факторизации.
Эти логарифмы, хотя и не влияют на сингулярность Ландау,
видоизменяют пертурбативную спектральную плотность,
используемую в АТВ для проведения аналитизации.
Именно этот факт привел Караникаса и Стефаниса~\cite{KS01,Ste02}
к предложению расширить концепцию аналитизации (дисперсионного представления)
с уровня эффективного заряда и его степеней
на уровень КХД-амплитуд в целом.\footnote{
Более точно, в работах~\cite{KS01,Ste02}
изучалась аналитизация объектов типа
$\int_{0}^{1}\!dx\!\int_{0}^{1}\!dy\,
  \alpha_\text{s}\left(Q^{2}xy\right) f(x)f(y)$,
которые можно трактовать как эффективный учет
логарифмических вкладов,
появляющихся в следующем за ведущим порядке
теории возмущений КХД.}
Применение такой концепции аналитизации требует
обобщения исходной АТВ на дробные степени эффективного заряда,
а также на их произведения со степенями логарифмов.

Стоит отметить, что дробные степени эффективного заряда
были рассмотрены неявно в~\cite{GLK84,BKM01}.\footnote{Здесь интересно
вспомнить ранние попытки изучать спектральную плотность,
отвечающую дробной степени эффективного заряда в КЭД~\cite{BLS59-rus}.
Подобная спектральная плотность была заново переоткрыта Оэме в КХД~\cite{Oeh90}.}
Необходимое для КХД обобщение АТВ на дробные степени эффективного заряда,
названное Дробно-Аналитической Теорией Возмущений (ДАТВ),
было недавно проведено в~\cite{BMS05,BMS06}
(в качестве краткого введения --- см.~\cite{AB07yalta}),
а затем применено в~\cite{BKS05} для анализа факторизуемого вклада
в электромагнитный формфактор пиона (см.~также недавний обзор~\cite{SK08}).
Кардинальным преимуществом ДАТВ в этом случае
стало уменьшение зависимости пертурбативных результатов
от выбора значения масштаба факторизации.
Это напоминает результаты применения АТВ
для анализа того же формфактора пиона
в $O(\alpha_\text{s}^2)$-порядке,
где результаты также практически перестали зависеть от
выбора схемы перенормировки и ее масштаба~\cite{BPSS04}
(краткое изложение см. в \cite{SBMPS04,Ste04mon,AB04pg,AB04ca,AB07yalta}).
В работе~\cite{BMS06} в рамках ДАТВ для области Минковского
было проведено изучение процесса распада хиггсовского бозона
в пару $b\bar{b}$-кварков.
Хотя область значений энергии в системе центра масс,
интересная для эксперимента,
здесь очень велика,
$\sqrt{s}\approx100$~ГэВ,
т.~е. здесь не следует ожидать заметных отличий
ДАТВ от стандартной ТВ КХД,
тем не менее, интересно было проверить
влияние переноса $\pi^2$-вкладов из коэффициентов,
посчитанных в стандартном подходе Четыркиным с соавторами~\cite{Che96,ChKS97,BCK05}
в $O(\alpha_\text{s}^{4})$-порядке,
в аналитизированный эффективный заряд ${\mathfrak A}_1$
и его степени ${\mathfrak A}_n$
(более точно, в ${\mathfrak A}_{1+\nu}$ и ${\mathfrak A}_{n+\nu}$,
 см. подробнее в~\cite{BMS06} и в разделе \ref{sec:Higgs.width}).
Результаты работы показали,
что отличия действительно невелики, меньше или порядка $2\%$.
\bigskip

В этой работе мы попытаемся последовательно рассказать
об основных элементах глобальной версии ДАТВ,
в которой учитываются пороги тяжелых кварков,
чтобы дать читателю возможность применять ее на практике
для расчетов реальных процессов.
При этом сама ДАТВ будет освещена достаточно сжато,
поскольку более полный обзор оснований ДАТВ
будет дан в готовящейся совместной публикации авторов ДАТВ~\cite{BMS05,BMS06}.
План изложения таков:
\begin{itemize}
  \item[\ding{183}] Во втором разделе дается краткий обзор
   основных положений АТВ на примере расчета $D$-функции Адлера
   в пространственноподобной области
   и ее двойника во временноподобной области ---
   $R$-отношения $e^+e^-$-аннигиляции.
   Мы кратко обсуждаем формализм АТВ в однопетлевом приближении
   для случая фиксированного числа ароматов,
   затем переходим к вопросу об учете порогов тяжелых кварков в АТВ
   и построению так называемой ``глобальной'' АТВ.
  \item[\ding{184}] В следующем разделе обсуждается неполнота АТВ
   и в качестве способа пополнения предлагается ДАТВ.
   Приводятся основные формулы ДАТВ, отвечающей $N_f=3$,
   и обсуждается построение глобальной версии ДАТВ,
   в которой учитываются пороги тяжелых кварков.
  \item[\ding{185}] Четвертый раздел посвящен
   расчету факторизуемой части формфактора пиона в глобальной ДАТВ.
   Кратко обсуждаются результаты полученные в АТВ (существенное снижение
   зависимости результатов от выбора схемы и масштаба перенормировки),
   после чего рассматривается зависимость результатов
   от масштаба факторизации в глобальной ДАТВ,
   а также переход в область Минковского и роль дисперсионного представления
   в этом переходе.
  \item[\ding{186}] В пятом разделе рассмотрен расчет
   ширины распада $H^0\to\bar{b}b$ в глобальной ДАТВ.
   Показывается, что в то время как результаты ДАТВ с $N_f=5$,
   отвечающие  переносу $\pi^2$-вкладов
   из коэффициентов теории возмущений в аналитические эффективные заряды,
   прекрасно согласуются с результатами стандартной теории возмущений
   уже на уровне двухпетлевого приближения,
   результат глобальной ДАТВ отличается от них на уровне 14\%.
  \item[\ding{187}] В следующем разделе обсуждается
   суммирование рядов $\sum_{n}d_n{\mathcal A}_{n}[L]$ теории возмущений
   в АТВ и ДАТВ:
   оказывается, что в случае однопетлевой АТВ
   такое суммирование можно провести точно~\cite{MS04} и выразить ответ
   в виде интеграла от ${\mathcal A}_{1}[L-t]$ по $t$ с весом $P(t)$,
   определяемыми коэффициентами пертурбативного ряда $d_n$.
   Мы показываем, что аналогичное суммирование можно провести
   и в случае однопетлевой ДАТВ.
   Получены все необходимые формулы для учета порогов кварков
   в этих методах суммирования для АТВ и ДАТВ.
  \item[\ding{74}] В заключении суммированы основные выводы работы,
   а важные технические детали собраны в пяти приложениях.
\end{itemize}

Прежде чем приступать к собственно изложению,
сделаем пояснение об используемых обозначениях.
По историческим причинам,
чтобы иметь прямую связь с работами~\cite{BMS05,BKS05,BMS06}
и чтобы упростить основные формулы,
в тех подразделах,
где мы будем обсуждать случаи фиксированного значения числа
активных кварков (или флейворов)
$N_f$,
мы будем считать основным элементом теории
нормализованный по Ширкову~\cite{Shi98} и Михайлову~\cite{MS04}
эффективный заряд
$a(Q^2)$=$b_0\,\alpha_\text{s}(Q^2)/(4\pi)$
и аналитические образы будут строиться для степеней
именно этого объекта:
\begin{subequations}
\begin{eqnarray}
 \label{eq:Non.Glo.Couplings}
 {\mathcal A}^{}_{n} =
   \textbf{A}_\text{E}\left[a^{n}_{}\right]\,;\qquad
 {\mathfrak A}^{}_{n} =
   \textbf{A}_\text{M}\left[a^{n}_{}\right]\,.
\end{eqnarray}
При переходе же к обсуждению глобальной версии теории,
когда $Q^2$ (или $s$) меняются во всей области значений
$[0,\infty)$ и $N_f$ эффективно становится зависящим от $Q^2$ (или $s$),
мы будем считать
основным объектом теории сам эффективный заряд
$\alpha_\text{s}(Q^2)$ и строить аналитизацию степеней этого объекта,
т.~е. мы будем считать
\begin{eqnarray}
\label{eq:Glo.Couplings}
 {\mathcal A}^\text{\tiny glob}_{n} =
   \textbf{A}_\text{E}\left[\alpha_\text{s}^{n}\right]\,;\qquad
 {\mathfrak A}^\text{\tiny glob}_{n} =
   \textbf{A}_\text{M}\left[\alpha_\text{s}^{n}\right]\,.
\end{eqnarray}
Чтобы различать два типа величин,
мы ввели верхний индекс $^\text{\tiny glob}$.
Нам также понадобятся заряды со стандартной нормировкой
при фиксированном значении $N_f$:
\begin{eqnarray}
 \label{eq:Bar.Couplings}
  \bar{\mathcal A}_{n}(Q^2;N_f)
  &\equiv& \frac{{\mathcal A}_{n}(Q^2)}{\beta_f^n}\,;\quad
  \bar{\mathfrak A}_{n}(s;N_f)
  \ =\     \frac{{\mathfrak A}_{n}(s)}{\beta_f^n}\,;\quad
  \beta_f
  \ =\ \frac{b_0(N_f)}{4\pi}\,,~
 \label{eq:A.U.n}
\end{eqnarray}
\end{subequations}
которые обладают дисперсионными интегральными представлениями
со спектральными плотностями
$\bar{\rho}_n(\sigma;N_f)={\rho}_n(\sigma)/{\beta_f^n}$.

Кроме того,
мы часто будем обсуждать эффективные заряды
как функции не $Q^2$ или $s$,
а логарифмов $L=\ln(Q^2/\Lambda^2)$ или $L_s=\ln(s/\Lambda^2)$.
В таких случаях мы будем использовать те же обозначения зарядов,
но аргумент ставить не в круглых скобках, а в квадратных,
т.~е. писать вместо
$\alpha_s^\nu(Q^2)$, ${\mathcal A}_{\nu}(Q^2)$ и ${\mathfrak A}_{\nu}(Q^2)$
следующие выражения:
$\alpha_s^\nu[L]$, ${\mathcal A}_{\nu}[L]$ и ${\mathfrak A}_{\nu}[L_s]$.

\section{АНАЛИТИЧЕСКАЯ ТЕОРИЯ ВОЗМУЩЕНИЙ}
 \label{sec:2}
В этом разделе мы обсуждаем основные положения и элементы  АТВ,
следуя в основном работам Ширкова и Соловцова~\cite{Shi98,SS99}.

\begin{figure}[bht]
 \centerline{\includegraphics[width=0.3\textwidth]{
  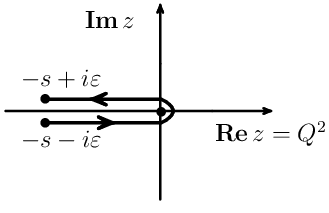}
  \vspace{-3mm}}
   \caption{Контур интегрирования для $\hat{R}$-преобразования
    в (\protect\ref{eq:R-operation}).
    \label{fig:contour}}
\end{figure}

Как уже говорилось во введении,
изначальной мотивацией введения новых эффективных зарядов
было желание связать рассчитываемую в евклидовой области
$D$-функцию Адлера
и величину
$R_{e^+e^-}=\sigma(e^+e^-\to\text{адроны})/\sigma(e^+e^-\to\mu^+\mu^-)$,
измеряемую в минковской области.
Для определения функции Адлера рассмотрим поляризационный оператор
\begin{eqnarray}
 \label{eq:Pi.mu.nu}
  \Pi_{\mu\nu}(q)
   = i\int e^{iqx}\,
    \langle|T\{J_\mu(x)J_\nu(0)\}|0\rangle\,
     d^{4}x
\end{eqnarray}
векторных кварковых токов
$J_\mu(x)=\bar{\psi}(x)\gamma_{\mu}\psi(x)$.
В киральном пределе,
когда массы легких кварков равны нулю,
эта функция зависит только
от одной переменной
$Q^2 =-q^2$ ($\ge0$ для пространственноподобных $q^2$)
\begin{eqnarray}
 \label{eq:Pi.Q^2}
  \Pi_{\mu\nu}(q)
   =\left(q_\mu q_\nu -g_{\mu\nu}q^2\right)
    \Pi(Q^2)\,.
\end{eqnarray}
В КХД поляризационный оператор $\Pi(Q^2)$
удовлетворяет дисперсионному соотношению
с одним вычитанием,
причем обычно вычитание производят в точке $Q^2=0$:
\begin{eqnarray}
 \label{eq:Pi.Q^2.Disp.Rel}
  \Pi(Q^2)
   = \Pi(0)
   - Q^2\!
     \int_0^{\infty}\!\!
      \frac{R(\sigma)\,d\sigma}
           {\sigma\left(\sigma+Q^2\right)}
\end{eqnarray}
с $R(\sigma)=\textbf{Im}\,\Pi(\sigma)/\pi$.
Функция Адлера~\cite{Adler74} определяется в виде логарифмической производной
поляризационного оператора $\Pi(Q^2)$,
т.~е.
\begin{eqnarray}
 \label{eq:Adler.Pi.Q^2}
  D(Q^2)=-Q^2\,\frac{\Pi(Q^2)}{dQ^2}\,,
\end{eqnarray}
так что дисперсионное представление для нее
\begin{subequations}
\begin{eqnarray}
  D(Q^2)
   = Q^2
    \int_0^{\infty}\!
     \frac{R(\sigma,\mu^2)}
          {(\sigma+Q^2)^2}\,
       d\sigma
 \label{eq:D.R}
\end{eqnarray}
не содержит вычитательной  константы $\Pi(0)$
и инвариантна относительно преобразований
ренормализационной
группы (РГ).
Функция $R(s)$, пропорциональная мнимой части
$\textbf{Im}\,\Pi(s)$,
в 1-петлевом приближении КЭД связана
с отношением сечений
$e^+e^-$-аннигиляции в адроны и в $\mu^+\mu^-$.
Из дисперсионного представления (\ref{eq:D.R}),
рассматриваемого в комплексной плоскости
переменной $z$,
которая на действительной положительной полуоси
совпадает с $Q^2$,
и связи (\ref{eq:Adler.Pi.Q^2})
немедленно следует
обратное представление
\begin{eqnarray}
 R(s)
  = \frac{1}{2 \pi i}
     \int_{-s-i\varepsilon}^{-s+i\varepsilon}\!
      \frac{D(z)}
           {z}\,dz\,,
 \label{eq:R.D}
\end{eqnarray}
\end{subequations}
где интеграл берется вдоль контура в комплексной плоскости $z$,
показанного на рис.~\ref{fig:contour}.

Обе функции изучаются в стандартной теории возмущений КХД:
\begin{subequations}
\begin{eqnarray}
 D(Q^2,\mu^2)
  &=& \sum_{n} d_n(Q^2/\mu^2)~a_\text{s}^n(\mu^2)
     \stackrel{\mu^2=Q^2}{\longrightarrow}
 D(Q^2)
  = \sum_{n}~d_n~a_\text{s}^n(Q^2)\,,
 \label{eq:D}\\
 R(s,\mu^2)
  &=& \sum_{m} r_m(s/\mu^2)~a_\text{s}^m(\mu^2)
     \stackrel{\mu^2=s}{\longrightarrow}
 R(s)
  = \sum_{m} ~r_m~a_\text{s}^m(s)\,,
\label{eq:R}
\end{eqnarray}
\end{subequations}
причем эффективные заряды
$a_\text{s}^m(\mu^2)$, $a_\text{s}^m(Q^2)$,
$a_\text{s}^m(\mu^2)$  и $a_\text{s}^m(s)$
определяются ренормгрупповым уравнением.
Однако, разложив по теории возмущений левые части
(\ref{eq:D}) и(\ref{eq:R}),
мы можем получить соотношения,
связывающие степени $\ln(s/\mu^2)$ и $\ln(Q^2/\mu^2)$
в коэффициентах $r_m(s/\mu^2)$ и $d_n(Q^2/\mu^2)$,
в то время как степени $a_\text{s}(\mu^2)$
остаются числовыми параметрами.
Полагая $\mu^2=Q^2$ в правой части (\ref{eq:D})
(или $\mu^2=s$ в (\ref{eq:R})),
коэффициенты $d_n=d_n(1)$ (аналогично $r_n=r_n(1)$)
становятся числовыми константами,
а степени зарядов $a^n_\text{s}(Q^2)$
(соответственно, $a^m_\text{s}(s)$)
теперь подвергаются интегральным преобразованиям (см.~ниже).

В случае, когда  эти заряды являются стандартными эффективными зарядами
пертурбативной КХД,
эта связь нарушается в любом петлевом порядке
из-за наличия сингулярности Ландау в евклидовой области.
Естественно возникает вопрос:
можно ли построить аналитические версии обоих типов зарядов,
для которых функции $D$ и $R$ могут быть
связаны дисперсионным соотношением?
АТВ как раз и дает \textit{положительный} ответ на этот вопрос,
вводя в рассмотрение
нестепенные (функциональные)
разложения~\cite{SS97,MS96,MiSol97,Shi98,SS99,SS06,Shi00b,Shi00,Shi01}.

Аналитические образы степеней нормированного эффективного заряда,
см.\ (\ref{eq:1}),
в евклидовой области определяются
с помощью формальной линейной операции $\textbf{A}_\text{E}$:
\begin{eqnarray}
 \textbf{A}_\text{E}\left[a^{n}_{(l)}\right]
  &=& {\mathcal A}^{(l)}_{n}\,,
    \text{~~где~~~}
   {\mathcal A}^{(l)}_{n}(Q^2)
  \equiv
   \int_0^{\infty}\!
    \frac{\rho^{(l)}_n(\sigma)}
         {\sigma+Q^2}\,
       d\sigma
\label{eq:A.E}
\end{eqnarray}
и спектральная плотность задается мнимой частью
пертурбативной величины:
\begin{eqnarray}
\label{eq:rho}
\rho^{(l)}_n(\sigma)
   \equiv \frac{1}{\pi}\,
           \textbf{Im}\,\big[a^{n}_{(l)}(-\sigma)\big]\,.
\end{eqnarray}
Степень (индекс) $n$ имеет здесь только положительные целые значения,
а петлевой порядок приближения указывается символом $l$ в скобках.

Аналогично, аналитические образы степеней нормированного эффективного заряда
в области Минковского определяются
с помощью другой линейной операции,
$\textbf{A}_\text{M}$,
а именно,
\begin{eqnarray}
 \textbf{A}_\text{M}\left[a^{n}_{(l)}\right]
  = {\mathfrak A}^{(l)}_{n}\,,
  \text{~~где~~~}
  {\mathfrak A}^{(l)}_{n}(s)
  \equiv
    \int_s^{\infty}\!
     \frac{\rho^{(l)}_n(\sigma)}
          {\sigma}\,
      d\sigma\,.
\label{eq:A.M}
\end{eqnarray}
Эти ``аналитизирующие'' операции определяются,
в свою очередь,
с помощью следующих двух интегральных преобразований
(мы сохраняем здесь терминологию Ширкова, см., например,~\cite{Shi00,Shi01,Shi05},
 а также примечание на стр.~\pageref{p.footnote.R.D}):
$\hat{D}$ --- преобразование из временноподобной области
в пространственноподобную область
\begin{eqnarray}
  \hat{D}\big[{\mathfrak A}^{(l)}_{n}\big]
  &=&  {\mathcal A}^{(l)}_{n}\,,
    \text{~~где~~~}
   {\mathcal A}^{(l)}_{n}(Q^2)
  \equiv
   Q^2 \int_0^{\infty}\!
     \frac{{\mathfrak A}^{(l)}_{n}(\sigma)}
          {\big(\sigma+Q^2\big)^2}\,
      d\sigma\,;
\label{eq:D-operation}
\end{eqnarray}
и $\hat{R}$ --- обратное преобразование
\begin{eqnarray}
  \hat{R}\big[{\mathcal A}^{(l)}_{n}\big]
  &=&
  {\mathfrak A}^{(l)}_{n}\,,
    \text{~~где~~~}
  {\mathfrak A}^{(l)}_{n}(s)
  \equiv \frac{1}{2\pi i}
   \int_{-s-i\varepsilon}^{-s+i\varepsilon}\!
    \frac{{\mathcal A}^{(l)}_{n}(\sigma)}
          {\sigma}\,
      d\sigma
\label{eq:R-operation}
\end{eqnarray}
и интеграл берется вдоль контура в комплексной плоскости $z$,
показанных на рис.~\ref{fig:contour}.
Заметим, что эти операции связаны  теперь друг с другом
соотношением взаимности
\begin{eqnarray}
 \label{eq:reciproc}
 \hat{D}\hat{R} = \hat{R}\hat{D} = 1\,,
\end{eqnarray}
справедливым для всего набора
$\big\{{\mathcal A}_n,{\mathfrak A}_n\big\}$
и в любом петлевом порядке
именно в силу регулярности новых эффективных зарядов
${\mathcal A}_n(Q^2)$ и ${\mathfrak A}_n(s)$.

\begin{figure}[t]\vspace*{0mm}
 \centerline{\includegraphics[width=0.45\textwidth]{
  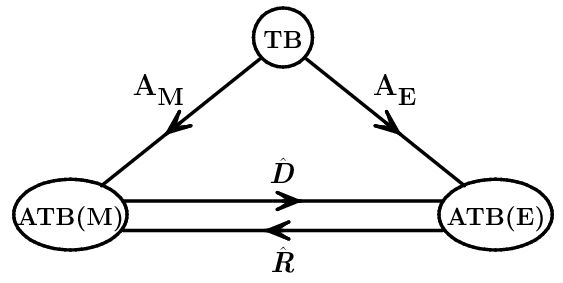}
  \vspace*{0mm}}
   \caption{Схема аналитизации как переход от стандартной
    теории возмущений (ТВ) к аналитической ТВ в областях
    Минковского (АТВ(М)) и Евклида (АТВ(Е)).\label{fig:APT-scheme}}
\end{figure}
Схематически операции $\textbf{A}_\text{E}$ и $\textbf{A}_\text{M}$,
определяющие аналитические эффективные заряды
в пространственно- ($Q^2>0$) и
временноподобной ($s>0$) областях,
представлены графически на рис.~\ref{fig:APT-scheme}.
Логика ``аналитизации'' приводит к схожим выражениям
для разложений КХД-амплитуд, зависящих от единственного масштаба $Q^2$,
и для их продолжений из евклидовой области в область Минковского.

В качестве примера рассмотрим $D$-функцию Адлера
в правой части (\ref{eq:D}),
которая разлагается в ряд по степеням $a_{(l)}^n(Q^2)$.
Применение к ней операции $\textbf{A}_\text{E}$
в соответствии с правой стрелкой рис.~\ref{fig:APT-scheme}
переводит ее в нестепенной ряд~\cite{Shi98,Shi00}
${\cal D}_\text{A}(Q^2)$
в евклидовой области:
\begin{eqnarray}
  D(Q^2)
  = \sum_{n}\,d_n\,a_{(l)}^n(Q^2)\, \Rightarrow
     \textbf{A}_\text{E}\left[D \right]
  \equiv   {\cal D}_\text{A}\,,
    \text{~где~~}
  {\cal D}_\text{A}(Q^2)
  = \sum_{n}\,d_n\,{\mathcal A}^{(l)}_{n}(Q^2)\, .
\label{eq:cal D}
\end{eqnarray}

Затем мы можем применить операцию $\hat R$,
задаваемую формулой (\ref{eq:R-operation}),
см.~нижнюю линию на рис.~\ref{fig:APT-scheme},
и получить величину ${\cal R}(s)$ в области Минковского
также в виде нестепенного разложения:\
\begin{eqnarray}
  \hat{R}\left[{\cal D}_\text{A} \right]
  \equiv {\cal R}\,,
    \text{~где~~}
  {\cal R}(s)
  = \sum_{n}\,d_n\,{\mathfrak A}^{(l)}_{n}(s)\, .
\label{eq:cal R}
\end{eqnarray}
С другой стороны, то же самое выражение для функции ${\cal R}(s)$
может быть получено~\cite{Rad82,KP82} в соответствии с левой стрелкой
рис.~\ref{fig:APT-scheme}
с помощью операции $\textbf{A}_\text{M}$:
\begin{eqnarray}
 D(Q^2)
 = \sum_{n}\,d_n\,a_{(l)}^n(Q^2)\,
 \Rightarrow
  \textbf{A}_\text{M}\left[D\right]
 = {\cal R}\,,
    \text{~где~~}
  {\cal R}(s)
 = \sum_{n}\,d_n\,{\mathfrak A}^{(l)}_{n}(s)\,.
\label{eq: DD}
\end{eqnarray}

 \subsection{Однопетлевая АТВ ($N_f$ фиксировано)}
  \label{subsec:APT.Nf3.1L}
Рассмотрим применение этого формализма в однопетлевом приближении с
фиксированным числом флейворов $N_f$.
Пользуясь простым видом эффективного заряда (\ref{eq:1})
и рецептом построения спектральной плотности (\ref{eq:rho}),
мы получаем очень простое выражение для
однопетлевой спектральной плотности
($L_\sigma=\ln\left(\sigma/\Lambda^2\right)$):
\begin{eqnarray}
 \rho_{1}^{(1)}(\sigma)
  = \frac{1}{\pi}\,\textbf{Im}\,\frac{1}{L_\sigma-i\pi}
  = \frac{1}{L^2_\sigma+\pi^2}\,.
 \label{eq:SpDen.1L.1}
\end{eqnarray}
Подстановка его в (\ref{eq:A.E}) и (\ref{eq:A.M})
дает однопетлевые аналитические эффективные заряды
${\mathcal A}^{(1)}_{1}$~\cite{SS96,SS97} и
${\mathfrak A}^{(1)}_{1}$~\cite{Rad82,JS95-349,JS95-357,MS96}\footnote{%
Заметим, что $\arccos(L/\sqrt{L^2+\pi^2})=\arctg(\pi/L)$ при $L>0$,
так что формула, полученная в~\cite{Rad82},
совпадает с (\ref{eq:U_1}).}
\begin{eqnarray}
  {\mathcal A}^{(1)}_{1}[L]
  &=&\frac1{L}- \frac1{e^L -1}\,;
 \label{eq:A_1}\\
{\mathfrak A}^{(1)}_{1}[L_s]
  &=& \frac{1}{\pi}\,
       \arccos\left(\frac{L_{s}}{\sqrt{L_{s}^2+\pi^2}}\right)\,,
 \label{eq:U_1}
\end{eqnarray}
где
\begin{eqnarray}
 \label{eq:L.s.L.sigma}
  L \,=\, \ln\left(Q^2/\Lambda^2\right)\,,\qquad
  L_s \,=\, \ln\left(s/\Lambda^2\right)\,.
\end{eqnarray}
Из этих уравнений мы заключаем,
что на однопетлевом уровне ``аналитизация'' ($\textbf{A}_\text{E}$)
в евклидовой области
означает вычитание полюса Ландау,
в то время как в области Минковского аналогичная операция
($\textbf{A}_\text{M}$)
означает суммирование $\pi^2/L^2$-вкладов во всех порядках:
${\mathfrak A}^{(1)}_{1}[L_s]$ при $L\gg1$
имеет разложение
$1/L(1-\pi^2/(3L^2)+\ldots)\,.$

Спектральные плотности высших степеней эффективного заряда,
определяемые в соответствии с (\ref{eq:rho}),
также можно записать достаточно компактно:
\begin{eqnarray}
 \rho_{n}^{(1)}(\sigma)
  = \frac{\sin[n\,\varphi_{(1)}[L_\sigma]]}
           {R_{(1)}^{\,n}[L_\sigma]}\,,~~
 \varphi_{(1)}[L_\sigma]
  = \arccos\left(\frac{L_\sigma}{\sqrt{L_\sigma^2+\pi^2}}\right)\,,~~
 R_{(1)}[L_\sigma]
  = \sqrt{L_\sigma^2+\pi^2}\,.~~~
 \label{eq:SpDen.1L.n}\\
\end{eqnarray}
Они при $n\geq2$ обладают рекуррентным свойством:
\begin{eqnarray}
 \rho_{n}^{(1)}(\sigma)
  = \frac{1}{n-1}\,
     \left(\frac{-d}{dL_\sigma}\right)
      \rho_{n-1}^{(1)}(\sigma)
  = \frac{1}{(n-1)!}\,
     \left(\frac{-d}{dL_\sigma}\right)^{n-1}
      \rho_{1}^{(1)}(\sigma)\,,
\label{eq:SpDen.1L.d.dL}
\end{eqnarray}
следующим из очевидного соотношения
$$ \frac{1}{(L-i\pi)^n}=%
   \frac{1}{n-1}\,%
   \left(\frac{-d}{dL}\right)\frac{1}{(L-i\pi)^{n-1}}=%
   \frac{1}{(n-1)!}\,%
   \left(\frac{-d}{dL}\right)^{n-1}\frac{1}{L-i\pi}\,.$$
Это свойство позволяет
записать выражения для аналитических образов
$n$-х степеней эффективных зарядов через сами аналитические заряды
в виде их $n$-х производных:
\begin{subequations}
 \label{eq:generator}
\begin{eqnarray}
  {\mathcal A}^{(1)}_{n}[L]
  &=& \frac{1}{(n-1)!}\,
       \left(\frac{-d}{dL}\right)^{n-1}
        {\mathcal A}^{(1)}_{1}[L]\,;
 \label{eq:A_n}\\
{\mathfrak A}^{(1)}_{n}[L_s]
  &=& \frac{1}{(n-1)!}\,
       \left(\frac{-d}{dL_s}\right)^{n-1}
        {\mathfrak A}^{(1)}_{1}[L_s]
 \ =\ \frac{\sin\left[(n-1)\arccos\left(L_s/\sqrt{L_s^2+\pi^2}\right)\right]}
           {(n-1)\,\pi\,\left[\sqrt{L_s^2+\pi ^2}\right]^{n-1}}
 \label{eq:U_n}~~~~~~~~~\\
  &=& \frac{\textbf{Im}{}\left[(L_s+i\pi)^{n-1}\right]}{(n-1)\,\pi\,(L_s^2+\pi^2)^{n-1}}
  ~~~\text{для}~~~n>1, n\in\mathbb{N} \,.
 \label{eq:U_n_Im}
\end{eqnarray}
\end{subequations}
Отметим,
что явное выражение для правой части в (\ref{eq:U_n})
было получено в~\cite{BMS06}
с помощью преобразования Лапласа,
см. раздел~III, формулу (3.2).

\begin{figure}[hbt]
 \centerline{\includegraphics[width=0.45\textwidth]{
  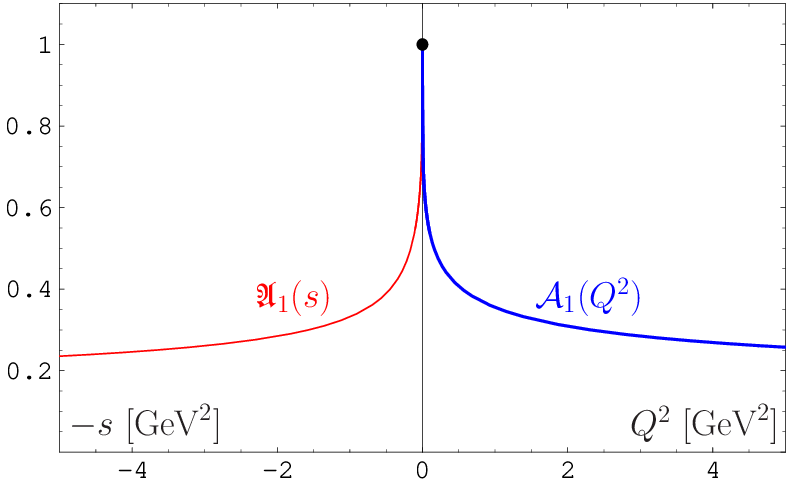}~~~
             \includegraphics[width=0.45\textwidth]{
  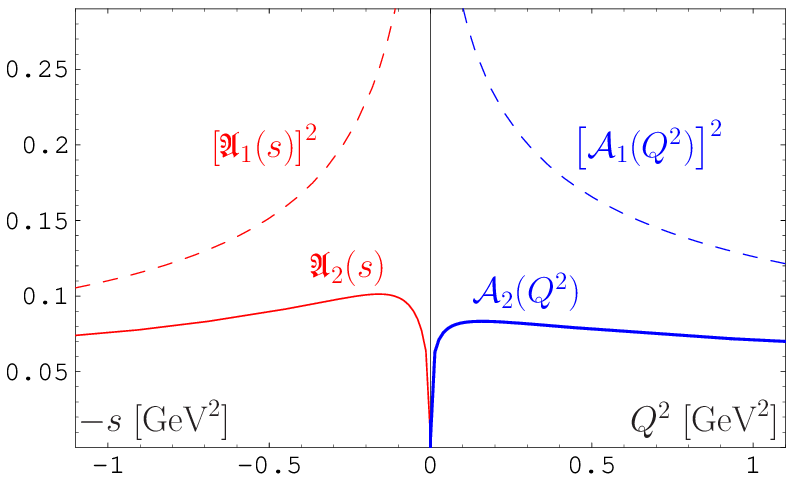}}
  \caption{Поведение однопетлевых аналитических зарядов
   ${\mathfrak A}_{1}(s)$ и ${\mathcal A}_{1}(Q^2)$ (слева) и
   ${\mathfrak A}_{2}(s)$ и ${\mathcal A}_{2}(Q^2)$ (справа).
   На правой панели штрихованными линиями показаны для сравнения
   квадраты соответствующих аналитических зарядов.
 \label{fig:U1_A1}}
\end{figure}
На рис.~\ref{fig:U1_A1} мы сравниваем поведение аналитических зарядов
в простран\-ственно- и временноподобных областях.
На левой панели показаны кривые для
${\mathfrak A}_{1}(s)$ и ${\mathcal A}_{1}(Q^2)$,
и для лучшей видимости проставлена черная точка,
указывающая предельное значение обоих зарядов при $Q^2=s=0$.
На правой панели кроме аналитических образов квадратов зарядов
${\mathfrak A}_{2}(s)$ и ${\mathcal A}_{2}(Q^2)$ (сплошные кривые)
для сравнения мы показываем штрихованными линиями
как ведут себя квадраты аналитизированных зарядов:
видно, что при $Q^2<1$~ГэВ$^2$ и $s<1$~ГэВ$^2$
аналитические квадраты существенно меньше,
причем в точке $Q^2=s=0$ они и вовсе обращаются в нуль,
в то время как ${\mathfrak A}_{1}^2(0)={\mathcal A}_{1}^2(0)=1$.
На обеих панелях рис.~\ref{fig:U1_A1}
виден эффект `искажающего зеркала'~\cite{SS97,SS06}:
при переходе от области Минковского к евклидовой области
зависимость заряда от аргумента меняется только слегка,
причем имеются 3 точки, в которых значения зарядов в обоих
областях совпадают:\\
а) $\displaystyle{\mathfrak A}_{1}(0)={\mathcal A}_{1}(0)=1$,
  причем $\displaystyle{\mathfrak A}_{1}(s)={\mathcal A}_{1}(s)=1+\frac{1}{\ln[s/\Lambda^2]}$
   при $s\to0$;\\
б) ${\mathfrak A}_{1}(\Lambda^2)={\mathcal A}_{1}(\Lambda^2)=1/2$;\\
в) ${\mathfrak A}_{1}(\infty)={\mathcal A}_{1}(\infty)=0$,
  причем $\displaystyle{\mathfrak A}_{1}(s)={\mathcal A}_{1}(s)=\frac{1}{\ln[s/\Lambda^2]}$
   при $s\to\infty$.\\
В то же время для аналитизированных квадратов
таких точек уже четыре:
кроме точек а) и в)  появляются еще 2 точки,
по одной между а) и б) и между б) и в),
причем в точке а) значения функций равны уже не $1$, а $0$.
Появление при переходе от $n$ к $n+1$ еще одной точки,
в которой функция разности
$\Delta_{n}(s)={\mathfrak A}_{n}(s)-{\mathcal A}_{n}(s)$
обращается в ноль,
вполне понятно, если вспомнить про рекуррентные соотношения
(\ref{eq:generator}), благодаря которым
$$\Delta_{n+1}[L] = \frac{1}{n}\,
   \left(\frac{-d}{dL}\right)\,
    \Delta_{n}[L]\,.$$
Функция $\Delta_{1}[L]$ имеет 3 нуля:
на концах ($L_{-}=-\infty$ и $L_{+}=+\infty$) и в центре ($L_{0}=0$).
Значит $\Delta_{2}[L]$ в добавок к двум нулям на концах
($L_{-}$ и $L_{+}$) будет иметь нули
в точках максимумов и минимумов функции $\Delta_{1}[L]$,
которые достигаются между ее нулями,
т.~е. один между $L_{-}$ и $L_{0}$,
другой между $L_{0}$ и $L_{+}$.
Рассуждая аналогично, мы убеждаемся,
что  $\Delta_{n}[L]$ обращается в ноль в $n+2$ точках.

Стоит отметить также следующие свойства однопетлевых функций
${\mathfrak A}_{1}[L]$ и ${\mathcal A}_{1}[L]$~\cite{SS97},
рассматриваемых как функции не $Q^2$ и $s$,
а их логарифмов (см. (\ref{eq:L.s.L.sigma}) и конец введения):
\begin{eqnarray}
  \label{eq:A.m.U.m.-L}
 {{\mathcal A}_{m}[-L]\choose{\mathfrak A}_{m}[-L]}
   &=& (-1)^{m}{{\mathcal A}_{m}[L]\choose{\mathfrak A}_{m}[L]}
       \text{~~для~~} m\geq2\,, m\in\mathbb{N}\,;\\
  \label{eq:A.m.U.m.infty}
 {\mathcal A}_{m}[\pm\infty]
   &=& {\mathfrak A}_{m}[\pm\infty]
  \ =\ 0 \text{~~для~~} m\geq2\,,\ m\in\mathbb{N}\,.
\end{eqnarray}

\begin{figure}[h]
 \centerline{\includegraphics[width=0.45\textwidth]{
  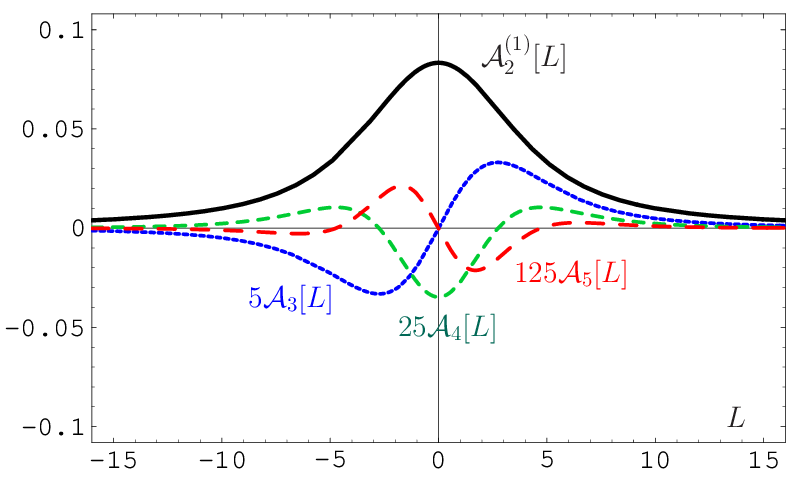}~~~%
             \includegraphics[width=0.45\textwidth]{
  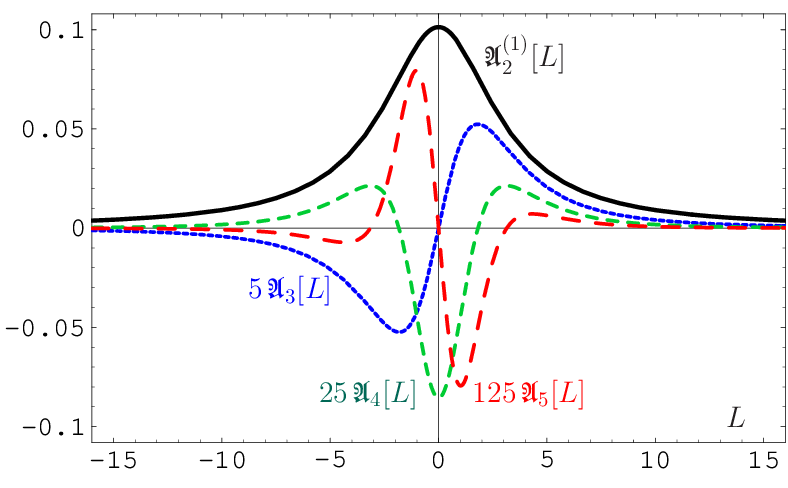}}
  \caption{Графики функций ${\mathcal A}_{n}^{(1)}[L]$ (слева) и
  ${\mathfrak A}_{n}^{(1)}[L]$ (справа) для $n=2, 3, 4, 5$.
  Чтобы показать все кривые вместе, мы масштабировали функции
  с помощью факторов $5^{n-2}$.
  \label{fig:UA_2345}}
\end{figure}
На рис.~\ref{fig:UA_2345} мы сравниваем
поведения аналитических образов высших степеней зарядов
в простран\-ственно- (слева) и временноподобных (справа) областях.
Эффект `искажающего зеркала' виден также и здесь,
но степень искажения растет с ростом степени заряда $n$:
значения ${\mathcal A}_{4}[0]$
и ${\mathfrak A}_{4}[0]$ различаются почти в 2 раза,
в то время как ${\mathcal A}_{2}[0]\simeq {\mathfrak A}_{2}[0]$.

Для получения двухпетлевых аналитических зарядов
необходимо использовать в качестве исходного пертурбативного
эффективного заряда $a^{n}_{(2)}(Q^2)$
точное решение двухпетлевого РГ уравнения (\ref{eq:beta.new.2L}),
которое выражается через функцию Ламберта,
как было показано в~\cite{Mag99},
см. (\ref{eq:App-Exactsolution.2L}).
Решения для эффективного заряда в высших петлевых приближениях
могут быть также выражены через функцию Ламберта,
см. (\ref{eq:App-Exactsolution.2L.3L}),
или аппроксимированы с использованием
приближенных выражений
для спектральных плотностей
с последующим их численным интегрированием
(см. Приложение~\ref{App:RG-solution-2L},
 где мы детально разобрали процедуру построения
 спектральных плотностей
 $\rho^\text{(2)it-1}_{1}[L_\sigma]$, формула (\ref{eq:Sp.Den.iteration-1}),
 и
 $\rho^\text{(2)it-2}_{1}[L_\sigma]$, формула (\ref{eq:Sp.Den.iteration-2}),
 отвечающие  итерационным решениям (\ref{eq:123.Iter})).

Как видно из этого краткого обсуждения,
все проблемы,
связанные с наличием ``призрачной'' сингулярности
у эффективного заряда в евклидовой области,
устраняются при аналитизации,
основанной на принципе причинности
(``спектральности''~\cite{Shi98})
и РГ инвариантности.
При этом устранение полюса Ландау
(аналогично, компенсация сингулярностей типа точек ветвления
 в высших петлевых приближениях)
\textbf{не вводится руками},
а \textbf{появляется как естественное следствие принципа аналитизации}
эффективного заряда,
без какой-либо апелляции к непертурбативным механизмам,
приводящим к степенным поправкам.
Тем не менее, такие поправки типа $\left(M^2/Q^2\right)^n$,
где $M$ может отвечать, например,
составной массе кварка,
иногда вводят в спектральную плотность,
чтобы учесть непертурбативные эффекты, см.~\cite{CV06,Ale05}.

 \subsection{Глобальная АТВ: учет порогов тяжелых кварков}
 \label{subsec:APT.Global}

Рассмотрим схему глобализации АТВ,
т.~е. учета порогов тяжелых кварков,
следуя подходу Ширкова--Соловцова~\cite{SM94,SS97,Mag99,Mag00},
в которых они используют следующие значения
полюсных масс $c$-, $b$- и $t$-кварков:
$m_c=1.2$~ГэВ, $m_b=4.3$~ГэВ и $m_t=175$~ГэВ.
При этом в стандартной теории возмущений КХД
в $\overline{\text{MS}\vphantom{^1}}$-схеме необходимо согласовывать значения
эффективных зарядов в евклидовой области на значениях $Q^2$,
отвечающих этим массам\footnote{%
Вопрос о том, почему в $\overline{\text{MS}\vphantom{^1}}$-схеме лучше согласовывать
эффективные заряды на значениях $m_q^2$,
а не $4 m_q^2$,
разбирается подробно в работе~\cite{SM94}.
Если говорить кратко,
то это связано с безмассовым характером ренормгрупповых генераторов
в $\overline{\text{MS}\vphantom{^1}}$-схеме,
в которой учет порогов массивных кварков производится
путем сравнения с MOM-схемой, где массы частиц учитываются явно.}:
$M_4=m_c$, $M_5=m_b$
и $M_6=m_t$.

Будем в дальнейшем определять логарифмы $L$ только по отношению
к трехфлейворному масштабу $\Lambda_3^2$:
$L(Q^2)=\ln\left(Q^2/\Lambda_3^2\right)$.
Пересчет к другим масштабам осуществляется путем
конечных добавок:
\begin{eqnarray}
 \label{eq:Log_Nf}
  \ln\left(Q^2/\Lambda_f^2\right)
    = L(Q^2)
    + \lambda_f\,,
   \quad \text{где}
   \quad \lambda_f\equiv\ln\left(\Lambda_3^2/\Lambda_f^2\right)\,,
\end{eqnarray}
а $\Lambda_f^2$ --- отвечающий данному $N_f$ КХД-масштаб.
Определим также соответствующие значения логарифмов на порогах $M_k$
($k=4, 5, 6$):
\begin{eqnarray}
 \label{eq:Log_Thresh_43}
 L_{k} &\equiv& \ln\left(M_{k}^2/\Lambda_3^2\right)\,.
\end{eqnarray}
Все КХД-масштабы $\Lambda_f$, $f=4, 5, 6$, мы будем трактовать
как функции одного параметра,
а именно трехфлейворного масштаба $\Lambda_3$:
\begin{eqnarray}
 \label{eq:Lambda_Nf}
 \Lambda_f \ =\ \Lambda_f(\Lambda_3)
 \quad \text{с}\quad \Lambda_3 >
                        \Lambda_4(\Lambda_3) >
                        \Lambda_5(\Lambda_3) >
                        \Lambda_6(\Lambda_3)\,,
\end{eqnarray}
которые должны определяться
из условий согласования на порогах для эффективного заряда
в обычной КХД.
Рассмотрим здесь для примера двухпетлевое приближение
с эффективным зарядом $\alpha_s^{(2)}[L_f;N_f]$
(см.~(\ref{eq:App-Exactsolution.2L});
 показатель петлевого порядка $^{(2)}$
 в этом примере далее будем опускать для экономии места)
\begin{eqnarray}
  \alpha_s[L_f;N_f]
  = \frac{-4\,\pi}{b_0(N_f)c_1(N_f)
                    \left[1+W_{-1}(z_W[L;N_f])\right]
                   }\,,
\end{eqnarray}
где $z_W[L;N_f]=\left(1/c_1(N_f)\right)\exp\left[-1+i\pi-L/c_1(N_f)\right]$.
Тогда условия согласования в евклидовой области $Q^2$
таковы
\begin{subequations}
\begin{eqnarray}
 \label{eq:matching-SS_43}
  \alpha_s\left[L_{4};3\right]
  &=&
  \alpha_s\left[L_{4}+\lambda_4;4\right]\,;\\
 \label{eq:matching-SS_54}
  \alpha_s\left[L_{5}+\lambda_4;4\right]
  &=&
  \alpha_s\left[L_{5}+\lambda_5;5\right]\,;\\
 \label{eq:matching-SS_65}
  \alpha_s\left[L_{6}+\lambda_5;5\right]
  &=&
  \alpha_s\left[L_{6}+\lambda_6;6\right]\,.
\end{eqnarray}
\end{subequations}
Они определяют константы $\lambda_f$ с $f=4, 5, 6$
как функции $\Lambda_3$
и непрерывный глобальный эффективный заряд КХД
\begin{eqnarray}
 \label{eq:global_PT_Euclid}
  \!\!\!\alpha_s^\text{\tiny glob}(Q^2)\!
   &=&\!\alpha_s\left[L(Q^2);3\right]\,
         \theta\left(Q^2\!<\!M_4^2\right)
     +  \alpha_s\left[L(Q^2)\!+\!\lambda_4;4\right]\,
         \theta\left(M_4^2\!\leq\!Q^2<\!M_5^2\right)
  \nonumber\\
   &+&\!\alpha_s\left[L(Q^2)\!+\!\lambda_5;5\right]\,
         \theta\left(M_5^2\!\leq\!Q^2<\!M_6^2\right)
     +  \alpha_s\left[L(Q^2)\!+\!\lambda_6;6\right]\,
         \theta\left(M_6^2\!\leq\!Q^2\right).~~~
\end{eqnarray}
Приведем для примера значения $\Lambda_f$, $\lambda_f$ и
  $L_{f}$ с $f=4, 5, 6$ для случая $\Lambda_3=400$ МэВ:
\begin{subequations}
\begin{eqnarray}
 \label{eq:Lambda_f.400MeV}
  \Lambda_4 &=& 344~\text{МэВ}\,,\quad
  \Lambda_5\ =\ 245~\text{МэВ}\,,\quad
  \Lambda_6\ =\ 103~\text{МэВ}\,;\\
 \label{eq:lambda_f.400MeV}
  \lambda_4 &=& 0.299\,,\qquad~~
  \lambda_5\ =\ 0.984\,,\qquad~~\
  \lambda_6\ =\ 2.716\,;\\
 \label{eq:L_f3.400MeV}
  L_{4} &=& 2.197\,,\qquad~
  L_{5}\ =\ 4.750\,,\qquad~
  L_{6}\ =\ 12.162\,.
\end{eqnarray}
\end{subequations}

\begin{figure}[t]
 \centerline{\includegraphics[width=0.33\textwidth]{
  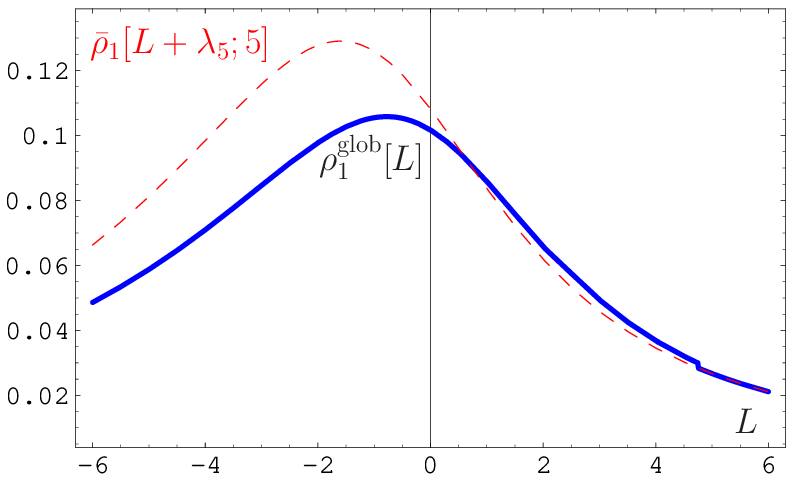}~%
             \includegraphics[width=0.33\textwidth]{
  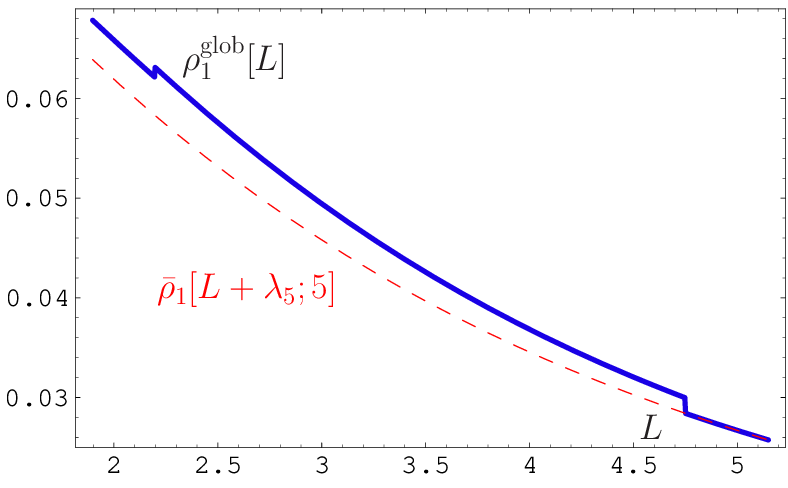}~%
             \includegraphics[width=0.33\textwidth]{
  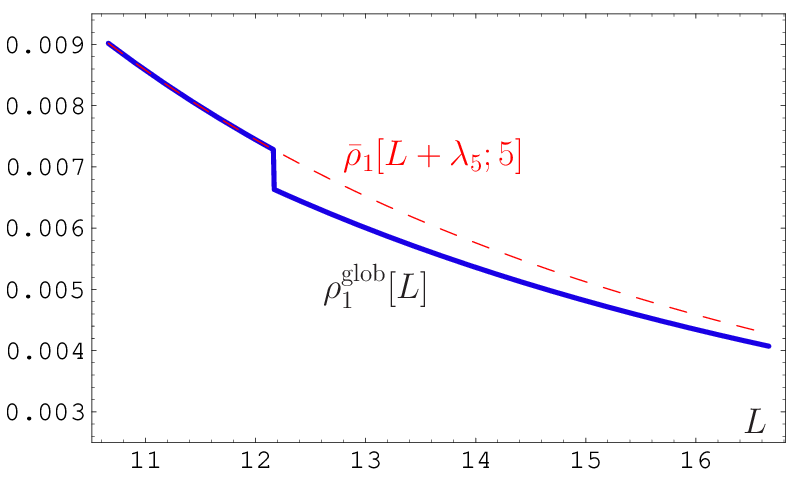}}
  \caption{Поведение глобальной спектральной плотности,
  $\rho_{1}^\text{\tiny glob}[L]$, для $\Lambda_3=400$~МэВ.
  Для сравнения штрихованными линиями мы показываем поведение
  локальной спектральной плотности, отвечающей 5 флейворам,
  $\bar{\rho}_{1}[L+\lambda_5;5]$. Слева показан интервал $L\in[-6;+6]$,
  где явно виден порог $L_{5}=4.75$,
  в центре --- в увеличенном по оси ординат масштабе
  интервал $L\in[+2;+5]$, так что видны оба порога
  $L_{4}=2.20$ и $L_{5}=4.75$, а справа --- снова в увеличенном
  по оси ординат масштабе интервал $L\in[+11;+16]$,
  так что становится явным разрыв на пороге $L_{6}=12.16$.
   \label{fig:rho_Glo}}
\end{figure}

Спектральная плотность аналитического образа
$n$-й степени эффективного заряда ${\mathcal A}_n^\text{\tiny glob}$
тогда есть кусочно-непрерывная функция
\begin{subequations}
\label{eq:global_PT_Rho}
\begin{eqnarray}
 \label{eq:global_PT_Rho(s)}
  \rho_n^\text{\tiny glob}(s)
  &=& \bar{\rho}_n\left[L(s);3\right]
       \theta\left(s<M_4^2\right)
    + \bar{\rho}_n\left[L(s)+\lambda_4;4\right]\,
       \theta\left(M_4^2\leq s<M_5^2\right)
  \nonumber\\
  &+& \bar{\rho}_n\left[L(s)+\lambda_5;5\right]\,
       \theta\left(M_5^2\leq s<M_6^2\right)
    + \bar{\rho}_n\left[L(s)+\lambda_6;6\right]\,
       \theta\left(M_6^2\leq s\right)\,,~~~
\end{eqnarray}
которую можно эквивалентно
представить как функцию логарифма $L=\ln(s/\Lambda_3^2)$
(см.~рис.\ \ref{fig:rho_Glo}, где для сравнения представлена также
 спектральная плотность $\bar{\rho}_{1}[L+\lambda_5;5]$ со сдвигом аргумента
 на $\lambda_5$ для того, чтобы в области $L\in[L_{5};L_{6}]$
 обе плотности совпадали):
\begin{eqnarray}
 \label{eq:Rho[L].Glo.n}
  \rho_n^\text{\tiny glob}[L]
  &=& \bar{\rho}_n\left[L;3\right]
       \theta\left(L<L_{4}\right)
    + \bar{\rho}_n\left[L+\lambda_4;4\right]\,
       \theta\left(L_{4}\leq L<L_{5}\right)
  \nonumber\\
  &+& \bar{\rho}_n\left[L+\lambda_5;5\right]\,
       \theta\left(L_{5}\leq L<L_{6}\right)
    + \bar{\rho}_n\left[L+\lambda_6;6\right]\,
       \theta\left(L_{6}\leq L\right)
\end{eqnarray}
\end{subequations}
со спектральными плотностями $\bar{\rho}_n\left[L;N_f\right]$,
отвечающими числу активных кварков $N_f$
и равными (см.~(\ref{eq:SpDen.nu.(2)}) для случая $\nu=n$)
\begin{eqnarray}
  \bar{\rho}_{n}\left[L;N_f\right]
  = \frac{\rho_{n}\left[L;N_f\right]}{\beta_f^n}
  \equiv
    \frac{\sin[n~\varphi_{(2)}(L;N_f)]}
         {\pi\,\left[\beta_f\,R_{(2)}(L;N_f)\right]^{n}}\,,
\label{eq:Spec.Dens.n.2L}
\end{eqnarray}
где $R_{(2)}(L;N_f)$ и $\varphi_{(2)}(L;N_f)$
определены в разделе~\ref{subsec:FAPT.Nf3.2L}
(см.~(\ref{eq:Lamb.R_(2)}) и (\ref{eq:Lamb.phi_(2)});
 заметим, что в этих формулах не указана явная зависимость
 от $N_f$, но она подразумевается,
 т.~к. параметр $c_1=c_1(N_f)=b_1(N_f)/b_0^2(N_f)$ явно зависит от $N_f$),
а $\beta_f=b_0(N_f)/(4\pi)$.
Тогда глобальные аналитические эффективные заряды
в евклидовой области и в области Минковского определяются
в соответствии с (\ref{eq:A.E}) и (\ref{eq:A.M})
\begin{subequations}
\begin{eqnarray}
 {\mathcal A}_{n}^\text{\tiny glob}\left(Q^2\right)
  = \int\limits_0^{\infty}\!
     \frac{\rho_n^\text{\tiny glob}(\sigma)}
          {\sigma+Q^2}\,
       d\sigma
 \quad\text{или}\quad
 {\mathcal A}_{n}^\text{\tiny glob}[L]
  = \int\limits_{-\infty}^{\infty}\!
     \frac{\rho_{n}^\text{\tiny glob}[L_\sigma]\,dL_\sigma}
          {1+\exp\left[L-L_\sigma\right]}\,;
 \label{eq:An.A_n.Glo.Q2.L}
 \\
 \label{eq:An.U_n.Glo.Q2.L}
 {\mathfrak A}_{n}^\text{\tiny glob}(s)
  = \int\limits_s^{\infty}\!
     \frac{\rho_n^\text{\tiny glob}\left(\sigma\right)}
          {\sigma}\,
      d\sigma
 \quad\text{или}\quad
 {\mathfrak A}_{n}^\text{\tiny glob}[L]
  = \int\limits_{L}^{\infty}
     \rho_n^\text{\tiny glob}\left[L_\sigma\right]\,
      dL_\sigma\,.
\end{eqnarray}
\end{subequations}
\begin{figure}[ht]
 \centerline{\includegraphics[width=0.45\textwidth]{
  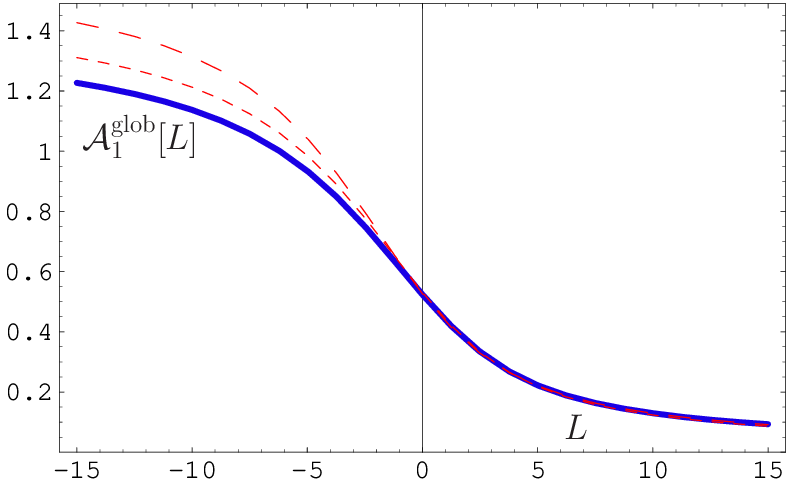}~~~%
             \includegraphics[width=0.45\textwidth]{
  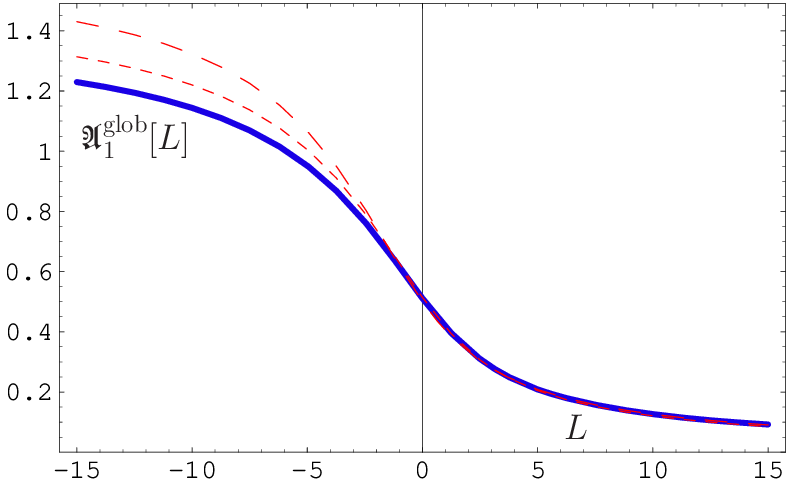}}
  \caption{Поведение глобальных аналитических зарядов
  ${\mathcal A}_{1}^{(2);\text{\tiny glob}}[L]$ (слева) и
  ${\mathfrak A}_{1}^{(2);\text{\tiny glob}}[L]$ (справа)
  для $\Lambda_3=400$~МэВ.
  Для сравнения штрихованными (крупно-штрихованными) линиями
  мы показываем поведение ${\mathcal A}_{1}^{(2)}[L+\lambda_4;N_f=4]$
  (${\mathcal A}_{1}^{(2)}[L+\lambda_5;N_f=5]$) и
  ${\mathfrak A}_{1}^{(2)}[L+\lambda_4;N_f=4]$
  (${\mathfrak A}_{1}^{(2)}[L+\lambda_5;N_f=5]$),
  поделенных на $\beta_f$ --- см.
  (\ref{eq:Non.Glo.Couplings})--(\ref{eq:Glo.Couplings}).
   \label{fig:UA_Glo}}
\end{figure}
\\\noindent
Несмотря на то, что спектральная плотность $\rho_n^\text{\tiny glob}\left(\sigma\right)$
является кусочно-непрерывной функцией
и в точках $\sigma=m_f^2$, $f=4, 5, 6$,
она терпит разрывы,
эффективный заряд ${\mathcal A}_n^\text{\tiny glob}\left(Q^2\right)$
оказывается аналитической функцией,
а ${\mathfrak A}_n^\text{\tiny glob}(s)$ --- просто непрерывной функцией,
которая в точках $s=m_f^2$, $f=4, 5, 6$,
имеет разрыв производной.

На рис.\ \ref{fig:UA_Glo} мы показываем
графики  ${\mathcal A}_{1}^{(2);\text{\tiny glob}}[L]$ и
${\mathfrak A}_{1}^{(2);\text{\tiny glob}}[L]$
для $\Lambda_3=400$~МэВ.
Мы видим, что аналитизация приводит к уменьшению
эффективных зарядов
${\mathcal A}_{1}^{(2);\text{\tiny glob}}[L]$
и
${\mathfrak A}_{1}^{(2);\text{\tiny glob}}[L]$
в сравнении с $\beta_4^{-1}{\mathcal A}_{1}^{(2)}[L+\lambda_4;4]$
($\beta_5^{-1}{\mathcal A}_{1}^{(2)}[L+\lambda_5;5]$)
и с $\beta_4^{-1}{\mathfrak A}_{1}^{(2)}[L+\lambda_4;4]$
($\beta_5^{-1}{\mathfrak A}_{1}^{(2)}[L+\lambda_5;5]$):
это понятно, поскольку соответствующая глобальная плотность
стала эффективно меньше, см. левую панель рис.\ \ref{fig:rho_Glo}.
Кроме того, создается впечатление,
что кривые слева и справа ничем не отличаются друг от друга.
Но это не так:
они действительно совпадают при $L\to\pm\infty$,
но в точке $L=0$ они различаются,
правда, разность имеет порядок 2\%.

 Отметим также, что удобные для численных приложений
простые параметризации двухпетлевых глобальных аналитических зарядов
получены в нашей работе~\cite{BPSS04}
(только для ${\mathcal A}_{1}^{(2);\text{\tiny glob}}(Q^2)$
          и ${\mathcal A}_{2}^{(2);\text{\tiny glob}}(Q^2)$)
и в работе Заякина--Ширкова~\cite{SZ05}
(как для ${\mathcal A}_{n}^{(2);\text{\tiny glob}}(Q^2)$,
 так и для ${\mathfrak A}_{n}^{(2);\text{\tiny glob}}(Q^2)$
 с $n=1,2,3,4$).

\section{ОТ АТВ К ДРОБНО-АНАЛИТИЧЕСКОЙ ТЕОРИИ ВОЗМУЩЕНИЙ}
 \label{sec:FAPT}

  Напомним, что амплитуды,
зависящие от единственной кинематической переменной $Q^2$,
которые в стандартной ТВ разлагаются в ряд по степеням эффективного заряда,
в АТВ представляются в виде нестепенного ряда~\cite{Rad82,Shi98,MS96},
см.~(\ref{eq:cal D}):
\begin{eqnarray}
 D[L] = \sum_n d_n a^{n}[L]
  \Rightarrow {\cal D}_\text{A}[L]
      = \sum_n d_n\,{\mathcal A}^{}_{n}[L]
 ~~~\text{и}~~~{\cal R}[L]
      = \sum_n d_n\,{\mathfrak A}^{}_{n}[L]\,,
\label{eq:initial}
\end{eqnarray}
где $d_n$ --- численные коэффициенты,
полученные в схеме минимальных вычитаний.
Казалось бы, сделав этот шаг мы уже построили
``аналитизацию'' всей пертурбативной КХД.

Но это не так и вот почему.
В стандартной теории возмущений мы также имеем:
 \begin{itemize}
  \item факторизационную КХД-процедуру,
   которая дает возможность разделить вклады больших и малых расстояний
   и фактически применять теорию возмущений КХД для описания
   вклада малых расстояний (область больших евклидовых $Q^2$).
   При этом естественным образом в жестких КХД-амплитудах
   возникают логарифмические факторы типа $a^\nu[L]\,L$;~\footnote{%
   Первое указание на необходимость определения специальной процедуры
   ``аналитизации'' для работы с такими логарифмами появилось
   в работе~\cite{KS01},
   где было предложено применять концепцию аналитичности
   к партонной амплитуде \textit{в целом}, а не только к
   эффективному заряду и его степеням.}
  \item ренормгрупповую эволюцию, генерирующую в партонных амплитудах
   факторы $B(Q^2)=\left[Z(Q^2)/Z(\mu^2)\right]\,B(\mu^2)$,
   которые в однопетлевом приближении сводятся
   к $Z[L] \sim a^\nu[L]$,
   где $\nu=\gamma_0/(2b_0)$ --- дробное число.
 \end{itemize}
Таким образом,
нам надо построить еще рецепты аналитизации
для новых объектов:
$\displaystyle a^\nu$,
$\displaystyle a^\nu L^m$, $\ldots$\,.
Наш набор $\{{\mathcal A}_{n}\}$
образует линейное пространство,
которое, однако, не снабжено операцией умножения элементов:
произведение
${\mathcal A}_{n} \cdot {\mathcal A}_{m}$
не принадлежит $\{{\mathcal A}_{n}\}$,
т.~к. не имеет дисперсионного представления (\ref{eq:A.E}),
а элемент ${\mathcal A}_{m+n}[L]$ при всех конечных $L$
серьезно отличается от
${\mathcal A}^{}_{n}[L] \cdot {\mathcal A}^{}_{m}[L]$.
В некотором смысле можно говорить о восстановлении
стандартной алгебры для указанного набора
базовых функций только в пределе $L \to \infty$,
когда $\{{\mathcal A}_{n}\} \to \{ a^{n}\}$~\cite{SS97,Shi98,SS99}
и, в частности, ${\mathcal A}_{n}[L]\cdot{\mathcal A}_{m}[L]\to {\mathcal A}_{m+n}[L]$.
Все то же самое справедливо и для произведения
${\mathfrak A}_{n} \cdot {\mathfrak A}_{m}$
в отношении набора $\{{\mathfrak A}_{n}\}$
и представления (\ref{eq:A.M}).

Мы пополняем эти наборы,
определяя отображения \textbf{A}$_{\textbf{E}}$
и \textbf{A}$_{\textbf{M}}$
\begin{subequations}
 \label{eq:An.AU.Disp.Repr}
\begin{eqnarray}
 \textbf{A}_{\textbf{E}}\left[a^\nu\right][L]
  \equiv {\mathcal A}_{\nu}[L]
  = \int_{-\infty}^{\infty}\!
     \frac{\rho_{\nu}[L_\sigma]\,dL_\sigma}
          {1+e^{L-L_\sigma}}\,,
 \label{eq:An.A.Disp.Repr}\\
 \textbf{A}_{\textbf{M}}\left[a^\nu\right][L_s]
  \equiv {\mathfrak A}_{\nu}[L_s]
  = \int_{L_s}^{\infty}
     \rho_{\nu}\left[L_\sigma\right]\,
      dL_\sigma\,,
 \label{eq:An.U.Disp.Repr}
\end{eqnarray}
\end{subequations}
для новых элементов так,
что выполняются следующие свойства:
\begin{enumerate}
 \item Они являются изоморфизмами,
 т.~е. сохраняют линейную структуру исходного
 набора функций:
 \begin{eqnarray}
  \textbf{A}_\textbf{E}\left[a^0\right] = {\mathcal A}^{}_{0}\equiv 1
  \quad\text{и}\quad
  \textbf{A}_\textbf{M}\left[a^0\right] = {\mathfrak A}^{}_{0}\equiv 1\,.
 \label{eq:lin-spa}
 \end{eqnarray}

 \item Они дают возможность проводить
  ренормгрупповое улучшение АТВ,
  т.~е. определены для
  \begin{eqnarray}
    f(a)= a^\nu,~\text{где}~\nu\in\mathbb{R}\,.
   \label{eq:RG}
  \end{eqnarray}
  Оказывается, что пополнение исходных наборов
  $\{{\mathcal A}_{n}; n\in\mathbb{N}\}$
  и $\{{\mathfrak A}_{n}; n\in\mathbb{N}\}$ такими объектами,
  т.~е. расширение их до наборов \
  $\{{\mathcal A}_{\nu}; \nu\in\mathbb{R}\}$
  и $\{{\mathfrak A}_{\nu}; \nu\in\mathbb{R}\}$,
  дает также возможность и дифференцировать
  по индексу $\nu$,
  а значит строить аналитизации для
  величин $a^{\nu} \ln(a) = (d/d\nu)a^{\nu}$:
  \begin{subequations}
  \begin{eqnarray}
   {\cal L}_{\nu}
    \equiv
     \textbf{A}_\textbf{E}\left[a^{\nu}\ln(a)\right]
    \ =\ {\cal D}\,
          \textbf{A}_\textbf{E}\left[a^{\nu}\right]
   ~&\text{и}&~
   {\mathfrak L}_{\nu}
    \equiv
     \textbf{A}_\textbf{M}\left[a^{\nu}\ln(a)\right]
    \ =\ {\cal D}\,
          \textbf{A}_\textbf{M}\left[a^{\nu}\right]\,,~~~
  \label{eq:a.nu.ln.a}\\
  ~&\text{где}&~
   {\cal D}
    \equiv \frac{d}{d\nu}\,.
  \label{eq:D.d/d.nu}
 \end{eqnarray}
 \end{subequations}
 \item Они дают возможность пользоваться
  факторизацией в КХД,
  т.~е. определены для
  \begin{subequations}
  \begin{eqnarray}
    f(a)= a^{\nu} L^{m},~~m\in\mathbb{N}\,,
   \label{eq:a.nu.log.m}
  \end{eqnarray}
  давая в результате аналитические образы
 \begin{eqnarray}
   {\cal L}_{\nu,m}
    \equiv
     \textbf{A}_\textbf{E}\left[a^{\nu}L^m\right]
   \quad\text{и}\quad
   {\mathfrak L}_{\nu,m}
    \equiv
     \textbf{A}_\textbf{M}\left[a^{\nu}L^m\right]\,.
 \label{eq:analy-images}
 \end{eqnarray}
 \end{subequations}
  Заметим, что в однопетлевом приближении такие объекты
  сводятся просто к $\left[a^{\nu-m}\right]_\text{an}$,
  так что этот пункт сводится к предыдущему.
  Но в высших порядках это не так,
  и решение здесь достигается
  определением спектральной плотности $\rho_{\nu,m}$,
  отвечающей $\left[a^{\nu}L^m\right]_\text{an}$.
\end{enumerate}
Принципиальная схема аналитизации в АТВ,
показанная на рис.\ \ref{fig:APT-scheme},
применима и к ДАТВ.
Различие между АТВ и ДАТВ
состоит в объектах,
на которые действуют операторы аналитизации $\textbf{A}_\textbf{E}$ и $\textbf{A}_\textbf{M}$,
а также операторы перехода от евклидовых объектов к минковским ($\hat{R}$)
и обратно ($\hat{D}$):
в ДАТВ набор шире за счет включения зарядов с вещественными (дробными) значениями индексов,
а также объектов типа $ {\cal L}_{\nu}$, $ {\mathfrak L}_{\nu}$, ${\cal L}_{\nu,m}$ и ${\mathfrak L}_{\nu,m}$.

На первый взгляд, предлагаемое нами пополнение набора базисных зарядов АТВ,
$\{{\mathcal A}_{n}\}$ и $\{{\mathfrak A}_{\nu}\}$,
может быть неединственным:
можно, например, преобразовать полученные наборы
$\{{\mathcal A}_{\nu}\}$ и $\{{\mathfrak A}_{\nu}\}$
в наборы $\{{\mathcal A}'_{\nu}[L]\}$ и $\{{\mathfrak A}'_{\nu}[L]\}$,
определяемые элементарными сдвигами
${\mathcal A}'_{\nu}[L]={\mathcal A}_{\nu}[L]+\sin\left(\pi\nu\right)f_{\nu}[L]$
и
${\mathfrak A}'_{\nu}[L]={\mathfrak A}_{\nu}[L]+\sin\left(\pi\nu\right)g_{\nu}[L]$,
которые при целых значениях $\nu\in\mathbb{Z}$ совпадают друг с другом
и с исходными целоиндексными наборами функций.
Однако нас спасает от этой неединственности
спектральное представление (\ref{eq:An.AU.Disp.Repr}):
оно серьезно ограничивает класс возможных функций,
в частности, исключая указанную многозначность.

 \subsection{Однопетлевая ДАТВ ($N_f=3$)}
  \label{subsec:FAPT.Nf3.1L}
 Реализация этих идей в однопетлевом приближении
с фиксированным числом флейворов $N_f$ была осуществлена
в работах~\cite{BMS05,BMS06}
полностью в
аналитическом виде.
Поскольку этот материал является предметом
подготавливаемого обзора всех троих авторов этих работ,
здесь мы не будем подробно останавливаться на нем
и дадим только краткую сводку результатов.

 В основе метода~\cite{BMS05,BMS06} лежат
рекуррентные соотношения (\ref{eq:generator}),
связывающие ${\mathcal A}_{n}[L]$ и ${\mathfrak A}_{n}[L]$ с
${\mathcal A}_{1}[L]$ и ${\mathfrak A}_{1}[L]$
и позволяющие получить явные выражения
для аналитических образов искомых дробных степеней зарядов:
\begin{subequations}
\label{eq:AU.nu.F}
\begin{eqnarray}
 {\mathcal A}_{\nu}[L]
  &=& \frac{1}{L^\nu}
    - \frac{F(e^{-L},1-\nu)}{\Gamma(\nu)}\,;
  \label{eq:A.nu.F}\\
 {\mathfrak A}_{\nu}[L]
  &=& \frac{\text{sin}\left[(\nu -1)\arccos\left(L/\sqrt{\pi^2+L^2}\right)\right]}
           {\pi(\nu -1) \left(\pi^2+L^2\right)^{(\nu-1)/2}}\,.
  \label{eq:U.nu.F}
\end{eqnarray}
\end{subequations}
Здесь $F(z,\nu)$ --- редуцированная трансцендентная функция Лерха~\cite{BE53}:
\begin{eqnarray}
 F(z,\nu)
  &=& \sum_{m=1}^{\infty}\,\frac{z^m}{m^{\nu}}\,.
\label{eq:Lerch}
\end{eqnarray}
Интересно отметить, что ${\mathcal A}_{\nu}[L]$ оказывается целой функцией по $\nu$,
а ${\mathfrak A}_{\nu}[L]$ выражается полностью через элементарные
(тригонометрические) функции.\footnote{%
Заметим, что это выражение было получено в работе~\cite{GLK84} в виде,
приспособленном для последующего разложения по положительным $1/L$,
т.~е. $\arccos\left(L/\sqrt{\pi^2+L^2}\right)$ был заменен на
$\arctg(\pi/L)$, что верно только при $L>0$.}
 При этом евклидовы ``обратные степени''
${\mathcal A}_{-m}[L]=L^m$ совпадают
с обратными степенями исходного эффективного заряда $a^{-m}[L]=L^m$,
в то время как для минковских ``обратных степеней'' возникают добавки
в виде низших степеней $L$ с $\pi^2$-коэффициентами:
\begin{eqnarray}
 \label{eq:A.-m}
  {\mathcal A }_{-m}[L]
   = L^m\,,\quad
  {\mathfrak A}_{-m}[L]
   =\frac{1}{\pi(m+1)}\,
     \textbf{Im}{}\left[(L + i \pi)^{m+1}\right]
   ~~~\text{~~для~~} m\in\mathbb{N}\,;~~~~~~~ \\
 {\mathfrak A}_{-1}[L] \!=\! L\,,\
 {\mathfrak A}_{-2}[L] \!=\! L^2-\frac{\pi^2}{3}\,,\
 {\mathfrak A}_{-3}[L] \!=\! L^3-\pi^2L\,,\
 {\mathfrak A}_{-4}[L] \!=\! L^4-2L^2\pi^2+\frac{\pi^4}{5}\,,\ldots~~~~~~
 \label{eq:U.-m}
\end{eqnarray}

\begin{table}[b]\vspace*{-1mm}
\caption{Сравнение обычной ТВ, АТВ и ДАТВ в евклидовой области
 ((Е), $\displaystyle L=\ln\left(Q^2/\Lambda^2\right)$) и
 в области Минковского
 ((М), $\displaystyle L=\ln\left(s/\Lambda^2\right)$).
 В строке `Обратные степени' записано ${\mathfrak A}_{-m}[L]=L^m+O(L^{m-2})$,
 что означает просто свойство (\ref{eq:U.-m}).\vspace*{1mm}
 \label{tab:PT.APT.FAPT}}
 \begin{tabular}{ccccc}\hline\hline
  Теория  & ТВ        & АТВ(Е,M) & ДАТВ(Е) & ДАТВ(М) $\vphantom{^{\int}}$
  \\ \hline
  Набор зарядов
          & $\Big\{a^\nu\Big\}_{\nu\in\mathbb{R}}\vphantom{^{\big|}_{\big|}}$
                      & ~~~$\Big\{{\mathcal A}_m, {\mathfrak A}_m\Big\}_{m\in\mathbb{N}}$
                                 & $\Big\{{\mathcal A}_{\nu}\Big\}_{\nu\in\mathbb{R}}$
                                         & $\Big\{{\mathfrak A}_{\nu}\Big\}_{\nu\in\mathbb{R}}$
  \\ \hline
  Разложение в ряд
          & $\sum\limits_{m}f_m\,a^m\vphantom{^{\big|}_{\big|}}$
                      & $\sum\limits_{m}f_m\,{\mathcal A}_m\left(f_m\,{\mathfrak A}_m\right)$
                                 & $\sum\limits_{m}f_m\,{\mathcal A}_m$
                                         & $\sum\limits_{m}f_m\,{\mathfrak A}_m$
  \\ \hline
  Разложение в ряд
          & $a^\nu\sum\limits_{m}f_m\,a^m\vphantom{^{\big|}_{\big|}}$
                      & ~\text{---}~
                                 & $\sum\limits_{m}f_m\,{\mathcal A}_{m+\nu}$
                                        & $\sum\limits_{m}f_m\,{\mathfrak A}_{m+\nu}$
  \\ \hline
  Обратные степени
          & $\left(a[L]\right)^{-m}\vphantom{^{\big|}_{\big|}}$
                      & ~\text{---}~
                                 & ~${\mathcal A}_{-m}[L]=L^m$~
                                        & ~${\mathfrak A}_{-m}[L]=L^m+O(L^{m-2})$~
  \\ \hline
  Производные по $\nu$
          & $a^{\nu}\ln{a}\vphantom{^{\big|}_{\big|}}$
                      & ~\text{---}~
                                 & ${\cal L}_{\nu}={\cal D}{\mathcal A}_{\nu}$
                                        & ${\mathfrak L}_{\nu}={\cal D}\,{\mathfrak A}_{\nu}$
  \\ \hline
  Степени логарифма
          & $a^{\nu} L^{m}\vphantom{^{\big|}_{\big|}}$
                      & ~\text{---}~
                                 & ${\cal L}_{\nu,m}={\mathcal A}_{\nu-m}$
                                        & ${\mathfrak L}_{\nu,m}={\mathfrak A}_{\nu-m}$
  \\ \hline\hline\vspace*{-1mm}
  \end{tabular}
\end{table}

Таким образом, линейные пространства
$\{{\mathcal A}^{}_{n}\}$ и $\{{\mathfrak A}^{}_{n}\}$
теперь пополнены  путем включения элементов
${\mathcal A}^{}_{\nu}$
и
${\mathfrak A}^{}_{\nu}$
с любыми вещественными значениями индексов $\nu$,
так что становится возможной
операция дифференцирования по индексу (\ref{eq:D.d/d.nu}).
Спектральные плотности $\rho_{\nu}[L]$,
отвечающие таким решениям,
обладают следующим важным свойством:
\begin{eqnarray}
 \label{eq:AU.nu.Laplace}
  \rho_{n+\nu}[L]
 = \frac{\Gamma(\nu)}{\Gamma(n+\nu)}
    \left(-\frac{d}{dL}\right)^{n}\rho_{\nu}[L]\,.
\end{eqnarray}

В таблице \ref{tab:PT.APT.FAPT} мы сравниваем
основные элементы обычной ТВ, АТВ и ДАТВ в евклидовой
и минковской областях.

\begin{figure}[th]
 \centerline{\includegraphics[width=0.47\textwidth]{
  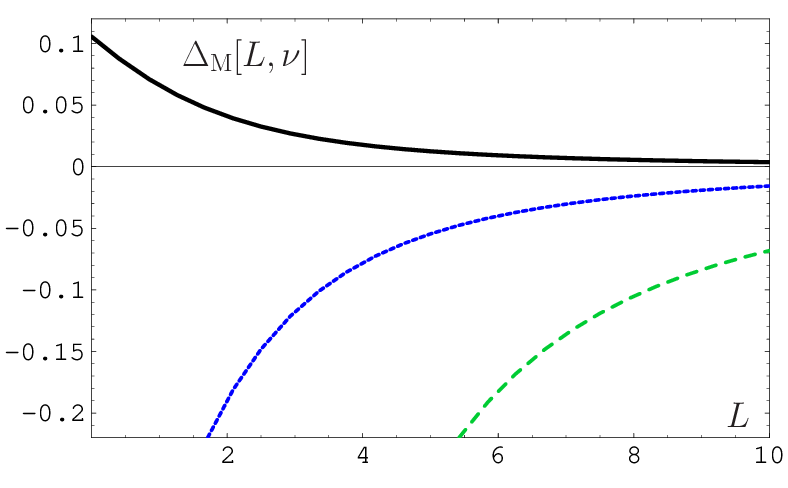}~~~%
             \includegraphics[width=0.47\textwidth]{
  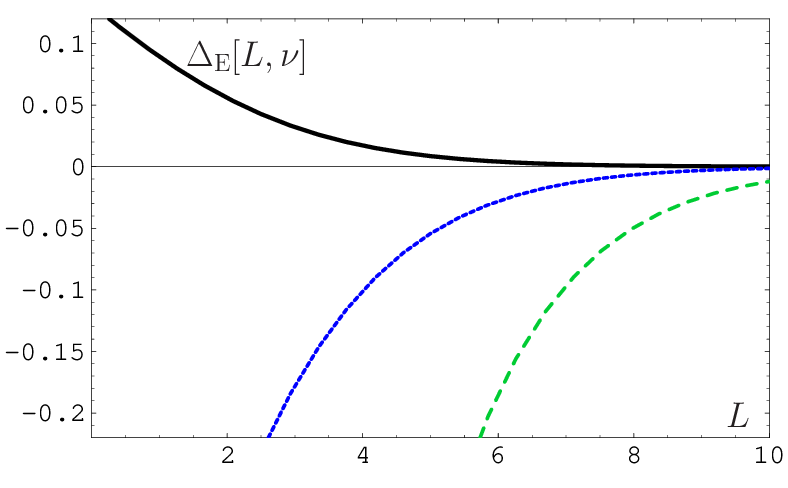}}%
     \caption{\label{fig:FAPT.nu.1^nu}
     Слева: Сравнение различных кривых для $\Delta_\text{M}(L,\nu)$
     как функций $L=\ln(s/\Lambda^2)$, отвечающих различным значениям $\nu$.
     Справа: То же сравнение для $\Delta_\text{E}(L,\nu)$
     как функций $L=\ln(Q^2/\Lambda^2)$.
     На обеих панелях сплошные (черные) линии отвечают значению
     $\nu=0.62$, синие пунктирные --- $\nu=1.62$
     и зеленые штрихованные --- $\nu=2.62$.}
\end{figure}

Прежде чем закончить этот раздел и перейти к обсуждению
двухпетлевого случая,
обсудим вопрос,
насколько сильно отличаются аналитические образы дробных степеней
эффективных зарядов от
дробных степеней аналитических зарядов
${\mathcal A}^{}_{1}[L]$
и
${\mathfrak A}^{}_{1}[L]$.
Для этого мы проанализируем
относительные отклонения в
минковской ($\Delta_\text{M}(L,\nu)$)
и евклидовой ($\Delta_\text{E}(L,\nu)$)
областях:
\begin{eqnarray}
 \label{eq:Ratio.nu}
  \Delta_\text{M}(L,\nu)
   = \frac{{\mathfrak A}^{}_{\nu}[L]
          -\left({\mathfrak A}^{}_{1}[L]\right)^{\nu}}
          {{\mathfrak A}^{}_{\nu}[L]}\,;\quad
  \Delta_\text{E}(L,\nu)
   = \frac{{\mathcal A}^{}_{\nu}[L]
          -\left({\mathcal A}^{}_{1}[L]\right)^{\nu}}
          {{\mathcal A}^{}_{\nu}[L]}\,.
\end{eqnarray}
На рис.\ \ref{fig:FAPT.nu.1^nu} показано
поведение $\Delta_\text{M}(L,\nu)$ (слева)
и $\Delta_\text{E}(L,\nu)$ (справа).
Видно, что как $\Delta_\text{M}(L,0.62)$,
так и $\Delta_\text{E}(L,0.62)$
меньше 5\%
при $Q^2,s\gtrsim1$~ГэВ$^2$
(что отвечает $L\gtrsim2.4$ при $\Lambda=300$~МэВ).
С другой стороны, и $|\Delta_\text{M}(L,1.62)|$,
и $|\Delta_\text{E}(L,1.62)|$
становятся меньше или порядка 5\%
только при $L\gtrsim5.1$,
т.~е. при $Q^2,s\gtrsim15.2$~ГэВ$^2$.
В то же время, $|\Delta_\text{M}(L,2.62)|>0.23$ и
$|\Delta_\text{E}(L,2.62)|>0.31$
при $L\lesssim5.1$,
т.~е. при $Q^2,s\lesssim15.2$~ГэВ$^2$.
Отсюда мы можем сделать вывод,
что аналитизация дробных степеней эффективных
зарядов особенно важна при $\nu>1$.
Кроме того, мы видим,
что при $0<\nu<1$ аналитизированные степени
${\mathfrak A}^{}_{\nu}[L]$ и ${\mathcal A}^{}_{\nu}[L]$
больше $({\mathfrak A}^{}_{1}[L])^\nu$ и $({\mathcal A}^{}_{1}[L])^\nu$,
а при $\nu>1$ аналитизированные степени
${\mathfrak A}^{}_{\nu}[L]$ и ${\mathcal A}^{}_{\nu}[L]$
становятся меньше
$({\mathfrak A}^{}_{1}[L])^\nu$ и $({\mathcal A}^{}_{1}[L])^\nu$,
причем с ростом $\nu$ это уменьшение становится все более заметным.

 \subsection{Двухпетлевая ДАТВ ($N_f=3$)}
  \label{subsec:FAPT.Nf3.2L}
Обобщение на случай высших петель можно проводить двумя путями.
В первом подходе используют разложения ДАТВ
для многопетлевых величин по однопетлевым зарядам.
Во втором, более мощном, подходе используют точные выражения
для многопетлевых спектральных плотностей,
с помощью которых,
пользуясь интегральными представлениями (\ref{eq:An.AU.Disp.Repr}),
численно восстанавливают сами аналитические заряды
и связанные с ними функции.
Мы будем использовать второй метод,
а использование его результатов для оценки точности ДАТВ-разложений,
получаемых в первом подходе,
будет изложено в готовящемся обзоре трех авторов ДАТВ~\cite{BMS05,BMS06}.

Рассмотрим $l$-петлевую спектральную плотность
$\rho_{\nu}^{(l)}(\sigma)$.
Она может быть представлена в том же виде,
что и в однопетлевом случае, см.~(\ref{eq:SpDen.1L.n}),
т.~е.
\begin{eqnarray}
 \label{eq:SpDen.nu.(l)}
  \rho_{\nu}^{(l)}(\sigma)
  = \frac{1}{\pi}\,
     \textbf{Im}\,\big[a^{\nu}_{(l)}[L_{-\sigma}]\big]
  = \frac{1}{\pi}\,\frac{\sin[\nu~
     \varphi_{(l)}[L_\sigma]]}{\left(R_{(l)}[L_\sigma]\right)^{\nu}}\,,
\end{eqnarray}
где фаза $\varphi_{(l)}$ и модуль $R_{(l)}$
имеют теперь $l$-петлевой смысл.
Поясним на примере двухпетлевого приближения,
где эффективный заряд точно выражается
через функцию Ламберта $W_{-1}[z_W[L_\sigma]]$,
см.~(\ref{eq:App-Exactsolution.2L}).
Нас интересует значение мнимой части для эффективного заряда
при $s=-\sigma<0$,
поэтому нам надо определять значение
$a_{(2)}[L]$
при $L=L_{-\sigma}=L_{\sigma}-i\pi$.
Представим $a_{(2)}[L_{\sigma}-i\pi]$
в виде:
\begin{subequations}
\label{eq:SpDen.nu.(2).R.phi}
\begin{eqnarray}
 \label{eq:a_(2).1.Lamb.R.phi}
  a_{(2)}\left[L_\sigma-i\pi\right]
   = \frac{\displaystyle e^{i\varphi_{(2)}[L_\sigma]}}{R_{(2)}[L_\sigma]}\,,
\end{eqnarray}
где
\begin{eqnarray}
 \label{eq:Lamb.R_(2)}
  R_{(2)}[L] &=& c_1\,\left|1+W_{\pm1}[z_W(L\mp i\pi)]\right|\,;~\\
 \label{eq:Lamb.phi_(2)}
   \varphi_{(2)}[L]
    &=& \arccos
       \left[
        \textbf{Re}\left(
                   \frac{-R_{(2)}[L]}
                        {c_1\,\left(1+W_{\pm1}[z_W(L\mp i\pi)]\right)}
                   \right)
      \right]\,.
\end{eqnarray}
\end{subequations}
Тогда мы сразу получим
\begin{eqnarray}
 \label{eq:a_(2).nu.Lamb.R.phi}
  \left[a_{(2)}[L_\sigma-i\pi]\right]^\nu
   = \frac{\displaystyle e^{i\nu\varphi_{(2)}[L_\sigma]}}
          {\left(R_{(2)}[L_\sigma]\right)^\nu}\,,
\end{eqnarray}
откуда немедленно следует формула (\ref{eq:SpDen.nu.(l)})
для случая $l=2$:
\begin{eqnarray}
 \label{eq:SpDen.nu.(2)}
  \rho_{\nu}^{(2)}[L_\sigma]
  = \frac{1}{\pi}\,
     \frac{\sin[\nu~\varphi_{(2)}[L_\sigma]]}
          {\left(R_{(2)}[L_\sigma]\right)^{\nu}}\,.
\end{eqnarray}

Для ``аналитизации'' более сложных выражений,
содержащих кроме степеней заряда
$\left(a_{(2)}\right)^\nu$ еще и степени логарифмов заряда,
нам потребуется следующее свойство
\begin{subequations}
\begin{eqnarray} \label{eq:log(a)}
 {\cal L}_{\nu}^{(l)}
   =  \left[ a_{(l)}^\nu\ln(a_{(l)})\right]_\text{an}
  &=& \frac{d}{d \nu}\,{\mathcal A}_{\nu}^{(l)}
  \equiv {\cal D}\,{\mathcal A}_{\nu}^{(l)}\,,\\
  \label{eq:log.m.(a)}
  {\cal L}_{\nu}^{m,(l)}
   =  \left[ a_{(l)}^\nu\ln^m(a_{(l)})\right]_\text{an}
  &=& {\cal D}^m\,{\mathcal A}_{\nu}^{(l)}\,.
\end{eqnarray}
\end{subequations}
В однопетлевом приближении этого самого по себе уже достаточно,
а в двухпетлевом приближении
мы используем интегральные представления типа
(\ref{eq:An.AU.Disp.Repr})
для интересующей нас величины ${\cal L}_{\nu}$
со спектральной плотностью
\begin{eqnarray}
 \rho^{(2)}_{{\cal L}_{\nu}}[L]
  &=& \textbf{Im}{}
       \left[a_{(2)}^\nu[L-i\pi]\ln(a_{(2)}[L-i\pi])
       \right]
   = \frac{d}{d\nu}\,
      \textbf{Im}{}
       \left[a_{(2)}^\nu[L-i\pi]
       \right]
   =  \frac{d}{d\nu}\,
       \rho^{(2)}_{\nu}[L]\nonumber\\
   &=& \frac{\cos\left[\nu\varphi_{(2)}[L]\right]}
            {R_{(2)}^\nu[L]}\,
        \varphi_{(2)}[L]
    -  \frac{\sin\left[\nu\varphi_{(2)}[L]\right]}
            {R_{(2)}^\nu[L]}\,
        \ln\left[R_{(2)}[L]\right]\,.~~~
 \label{eq:Sp.Den.L.nu.(2)}
\end{eqnarray}

 Спектральные плотности степеней заряда,
умноженных на логарифмы передачи импульса,
т.~е. $\left(a_{(2)}\right)^\nu L^m$,
\begin{eqnarray*}
 \rho^{(2)}_{{\cal L}_{\nu,m}}[L]
  = \textbf{Im}{}
       \left[a_{(2)}^\nu[L-i\pi]\,[L-i\pi]^m\right]
  = \textbf{Im}{}
     \left[\frac{a_{(2)}^\nu[L-i\pi]}
                {a_{(1)}^m[L-i\pi]}
     \right]
\end{eqnarray*}
тоже можно получить достаточно просто,
если использовать представления (\ref{eq:SpDen.nu.(l)})
для $l=1$ и $l=2$:
\begin{eqnarray}
 \rho^{(2)}_{{\cal L}_{\nu,m}}[L]
  &=& \textbf{Im}{}
       \left[\frac{R_{(1)}^m[L]}
                  {R_{(2)}^\nu[L]}
             e^{i\left[\nu\varphi_{(2)}[L]-m\varphi_{(1)}[L]\right]}
       \right]
 \ =\ \frac{R_{(1)}^m[L]}
           {R_{(2)}^\nu[L]}
  \sin\left[\nu\varphi_{(2)}[L]
            - m\varphi_{(1)}[L]
      \right].~~~
 \label{eq:Sp.Den.L.nu.m.(2)}
\end{eqnarray}
Отметим здесь, кстати,
что сходимость интеграла по $L_\sigma$ в
$\int\rho_{{\cal L}_{\nu,m}}[L_\sigma]dL_\sigma$
при больших $L_\sigma$
имеется не при любых значениях $m$ и $\nu$:
т.~к. $\big|\rho_{{\cal L}_{\nu,m}}[L_\sigma]\sim L_\sigma^{m-\nu}\big|$,
то интеграл сходится только при $m<\nu$
(при $\nu-1\leq m<\nu$ он сходится благодаря знакопеременности
 спектральной плотности (\ref{eq:Sp.Den.L.nu.m.(2)})).
К счастью, в ТВ КХД такие объекты возникают именно
в виде `допустимых' комбинаций, например,
$\alpha_s^{2+\nu}\,L$ или $\alpha_s^{3+\nu}\,L^2$ с $\nu>0$,
идущими от эволюционных факторов,
следующих из уравнений ЕРБЛ
(Ефремова--Радюшкина--Бродского--Лепажа)
или ДГЛАП
(Докшицера--Грибова--Липатова--Альтарелли-Паризи).

 \subsection{Глобальная ДАТВ: учет порогов тяжелых кварков}
  \label{subsec:FAPT.Global}
  Как было показано в разделе \ref{subsec:APT.Global},
главный объект в глобальной теории ---
спектральная плотность $\rho_n^\text{\tiny glob}[L]$,
см. (\ref{eq:Rho[L].Glo.n}),
а все аналитизированные заряды и другие величины определяются
по спектральным плотностям через формулы
типа (\ref{eq:An.A_n.Glo.Q2.L})--(\ref{eq:An.U_n.Glo.Q2.L}).
Точно так же главным объектом в глобальной ДАТВ
является спектральная плотность
$\rho_{\nu}^\text{\tiny glob}[L]$
\begin{eqnarray}
 \rho_{\nu}^{\text{\tiny glob};(l)}[L]
  &=& \bar{\rho}_{\nu}^{(l)}\left[L;3\right]
       \theta\left(L<L_{4}\right)
    + \bar{\rho}_{\nu}^{(l)}\left[L+\lambda_4;4\right]\,
       \theta\left(L_{4}\leq L<L_{5}\right)
  \nonumber\\
  &+& \bar{\rho}_{\nu}^{(l)}\left[L+\lambda_5;5\right]\,
       \theta\left(L_{5}\leq L<L_{6}\right)
    + \bar{\rho}_{\nu}^{(l)}\left[L+\lambda_6;6\right]\,
       \theta\left(L_{6}\leq L\right)~~~
 \label{eq:Rho[L].Glo.nu}
\end{eqnarray}
со спектральными плотностями $\bar{\rho}_{\nu}[L;N_f]$,
определяемыми в полной аналогии с (\ref{eq:Spec.Dens.n.2L}):
\begin{eqnarray}
  \bar{\rho}_{\nu}^{(l)}\left[L;N_f\right]
  = \frac{\rho_{\nu}^{(l)}\left[L;N_f\right]}{\beta_f^n}
  \equiv
    \frac{\sin[\nu~\varphi_{(l)}(L;N_f)]}
         {\pi\,\left[\beta_f\,R_{(l)}(L;N_f)\right]^{\nu}}\,.
\label{eq:Spec.Dens.nu.2L}
\end{eqnarray}
Благодаря наличию ступенчатой спектральной плотности
даже в однопетлевом случае
основное рекуррентное соотношение (\ref{eq:generator}) нарушается
и заменяется на более сложное.
Поэтому изящные формулы (\ref{eq:AU.nu.F})
раздела \ref{subsec:FAPT.Nf3.1L}
здесь неприменимы и приходится пользоваться
интегральными представлениями
\begin{subequations}
\label{eq:Disp.Repr.(l)}
\begin{eqnarray}
 {\mathcal A}_{\nu}^{\text{\tiny glob};(l)}[L]
  = \int\limits_{-\infty}^{\infty}\!
     \frac{\rho_{\nu}^{\text{\tiny glob};(l)}[L_\sigma]\,dL_\sigma}
          {1+e^{L-L_\sigma}}
 \quad\text{и}\quad
 {\mathfrak A}_{\nu}^{\text{\tiny glob};(l)}[L_s]
  = \int\limits_{L_s}^{\infty}
     \rho_{\nu}^{\text{\tiny glob};(l)}\left[L_\sigma\right]\,
      dL_\sigma\,,
 \label{eq:An.AU_nu.Glo.L}
\end{eqnarray}
где логарифмы $L=\ln Q^2/\Lambda_3^2$ и $L_s=\ln s/\Lambda_3^2$
определяются по отношению
к трехфлейворному масштабу $\Lambda_3$.

Благодаря тому, что спектральные плотности с фиксированным числом
флейворов
просто связаны с аналитизированными зарядами
в минковской области
\begin{eqnarray}
 \label{eq:rho.bar.U.bar}
 \bar{\rho}_{\nu}^{(l)}\left[L;N_f\right]
  &=& \left(\frac{-d}{dL}\right)\,
       \bar{{\mathfrak A}}_{\nu}^{(l)}[L;N_f]\,,
\end{eqnarray}
что следует из (\ref{eq:An.U.Disp.Repr}),
мы можем преобразовать выражение для
${\mathfrak A}_{\nu}^{\text{\tiny glob};(l)}[L_s]$
к явному виду:
\begin{eqnarray}
 {\mathfrak A}_{\nu}^{\text{\tiny glob};(l)}[L_s]
  \!&\!=\!&\!
    \theta\left[L_s\!<\!L_4\right]
     \Bigl(\bar{{\mathfrak A}}_{\nu}^{(l)}[L_s;3]
          -\bar{{\mathfrak A}}_{\nu}^{(l)}[L_4;3]
          +\bar{{\mathfrak A}}_{\nu}^{(l)}[L_4+\lambda_4;4]
          -\bar{{\mathfrak A}}_{\nu}^{(l)}[L_5+\lambda_4;4]
  \nonumber\\
  \!&\!\!&~~~~~~~~~~~~
          +\bar{{\mathfrak A}}_{\nu}^{(l)}[L_5+\lambda_5;5]
          -\bar{{\mathfrak A}}_{\nu}^{(l)}[L_6+\lambda_5;5]
          +\bar{{\mathfrak A}}_{\nu}^{(l)}[L_6+\lambda_6;6]
     \Bigr)
  \nonumber\\
  \!&\!+\!&\!
    \theta\left[L_4\!\leq\!L_s\!<\!L_5\right]
     \Bigl(\bar{{\mathfrak A}}_{\nu}^{(l)}[L_s+\lambda_4;4]
          -\bar{{\mathfrak A}}_{\nu}^{(l)}[L_5+\lambda_4;4]
          +\bar{{\mathfrak A}}_{\nu}^{(l)}[L_5+\lambda_5;5]
  \nonumber\\
  \!&\!\!&~~~~~~~~~~~~~~~~~~~
          -\bar{{\mathfrak A}}_{\nu}^{(l)}[L_6+\lambda_5;5]
          +\bar{{\mathfrak A}}_{\nu}^{(l)}[L_6+\lambda_6;6]
     \Bigr)
  \nonumber\\
  \!&\!+\!&\!
    \theta\left[L_5\!\leq\!L_s\!<\!L_6\right]
     \Bigl(\bar{{\mathfrak A}}_{\nu}^{(l)}[L_s+\lambda_5;5]
          -\bar{{\mathfrak A}}_{\nu}^{(l)}[L_6+\lambda_5;5]
          +\bar{{\mathfrak A}}_{\nu}^{(l)}[L_6+\lambda_6;6]
     \Bigr)
  \nonumber\\
  \!&\!+\!&\!
    \theta\left[L_6\!\leq\!L_s\right]\,
     \bar{{\mathfrak A}}_{\nu}^{(l)}[L_s+\lambda_6;6]\,.
 \label{eq:An.U_nu.Glo.Expl}
\end{eqnarray}
Для евклидовой области таких простых формул нет,
но мы также можем все свести к эффективному заряду при $N_f=6$
с конечными поправками:
\begin{eqnarray}
 {\mathcal A}_{\nu}^{\text{\tiny glob};(l)}[L]
  &=& \bar{{\mathcal A}}_{\nu}^{(l)}[L+\lambda_6;6]
    + \Delta\bar{{\mathcal A}}_{\nu}[L]\,;
  \label{eq:An.A_nu.Glo.Expl}\\
 \Delta\bar{{\mathcal A}}_{\nu}^{(l)}[L]
  &\equiv& \sum_{f=3}^{5}
      \int\limits_{L_{f}}^{L_{f+1}}\!
       \frac{\bar{\rho}_{\nu}^{(l)}[L_\sigma+\lambda_{f};N_f]
            -\bar{\rho}_{\nu}^{(l)}[L_\sigma+\lambda_{6};6]}
            {1+e^{L-L_\sigma}}\,
         dL_\sigma\,,
  \label{eq:Delta_f.A_nu}
\end{eqnarray}
где мы определили $L_3=-\infty$ и $\lambda_3=0$.
На левой панели рис.~\ref{fig:Comp.Delta-AGlo}
мы сравниваем зависимости
$\Delta\bar{{\mathcal A}}_{1}^{(2)}[L]$ (сплошная линия)
и
${\mathcal A}_{1}^{\text{\tiny glob};(2)}[L]$ (штрихованная линия),
а на правой панели этого рисунка показано,
как ведет себя отношение
$\Delta\bar{{\mathcal A}}_{1}^{(2)}[L]/{\mathcal A}_{1}^{\text{\tiny glob};(2)}[L]$
в зависимости от значения аргумента $L$:
оно меняется от $-20\%$ при больших отрицательных значениях $L\approx-10$,
затем в районе $L\approx-5$ проходит через ноль и растет до значения
 $+20\%$, достигаемого при  $L\approx0$, и затем падает до 0 при $L\to\infty$.
\begin{figure}[ht]
 \centerline{\includegraphics[width=0.47\textwidth]{
  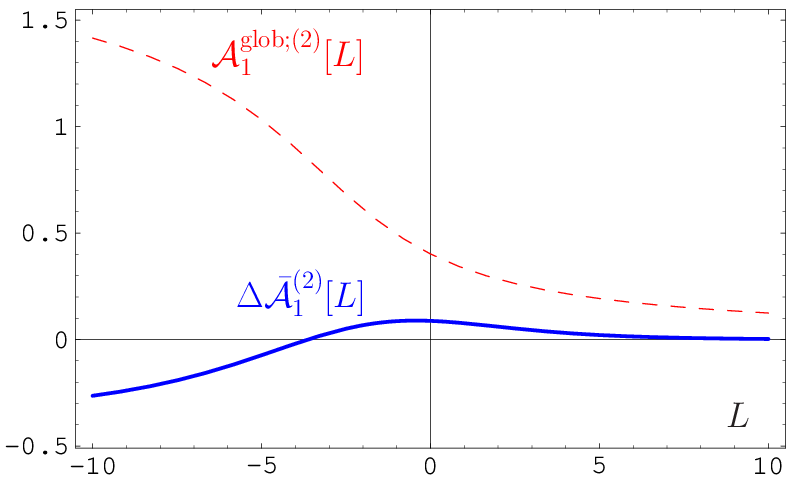}~~~%
             \includegraphics[width=0.47\textwidth]{
  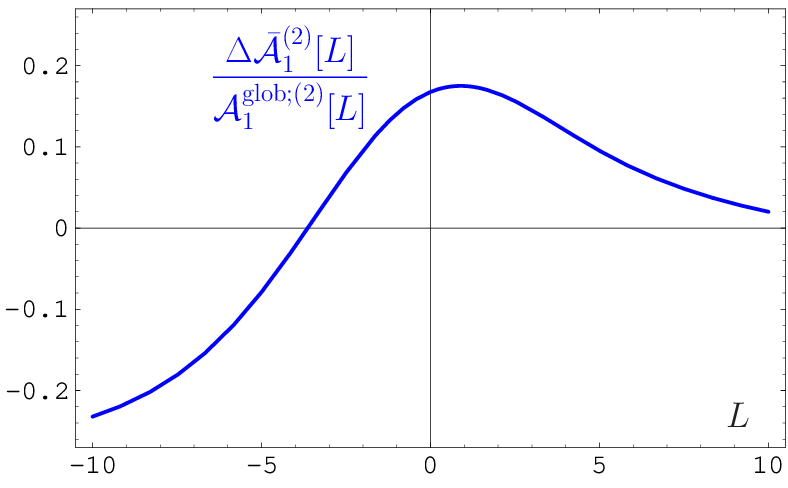}}
   \caption{\label{fig:Comp.Delta-AGlo}\footnotesize
    Слева показаны зависимости $\Delta\bar{{\mathcal A}}_{1}^{(2)}[L]$
    (сплошная линия) и
    ${\mathcal A}_{1}^{\text{\tiny glob};(2)}[L]$ (штрихованная линия),
    справа --- отношение
    $\Delta\bar{{\mathcal A}}_{1}^{(2)}[L]/{\mathcal A}_{1}^{\text{\tiny glob};(2)}[L]$
    в зависимости от значения аргумента $L$.}
\end{figure}

Аналогичные интегральные представления
используются и для аналитизации более сложных выражений
с логарифмами
($[\alpha_s^{\text{\tiny glob};(l)}]^\nu\,\ln[\alpha_s^{\text{\tiny glob};(l)}]$
и $[\alpha_s^{\text{\tiny glob};(l)}]^\nu\,L^m$):
\begin{eqnarray}
 {\cal L}_{\nu}^{\text{\tiny glob};(l)}[L]
  = \int\limits_{-\infty}^{\infty}\!
     \frac{\rho_{{\cal L}_{\nu}}^{\text{\tiny glob};(l)}\left[L_\sigma\right]\,
            dL_\sigma}
          {1+e^{L-L_\sigma}}
 \quad\text{и}\quad
 {\mathfrak L}_{\nu}^{\text{\tiny glob};(l)}[L]
  = \int\limits_{L}^{\infty}
     \rho_{{\cal L}_{\nu}}^{\text{\tiny glob};(l)}\left[L_\sigma\right]\,
      dL_\sigma\,;
 \label{eq:An.AUL_nu.Glo.L}\\
 {\cal L}_{\nu;m}^{\text{\tiny glob};(l)}[L]
  = \int\limits_{-\infty}^{\infty}\!
     \frac{\rho_{{\cal L}_{\nu;m}}^{\text{\tiny glob};(l)}\left[L_\sigma\right]\,
            dL_\sigma}
          {1+e^{L-L_\sigma}}
 \quad\text{и}\quad
 {\mathfrak L}_{\nu;m}^{\text{\tiny glob};(l)}[L]
  =  \int\limits_{L}^{\infty}
     \rho_{{\cal L}_{\nu;m}}^{\text{\tiny glob};(l)}\left[L_\sigma\right]\,
      dL_\sigma\,.
 \label{eq:An.AUL_nu;m.Glo.L}
\end{eqnarray}
\end{subequations}
Глобализованные спектральные плотности
$\rho_{{\cal L}_{\nu}}^{\text{\tiny glob};(l)}\left[L_\sigma\right]$
строятся по аналогии с (\ref{eq:Rho[L].Glo.nu})
из плотностей (\ref{eq:Sp.Den.L.nu.(2)}):
\begin{eqnarray}
 \bar{\rho}^{(l)}_{{\cal L}_{\nu}}[L;N_f]
  &=& \frac{\cos\left[\nu\varphi_{(l)}[L]\right]}
           {\left[\beta_f\,R_{(l)}[L]\right]^\nu}\,
        \varphi_{(l)}[L]
    -  \frac{\sin\left[\nu\varphi_{(l)}[L]\right]}
            {\left[\beta_f\,R_{(l)}[L]\right]^\nu}\,
        \ln\left[\beta_f\,R_{(l)}[L]\right]\,;~~~
 \label{eq:Bar.Sp.Den.L.nu.(l)}
\end{eqnarray}
нормированных введением необходимых степеней $\beta_f$.
В то же время глобализованные спектральные плотности
$\rho_{{\cal L}_{\nu;m}}^{\text{\tiny glob};(l)}\left[L_\sigma\right]$
имеют более сложную структуру,
так как аналитизируется произведение двух объектов,
один из которых, а именно $\alpha_s^\nu[L']$ непрерывен
за счет использования разных $\Lambda_f$ в аргументе
$L'=L+\lambda_f$, где $L=\ln(Q^2/\Lambda_3^2)$,
см.~(\ref{eq:Log_Nf});
второй же фактор $L^m$ непрерывен сразу с самого начала,
т.~к. он определяется по отношению к фиксированному масштабу
$\Lambda_3$.
Поэтому:
\begin{eqnarray}
 \rho_{{\cal L}_{\nu,m}}^{\text{\tiny glob};(l)}[L]
  &=& \bar{\rho}_{{\cal L}_{\nu,m}}^{(l)}\left[L,L;3\right]
       \theta\left(L<L_{4}\right)
    + \bar{\rho}_{{\cal L}_{\nu,m}}^{(l)}\left[L,L+\lambda_4;4\right]\,
       \theta\left(L_{4}\leq L<L_{5}\right)
  \nonumber\\
  &+& \bar{\rho}_{{\cal L}_{\nu,m}}^{(l)}\left[L,L+\lambda_5;5\right]\,
       \theta\left(L_{5}\leq L<L_{6}\right)
    + \bar{\rho}_{{\cal L}_{\nu,m}}^{(l)}\left[L,L+\lambda_6;6\right]\,
       \theta\left(L_{6}\leq L\right)\,,~~~
 \label{eq:Rho_L[nu.m].Glo}
\end{eqnarray}
где
\begin{eqnarray}
 \bar{\rho}^{(l)}_{{\cal L}_{\nu,m}}[L,L';N_f]
  &=& \frac{R_{(1)}^m[L]}
           {\left[\beta_f\,R_{(l)}[L']\right]^\nu}
  \sin\left[\nu\varphi_{(l)}[L']
            - m\varphi_{(1)}[L]
      \right]\,.~~~
 \label{eq:Bar.Sp.Den.L.nu.m.(l)}
\end{eqnarray}
Отметим здесь сразу полезное свойство этой плотности,
следующее из ее определения
\begin{eqnarray}
 \bar{\rho}^{(l)}_{{\cal L}_{\nu,1}}[L-\lambda,L;N_f]
 &=&\textbf{Im}
     \left[(L-\lambda-i\pi)
            \frac{\exp\left(i\nu\varphi_{(l)}[L]\right)}
                 {\left[\beta_f\,R_{(l)}[L]\right]^\nu}
     \right]\nonumber\\
 &=&\bar{\rho}^{(l)}_{{\cal L}_{\nu,1}}[L,L;N_f]
  - \lambda\,
     \bar{\rho}^{(l)}_{\nu}[L;N_f]\,.~~~
 \label{eq:Prop.SD.L.nu.1.(l)}
\end{eqnarray}
Понятно, что аналогичные соотношения можно записать
и для $\bar{\rho}^{(l)}_{{\cal L}_{\nu,m}}[L-\lambda,L;N_f]$
при других целых значениях $m>1$:
\begin{eqnarray}
 \bar{\rho}^{(l)}_{{\cal L}_{\nu,m}}[L-\lambda,L;N_f]
 &=&\textbf{Im}
     \left[(L-\lambda-i\pi)^m
            \frac{\exp\left(i\nu\varphi_{(l)}[L]\right)}
                 {\left[\beta_f\,R_{(l)}[L]\right]^\nu}
     \right]\nonumber\\
 &=& \sum\limits_{k=0}^{m-1}
      C^{k}_{m}\,
        (-\lambda)^k\,
       \bar{\rho}^{(l)}_{{\cal L}_{\nu,m-k}}[L,L;N_f]
  + (-\lambda)^m\,
     \bar{\rho}^{(l)}_{\nu}[L;N_f]\,.~~~
 \label{eq:Prop.SD.L.nu.m.(l)}
\end{eqnarray}

Интересно сравнить результат такой полной аналитизации
с аналитизацией,
применявшейся в~\cite{BKS05},
где использовался аналитизированный образ
${\cal L}_{2;1}^{\text{\tiny glob};(l)}[L]$
в виде наивного разложения
с учетом только $O(c_1)$-вклада:
\begin{subequations}
\label{eq:Glo.alpha^2.Log}
\begin{eqnarray}
 \label{eq:Glo.alpha^2.Log.nai}
  {\cal L}_{\nu;1}^{\text{Nai}}[L]
   &=& \beta_3^{-1}\,
        {\mathcal A}_{\nu-1}^{\text{\tiny glob};(2)}[L]
     + c_1[3]\,
        \beta_3^{-\nu}\,
         {\cal D}\,
          {\mathcal A}_{\nu}^{(1)}[L]\,.
\end{eqnarray}
Это разложение можно улучшить,
если вместо однопетлевого заряда с $N_f=3$ в слагаемом с фактором $c_1$
использовать двухпетлевой глобальный заряд:
\begin{eqnarray}
 \label{eq:Glo.alpha^2.Log.2-loop}
  {\cal L}_{\nu;m}^{\text{2-loop}}[L]
   &=& \beta_3^{-1}\,
        {\mathcal A}_{\nu-1}^{\text{\tiny glob};(2)}[L]
     + c_1[3]\,
         {\cal D}\,
          {\mathcal A}_{\nu}^{(2);\text{\tiny glob}}[L]\,.
\end{eqnarray}
\end{subequations}
При этом, для работы~\cite{BKS05} важна область
$Q^2=0.01-50$~ГэВ$^2$,
что соответствует $L\in[-2.8;6.7]$,
причем область $L\in[-2.8;1.1]$ необходима для анализа BLM-схемы,
а $L\in[-1.1;2.8]$ --- для анализа $\alpha_V$-схемы.

\begin{figure}[ht]
 \centerline{\includegraphics[width=0.47\textwidth]{
  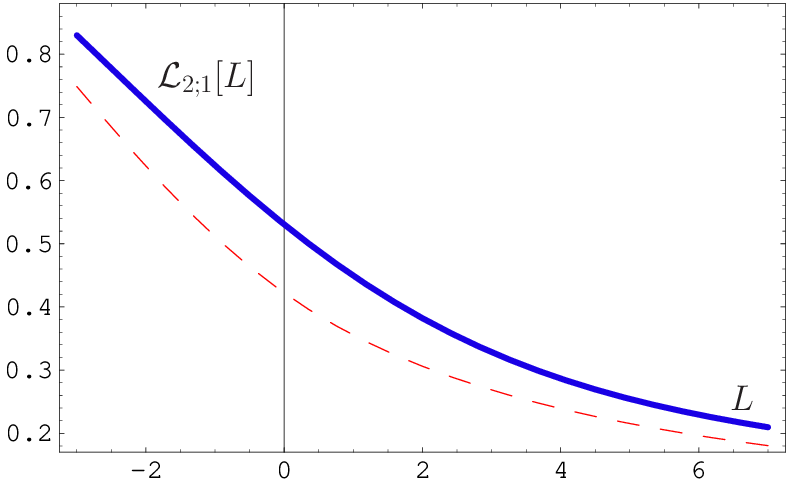}~~~
             \includegraphics[width=0.47\textwidth]{
  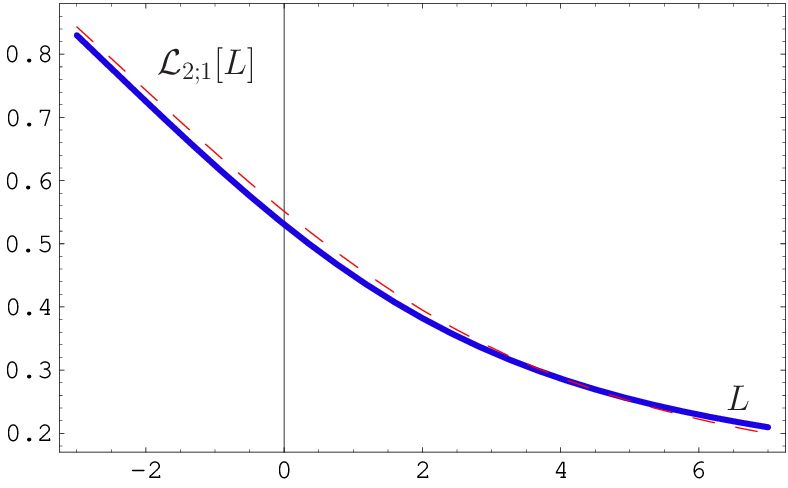}}
   \caption{\label{fig:L.2.1.Exa.Nai}\footnotesize
    Сравнение различных ``аналитизаций'' величины
    $\alpha_{s;(2)}^2[L]\,L$.
    Сплошная кривая отвечает точной аналитизации ${\cal L}_{2;1}^{\text{\tiny glob};(l)}[L]$,
    см. (\ref{eq:An.AUL_nu;m.Glo.L}) со спектральной плотностью
    (\ref{eq:Rho_L[nu.m].Glo}), а штрихованная ---
    наивному разложению,
    ${\cal L}_{\nu;1}^{\text{Nai}}[L]$,
    применявшемуся в работе~\cite{BKS05} (слева),
    или разложению  через двухпетлевые заряды,
    ${\cal L}_{\nu;1}^{\text{2-loop}}[L]$ (справа).
}
\end{figure}
На рис.~\ref{fig:L.2.1.Exa.Nai}
мы приводим сравнение этих функций во всей области $L\in[-3;7]$.
Относительная ошибка наивной аппроксимации (\ref{eq:Glo.alpha^2.Log.nai})
меняется от $-10$\% при $L=-3$,
проходит через минимум $-20$\% при $L\simeq+1$
и потом монотонно растет до $-14$\%  при $L=+7$.
В то же время, относительная ошибка двухпетлевой аппроксимации
(\ref{eq:Glo.alpha^2.Log.2-loop})
меняется от $+2$\% при $L=-3$,
проходит через максимум $+4$\% при $L\simeq+1$
и потом монотонно падает до $-4$\%  при $L=+7$.
Из этого обсуждения заключаем,
что использованная в~\cite{BKS05} аппроксимация (\ref{eq:Glo.alpha^2.Log.nai})
слишком груба и требуется перепроверить выводы этой работы,
используя точную формулу аналитизации (\ref{eq:An.AUL_nu;m.Glo.L}) и
(\ref{eq:Rho_L[nu.m].Glo}),
что и будет сделано в разделе~\ref{subsec:pionFF.FAPT}.

Перед тем как закончить этот раздел,
скажем несколько слов о численном интегрировании
в (\ref{eq:Disp.Repr.(l)}).
Проблема здесь заключается в медленной сходимости интегралов
на `плюс'-бесконечности, $L_\sigma\to+\infty$,
которая обеспечивается только убыванием спектральной плотности,
в то время как $1/(1+e^{-L_\sigma})\simeq 1+O(e^{-L_\sigma})$.
Положение спасает то,
что в однопетлевом случае интеграл берется точно,
а при больших $L_\sigma\gg L_{*}\sim10^2$
$l$-петлевые спектральные плотности очень близки к однопетлевым:
\begin{eqnarray*}
 \bar{\rho}_{\nu}^{(l)}[L,N_f]
  = \bar{\rho}_{\nu}^{(1)}[L,N_f]
    \left(1+ \varepsilon_{\nu}[L]\right)\,,
\end{eqnarray*}
где ошибка $\varepsilon_{\nu}[L_{*}]$ легко оценивается
для каждого значения $\nu$.
Однопетлевые же спектральные плотности
являются производными однопетлевых аналитических зарядов
в пространстве Минковского,
см. (\ref{eq:rho.bar.U.bar}),
поэтому, например, в случае евклидова глобального заряда
будем иметь:
\begin{eqnarray}
 {\mathcal A}_{\nu}^{\text{\tiny glob};(l)}[L]
  =
    \int\limits_{-\infty}^{L_{*}}\!
     \frac{\rho_{\nu}^{\text{\tiny glob};(l)}[L+L_\sigma]}
          {1+e^{-L_\sigma}}\,dL_\sigma
  + \bar{{\mathfrak A}}_{\nu}^{(1)}[L+L_{*}+\lambda_6,6]
  + O\left(\varepsilon_{\nu}[L+L_{*}+\lambda_6]\right)\,.
 \label{eq:An.A_nu.Glo.Approx}
\end{eqnarray}
Для глобальных аналитических зарядов в минковской области
это представление упрощается еще больше:
\begin{eqnarray}
 {\mathfrak A}_{\nu}^{\text{\tiny glob};(l)}[L]
  &=& \int\limits_{L}^{L_{*}}\!
       \rho_{\nu}^{\text{\tiny glob};(l)}[L_\sigma]\,
        dL_\sigma
   + \bar{{\mathfrak A}}_{\nu}^{(1)}[L_{*}+\lambda_6;6]
   + O\left(\varepsilon_{\nu}[L_{*}+\lambda_6]\right)\,.
 \label{eq:An.U_nu.Glo.Approx}
\end{eqnarray}

\section{РАСЧЕТ ФАКТОРИЗУЕМОЙ ЧАСТИ ФОРМФАКТОРА ПИОНА В АТВ И ДАТВ}
 \label{sec:pionFF}

В этой главе мы рассмотрим расчет электромагнитного формфактора (ФФ)
пиона
в теории возмущений КХД.
Как хорошо известно,
см., например,~\cite{IS82,NR82},
он определяется матричным элементом
\begin{equation}
 \langle{\pi^{+}(P^{\prime})| J_{\mu}(0) | \pi^{+}(P)}\rangle
 =  {\left( P + P^{\prime}\right)}_{\mu} F_{\pi}(Q^{2})\,,
\label{eq:pionvertex}
\end{equation}
где $J_{\mu}$ --- электромагнитный ток,
записанный через операторы кварковых полей,
$(P^{\prime}-P)^2=q^2\equiv-Q^2$
--- виртуальность фотона,
т.~е. квадрат большого импульса,
передаваемого пиону от фотона.
ФФ нормирован на единицу $F_\pi(0)=1$
за счет сохранения векторного тока в КХД,
что гарантируется электромагнитным тождеством Уорда.
Исторически именно при расчете формфакторов адронов в КХД
возникло понятие о факторизации жестких (пертурбативных)
и мягких (непертурбативных)
вкладов~\cite{CZ77,CZS77,ER80,ER80tmf,LB79,LB80}
и были определены так называемые амплитуды распределений (АР)
кварков и глюонов в адронах~\cite{Rad77,Rad84,Rad90}.
При этом доказанные теоремы факторизации
определяли каким образом нужно выделять вклады
жестких партонных подпроцессов,
которые описываются теорией возмущений КХД.
Именно, ФФ пиона представляется
в виде свертки
жесткой КХД-амплитуды партонного рассеяния
с двумя матричными элементами
операторов ведущего твиста,
которые параметризуются АР кварков
в начальном и конечном пионах,
$\varphi_{\pi}(x,\mu_\text{F}^2)$:
\begin{subequations}
\begin{eqnarray}
 F_{\pi}(Q^{2})
  &=& F_{\pi}^\text{Fact}(Q^{2})
   +  F_{\pi}^\text{non-Fact}(Q^{2})\,,
 \label{eq:pff}\\
 F_{\pi}^\text{Fact}(Q^{2}; \mu_\text{F}^2,\mu_\text{R}^2)
  &=& f_\pi^2\,
      \varphi_{\pi}(x,\mu_\text{F}^2)\mathop{\otimes}\limits_{x}
       T_\text{H}(x,y,Q^{2};\mu_\text{F}^2,\mu_\text{R}^2)\mathop{\otimes}\limits_{y}
        \varphi_{\pi}(y,\mu_\text{F}^2)\,,
 \label{eq:pff-Fact}
\end{eqnarray}
\end{subequations}
где $F_{\pi}^\text{Fact}(Q^{2})$ --- факторизуемая часть ФФ,
описываемая пертурбативной КХД,
символ $\otimes$ обозначает обычную свертку
($A(z)\mathop{\otimes}\limits_{z}B(z) \equiv \int_0^1 dz A(z) B(z)$)
по долям продольного импульса $x$ ($y$),
$\mu_\text{F}$ обозначает масштаб факторизации
больших и малых импульсов,
$\mu_\text{R}$ --- масштаб перенормировки константы связи,
а $F_{\pi}^\text{non-Fact}(Q^{2})$ представляет
нефакторизуемую часть,
называемую обычно ``мягким вкладом''~\cite{NR82},
которая генерируется непертурбативными поправками
и при больших $Q^2$
подавлена дополнительной степенью $1/Q^2$ по сравнению
с $F_{\pi}^\text{Fact}(Q^{2})$.
Важно подчеркнуть, что представление (\ref{eq:pff})
становится все более неадекватным по мере того,
как $Q^2$ приближается к $m_\rho^2$,
играющему роль характерного адронного масштаба в этом случае.
Это связано с нарушением представления
о коллинеарности кварков
в пионе:
в этой области поперечные импульсы кварков в пионе
становятся сравнимыми с $Q^2$
и необходимо каким-то образом модифицировать
коллинеарный факторизуемый вклад
$F_{\pi}^\text{Fact}(Q^{2})$,
как это сделано, например, в~\cite{BPSS04}.
Кроме того, в обычной ТВ КХД масштаб перенормировки
$\mu_\text{R}^2\sim Q^2$
при столь малых $Q^2$
приближается к сингулярности Ландау,
так что сам подход теории возмущений
становится необоснованным.

В представлении (\ref{eq:pff-Fact})
$T_\text{H}(x,y,Q^{2};\mu_\text{F}^2,\mu_\text{R}^2)$ --- амплитуда жесткого рассеяния,
описывающая взаимодействие коллинеарной кварк-антикварковой пары
с полным импульсом $P$,
ударяемой виртуальным фотоном с импульсом $q$,
в результате чего она переходит опять же в
коллинеарную кварк-антикварковую пару
с импульсом $P'=P+q$.
Эта амплитуда была рассчитана в теории возмущений КХД
при больших $Q^{2}$
в виде разложения по степеням эффективного
заряда~\cite{DR81,RK85,KMR86,MNP99a}:
\begin{eqnarray}
   T_\text{H}(x,y,Q^2;\mu_\text{F}^2,\mu_\text{R}^2) =
       \alpha_s(\mu_\text{R}^2)\,  T_\text{H}^{(0)}(x,y,Q^2)
         + \frac{\alpha_s^2(\mu_\text{R}^2)}{4 \pi} \,
             T_\text{H}^{(1)}(x, y, Q^2;\mu_\text{F}^2, \mu_\text{R}^2)
                + \ldots
 \label{eq:TH}
\end{eqnarray}
Явный вид вкладов $T_\text{H}^{(0)}$ и $T_\text{H}^{(1)}$
нам здесь не очень важен
(за деталями отсылаем читателя к нашей работе~\cite{BPSS04},
 а также к работе~\cite{Mul98}),
мы только хотим отметить появление характерных логарифмов
$\ln(Q^2/\mu_\text{R}^2)$ и $\ln(Q^2/\mu_\text{F}^2)$
в $ T_\text{H}^{(1)}(x, y, Q^2;\mu_\text{F}^2, \mu_\text{R}^2)$,
причем первый логарифм появляется,
как и следовало ожидать, в сочетании с $b_0$-фактором,
что важно для реализации схемы Бродского--Лепажа--Маккензи
(BLM)~\cite{BLM83}.

 Амплитуда распределения кварков в пионе
параметризует следующий матричный элемент
калибровочно-инвариантного раздвинутого аксиального кваркового тока
между физическим вакуумом $\langle0\mid$
и физическим пионом $\mid\pi(P)\rangle$ с импульсом $P$~\cite{Rad77}
\begin{eqnarray}
 \label{eq:pi-DA-ME}
 \langle{0\mid\bar{d}(z)\gamma^{\mu}\gamma_5\,
 {\cal C}(z,0) u(0)\mid\pi(P)}
  \Big|_{z^2=0}\rangle
 &=& i\,f_\pi\,P^{\mu}
     \int^1_0 dx e^{ix(zP)}\
      \varphi_{\pi}\left(x,\mu_0^2\right)\,;\\
 \int_0^1 \varphi_{\pi}(x,\mu_0^2)\,dx
  &=& 1\,,
 \label{eq:pi-DA-fpi}
\end{eqnarray}
где $f_{\pi} = 130.7 \pm 0.4$~МэВ~\cite{PDG2002} ---
константа распада пиона,
а
\begin{equation}
  {\cal C}(0,z)
  = {\cal P}
  \exp\!\left[-ig_s\!\!\int_0^z t^{a} A_\mu^{a}(y)dy^\mu\right]
\label{eq:pexponent}
\end{equation}
есть фазовый струнный фактор Фока--Швингера,
упорядоченный вдоль прямого пути, соединяющего точки $0$ и $z$,
и вводимый для обеспечения калибровочной инвариантности
раздвинутого кваркового тока.
Определенная таким образом АР пиона
имеет физический смысл амплитуды перехода физического пиона
в валентные кварк (с импульсом $xP$)
и антикварк (с импульсом $\bar{x}P$, $\bar{x}\equiv1-x$).
Масштаб $\mu_0^2$ является масштабом нормировки пионной АР
и связан с УФ регуляризацией кварковых полевых операторов,
раздвинутых на световом конусе.
В работе~\cite{BPSS04} мы подробно обсудили вопросы,
связанные с двухпетлевой эволюцией ЕРБЛ~\cite{ER80,ER80tmf,LB80}
АР пиона
и показали,
что для диктуемой непертурбативной КХД
двугегенбауэровской модели~\cite{BMS01,BMS01c}
\begin{eqnarray}
 \varphi_{\pi}(x,\mu^2)
  = 6 x (1-x)
     \left[ 1
          + a_2(\mu^2) \, C_2^{3/2}(2 x -1)
          + a_4(\mu^2) \, C_4^{3/2}(2 x -1)
     \right]
\label{eq:AR.2-Geg}
\end{eqnarray}
в рассматриваемой задаче с 1\%-ной точностью
можно использовать однопетлевую эволюцию:
\begin{eqnarray}
 a_n^{\text{LO}}(\mu^2)
  &=& a_n(\mu_0^2) \, E_n^{\text{LO}}(\mu^2,\mu_0^2)
 \label{eq:anLO}
 \quad\text{и}\quad
 E_n^{\text{LO}}(\mu_\text{F}^2,\mu_0^2)
  = \left[\frac{\alpha_\text{s}(\mu_\text{F}^2)}{\alpha_\text{s}(\mu_0^2)}
      \right]^{\gamma_n^{(0)}/(2b_0)}\,,
\end{eqnarray}
где $\gamma_n^{(0)}$ --- аномальные размерности ведущего порядка,
см. Приложение~\ref{app:DA.Evo}, (\ref{eq:gamma0}).

Запишем теперь выражение для факторизуемой части ФФ пиона
(\ref{eq:pff-Fact}),
получаемое с использованием двугегенбауэровской модели для
амплитуды распределения кварков в пионе (\ref{eq:AR.2-Geg})--(\ref{eq:anLO}):
\begin{eqnarray}
 F_{\pi}^\text{Fact}(Q^2;\mu_\text{F}^2,\mu_\text{R}^2)
  = \alpha_s(\mu_\text{R}^2)\,
     {\cal F}_{\pi}^\text{LO}(Q^2;\mu_\text{F}^2)
  + \frac{\alpha_s^2(\mu_\text{R}^2)}{\pi}\,
     {\cal F}_\pi^\text{NLO}(Q^2;\mu_\text{F}^2,\mu_\text{R}^2)\,,
 \label{eq:Q2pff}
\end{eqnarray}
где каллиграфические выражения используются для обозначения
величин без соответствующих $\alpha_s$-факторов,
ведущий вклад есть
\begin{eqnarray}
 {\cal F}_{\pi}^\text{LO}(Q^2;\mu_\text{F}^2)
  &\equiv& \frac{8\,\pi\,f_{\pi}^2}{Q^2}\,
            \left[1 + a_2^\text{LO}(\mu_\text{F}^2) + a_4^\text{LO}(\mu_\text{F}^2)\right]^2\,,
 \label{eq:Q2pffLO}
\end{eqnarray}
а неведущий
\begin{subequations}
 \label{eq:Q2.pff.NLO}
\begin{eqnarray}
 {\cal F}_\pi^\text{NLO}(Q^2;\mu_\text{F}^2,\mu_\text{R}^2)
 &\equiv& b_0\, {\cal F}_{\pi}^{(1,\beta)}(Q^2;\mu_\text{F}^2,\mu_\text{R}^2)
         + {\cal F}_{\pi}^{(1,\text{FG})}(Q^2;\mu_\text{F}^2)
         + C_\text{F}\,
           {\cal F}_\pi^{(1,\text{F})}(Q^2;\mu_\text{F}^2)~~~~~
 \label{eq:Q2.pff.NLO.Dia}
\end{eqnarray}
представлен в виде разложения по характерным цветовым структурам
($b_{0}$ --- первый коэффициент разложения $\beta$-функции,
 а $C_\text{F}$ --- глюонный цветовой фактор,
 см. Приложение~\ref{App:RG-solution-2L},
 (\ref{eq:beta0&1&2})--(\ref{eq:beta.new.2L})):
\begin{eqnarray}
 {\cal F}_{\pi}^{(1,\beta)}(Q^2;\mu_\text{F}^2,\mu_\text{R}^2)
  = \frac{2\,\pi\,f_{\pi}^2}{Q^2}
    &&\!\!\!\!
      \left[\frac{5}{3}
          + \frac{3 + \displaystyle (43/6) a_2^\text{LO}(\mu_\text{F}^2)
                    + (136/15) a_4^\text{LO}(\mu_\text{F}^2)}
                     {1 + a_2^\text{LO}(\mu_\text{F}^2) + a_4^\text{LO}(\mu_\text{F}^2)}
          - \ln \frac{Q^2}{\mu_\text{R}^2}
      \right]~~~~~\nonumber\\
    &&\!\!\!\!\times\Bigl[1+a_2^\text{LO}(\mu_\text{F}^2)+a_4^\text{LO}(\mu_\text{F}^2)\Bigr]^2\,,
 \label{eq:Q2.pff.NLO.beta}\\
 {\cal F}_{\pi}^{(1,\text{FG})}(Q^2;\mu_\text{F}^2)
  = -\frac{2\,\pi\,f_{\pi}^2}{Q^2}
  &&\!\!\!\!\Bigl\{ 15.67
                + a_2^\text{LO}(\mu_\text{F}^2)
                 \left[21.52 - 6.22\, a_2^\text{LO}(\mu_\text{F}^2)
                 \right]\nonumber\\
  &&+ a_4^\text{LO}(\mu_\text{F}^2)
                 \left[7.37 - 37.40\, a_2^\text{LO}(\mu_\text{F}^2)
                       - 33.61\, a_4^\text{LO}(\mu_\text{F}^2)
                 \right]
        \Bigr\}\,,~~~~~
 \label{eq:Q2.pff.NLO.FG}\\
 {\cal F}_\pi^{(1,\text{F})}(Q^2;\mu_\text{F}^2)
  = -\frac{2\,\pi\,f_{\pi}^2}{Q^2}
  &&\!\!\!\!
        \ln\left[\frac{Q^2}{\mu_\text{F}^2}\right]\,
        \left[\frac{25}{3}\,a_2^\text{LO}(\mu_\text{F}^2)
            + \frac{182}{15}\,a_4^\text{LO}(\mu_\text{F}^2)
        \right]~~~~~\nonumber\\
  &&\!\!\!\!\times
    \Bigl[1 + a_2^\text{LO}(\mu_\text{F}^2)+ a_4^\text{LO}(\mu_\text{F}^2)\Bigr]\,.~~~~~~~
 \label{eq:Fcal.(1)F}
\end{eqnarray}
\end{subequations}

  В разделе~\ref{subsec:pionFF.APT} мы будем исследовать применение
АТВ в задаче расчета факторизуемой части пионного ФФ
и выбирать при этом масштаб факторизации $\mu_\text{F}^2=Q^2$,
что приведет к обращению ${\cal F}_\pi^{(1,\text{F})}(Q^2;\mu_\text{F}^2)$
в ноль.
  В разделе же~\ref{subsec:pionFF.FAPT}
мы будем выбирать фиксированный масштаб факторизации $\mu_\text{F}^2=\textsl{const}$
и обсуждать зависимость результатов анализа
от выбора этого постоянного значения масштаба факторизации
в интервале $\mu_\text{F}^2=1-10$~ГэВ$^2$.

 \subsection{АТВ: выбор схемы и масштаба перенормировки}
  \label{subsec:pionFF.APT}
Рассмотрим сначала, что дает применение АТВ для факторизуемой части
ФФ пиона, следуя в основном работе~\cite{BPSS04},
где с самого начала выбирался масштаб факторизации
$\mu_\text{F}^2=Q^2$.
При таком выборе логарифм $\ln(Q^2/\mu_\text{F}^2)$ обращается в ноль
и не возникает проблем с его аналитизацией.
Однако при этом возникают эволюционные факторы
$E_n^{\text{LO}}(Q^2,\mu_0^2)\sim \alpha_s^{\nu_n}(Q^2)$,
см.(\ref{eq:anLO}),
где $\nu_n=\gamma_n^{(0)}/(2b_0)$ --- дробное число,
проводить аналитизацию которых в АТВ нет возможности.
Поэтому в~\cite{BPSS04} мы проводили численное решение
уравнений эволюции ЕРБЛ
с аналитической константой связи.
Таким способом были исследованы стандартная схема перенормировки
$\overline{\text{MS}}$ с несколькими способами
выбора масштаба перенормировки
(стандартный $\mu_\text{F}^2=Q^2$, по принципу минимальной чувствительности,
 по методу быстрейшей сходимости,
 простое и модифицированное BLM-предписания),
а также $\alpha_\text{V}$-схема~\cite{BJPR98}.

\subsubsection{Схема $\overline{\text{MS}}$}
 \label{sect:MS-bar}
Итак, займемся различными способами выбора масштаба перенормировки
$\mu_\text{R}^2$ в $\overline{\text{MS}}$-схеме.
Формулы (\ref{eq:Q2pff})--(\ref{eq:Q2.pff.NLO.Dia})
показывают нам,
что эта зависимость имеется в эффективном заряде $\alpha_s(\mu_\text{R}^2)$,
а также в неведущей поправке ${\cal F}_{\pi}^{(1,\beta)}$,
которая пропорциональна коэффициенту $b_0$ и, таким образом,
зависит явно от $N_f$.
В результате, естественно возникает вопрос:
как определить правильное значение $N_f$ в
формулах (\ref{eq:Q2pff})--(\ref{eq:Q2.pff.NLO.Dia})
для ФФ?

Имеется несколько рецептов. Разберем их по порядку.
\begin{enumerate}
  \item[(i)] В первом из них предлагается основываться на стандартном выборе
 $\mu_\text{R}^2=Q^2$ (или $\mu_\text{R}^2=Q^2/\lambda_{\text{PMS}}$,
 или $\mu_\text{R}^2=Q^2/\lambda_{\text{FAC}}$)
 и сдвигать $\mu_\text{R}^2$ на порогах тяжелых кварков так,
 чтобы обеспечить непрерывность самого ФФ.
  \item[(ii)] Во втором подходе используется BLM-предписание для выбора
  масштаба перенормировки, $\mu_\text{R}^2=\mu_\text{BLM}^2$,
  при котором $b_0$-вклад полностью обращается в ноль.
  В этом случае единственной проблемой является малые значения
  масштаба $\mu_\text{R}^2\approx Q^2/100$,
  поскольку $b_0$-вклад полностью отсутствует
  и $N_f$-зависящие вклады не появляются.
  \item [(iii)] Модифицированное BLM-предписание ($\overline{\text{BLM}}$)
  предлагает использовать $\mu_\text{R}^2=\mu_\text{BLM}^2$ только,
  когда этот масштаб больше или порядка
  некоторого минимального масштаба, $\mu_\text{min}$,
  имеющего значение некоторого характерного адронного масштаба,
  например, $m_\rho^2$.
  Ниже этого значения BLM-масштаб замораживается на величине
  $\mu_\text{min}$, где только легкие кварки дают вклад,
  и, таким образом,  $N_f=3$.
\end{enumerate}

Обрыв ряда теории возмущений на конечном порядке
(в нашем случае --- на третьем, второй порядок мы учитываем)
индуцирует остаточную зависимость результатов анализа
от масштаба $\mu_\text{R}$,
в  то время как учет более высоких порядков теории возмущений
должен уменьшать эту зависимость.
Мы проанализируем несколько способов выбора этого масштаба,
предложенных как раз с целью ослабить зависимость получаемых
результатов от отбрасывания неизвестных поправок:
\begin{enumerate}
  \item Простейший и широко используемый стандартный выбор
  \begin{eqnarray}
    \mu_\text{R}^2 = Q^2
   \label{eq:muRQ2}
  \end{eqnarray}
  основан на допущении,
  что отбрасываемые вклады имеют порядок $\alpha_s^3$
  и при $Q^2\gtrsim10$~ГэВ$^2$ малы.
  Однако в области $Q^2\lesssim10$~ГэВ$^2$
  этот выбор не столь очевиден.
  \item Принцип быстрейшей сходимости (FAC),
  предложенный в~\cite{Grunberg80,Grunberg84},
  предлагает фиксировать $\mu_\text{R}$ из требования,
  чтобы следующая за ведущей поправка полностью обращалась
  в ноль. В нашем случае это означает
  \begin{eqnarray}
    {\cal F}_{\pi}^{\text{NLO}}(Q^2;\mu_\text{F}^2,\mu_\text{R}^2=\mu_{\text{FAC}}^2)=0\,.
   \label{eq:FACpff}
   \end{eqnarray}
  \item Принцип минимальной чувствительности
  (PMS)~\cite{Stevenson81a,Stevenson81,Stevenson82,Stevenson84},
   с другой стороны, минимизирует зависимость результатов от выбора
   масштаба требованием совпадения $\mu_\text{R}$ со стационарной точкой
   оборванного пертурбативного ряда, т.~е.
   \begin{eqnarray}
    \frac {d} {d\mu_\text{R}^2}
     \left[\alpha_s(\mu_\text{R}^2)\,
            {\cal F}_{\pi}^\text{LO}(Q^2;\mu_\text{F}^2)
         + \frac{\alpha_s^2(\mu_\text{R}^2)}{\pi}\,
            {\cal F}_\pi^\text{NLO}(Q^2;\mu_\text{F}^2,\mu_\text{R}^2)
     \right]_{\displaystyle\mu_\text{R}^2=\displaystyle\mu_{\text{PMS}}^2}
       = 0\,.
      \label{eq:PMSpff}
    \end{eqnarray}
  \item BLM-предписание~\cite{BLM83} предлагает
  все эффекты поляризации вакуума,
  пропорциональные первому коэффициенту $\beta$-функции КХД $b_0$,
  перевести в изменение масштаба эффективного заряда КХД:
   \begin{eqnarray}
     {\cal F}_{\pi}^{(1,\beta)}(Q^2;\mu_\text{F}^2,\mu_\text{R}^2=\mu_{\text{BLM}}^2)=0\,.
    \label{eq:BLMpff}
   \end{eqnarray}
\end{enumerate}

\begin{table}[t]
\caption{Масштабы $\mu_{\text{PMS}}$, $\mu_{\text{FAC}}$,
 $\mu_{\text{BLM}}$ и $\mu_{V}$
 для АР пиона: асимптотической, БМС и ЧЖ.
 \label{tab:scales}}
\begin{ruledtabular}
\begin{tabular}{cccccc}
АР   & $\lambda_\text{FAC}=Q^2/\mu_{\text{FAC}}^2$
        & $\lambda_\text{PMS}=Q^2/\mu_{\text{PMS}}^2$
           & $\lambda_\text{BLM}=Q^2/\mu_{\text{BLM}}^2$
              & $\lambda_\text{V}=Q^2/\mu_{V\vphantom{_{|}}}^2$
                 & $Q^2$ \\ \hline \hline
Асимп.   & $\displaystyle 18$
        & $\displaystyle 27$
           & $\displaystyle 106$
              & $\displaystyle 20$
                 & любое \\
БМС  & $\displaystyle
       16-20$
        & $\displaystyle 24-29$
           & $\displaystyle 105-117$
              & $\displaystyle 20-22$
                 & $1-50$ ГэВ$^2$  \\
ЧЖ  & $\displaystyle 146-62$
        & $\displaystyle 217-92$
           & $\displaystyle 475-278$
              & $\displaystyle 90-52$
                 & $1-50$ ГэВ$^2$
\end{tabular}
\end{ruledtabular}
\end{table}
В таблице~\ref{tab:scales} мы показываем численные значения
обсуждаемых масштабов для различных АР пиона:
экстремальных
(асимптотической~\cite{ER80,ER80tmf} и Черняка--Житницкого(ЧЖ)~\cite{CZ82}),
и реалистической --- Бакулева--Михайлова--Стефаниса (БМС)~\cite{BMS01,BMS01c}.
Это ставит под сомнение применимость
предписания BLM при экспериментально доступных значениях $Q^2$
в обычной ТВ КХД.
Однако в АТВ оно может применяться безо всяких ограничений,
и, как мы увидим, дает вполне разумные результаты.

\subsubsection{$\alpha_V$-Схема}
 \label{sect:V-Scheme}
В работе~\cite{BJPR98} была предложена модификация
BLM-предписания в так называемой $\alpha_V$-схеме,
в которой эффективный заряд $\alpha_V(\mu^2)$
определяется с использованием потенциала в системе тяжелых кварков
$V(\mu^2)$.
Соотношение между эффективными зарядами $\alpha_{\overline{\text{MS}}}$
и $\alpha_V$ дается~\cite{BJPR98}
\begin{eqnarray}
 \alpha_s(\mu_{\text{BLM}}^2)
  = \alpha_V(\mu_V^2)
     \left[1
         + \frac{\alpha_V(\mu_V^2)}{4\pi}\,
            \frac{8 C_\text{A}}{3}
         + \cdots
     \right]\,,\quad
 \text{где}\quad
 \mu_V^2 = e^{5/3}\,\mu_{\text{BLM}}^2\,.
\label{eq:alphaV}
\end{eqnarray}
Численные значения масштаба $\mu_V$ для трех выбранных пионных АР
показаны в таблице~\ref{tab:scales}.
С учетом (\ref{eq:alphaV}) $O(\alpha_s)$-вклад для ФФ пиона,
задаваемый (\ref{eq:Q2.pff.NLO}),
модифицируется в этой схеме следующим образом
\begin{eqnarray}
 \alpha_s(\mu_\text{R}^2)
   & \rightarrow & \alpha_V(\mu_V^2)
  \nonumber \\   \label{eq:FpialphaV} \\
  {\cal F}_{\pi}^\text{NLO}(Q^2;\mu_\text{F}^2)
   & \rightarrow &
  {\cal F}_{\pi}^{(1,\text{FG})}(Q^2;\mu_\text{F}^2)
  + 2 {\cal F}_{\pi}^{\text{LO}}(Q^2;\mu_\text{F}^2)\,.
          \nonumber
\end{eqnarray}
При этом в качестве $\alpha_V(\mu^2)$ в АТВ
мы будем использовать сам аналитический эффективный заряд
${\mathcal A}_{1}(\mu^2)$,
который всюду конечен и по этой причине вполне подходит на роль
низкоэнергетического физического эффективного заряда.

\subsubsection{Численные результаты для ФФ пиона: триумф АТВ}
\label{subsec:Num.FF.Pion}
Процедура аналитизации пионного формфактора в $O(\alpha_s^2)$-порядке
ведет к неоднозначности, впервые обсуждавшейся в~\cite{SSK00}.
Именно, возникает вопрос:
если мы заменяем $\alpha_s(\mu^2)\to{\mathcal A}_{1}(\mu^2)$,
то как поступать с квадратом эффективного заряда,
$\alpha_s^2(\mu^2)$?
Исторически было предложено два рецепта действий:

(i)  В работе~\cite{SSK00} использовалась схема ``наивной аналитизации'',
в которой $\alpha_s^2(\mu^2)\to\left[{\mathcal A}_{1}(\mu^2)\right]^2$:
\begin{subequations}
 \begin{eqnarray}
  \left[F_{\pi}^\text{Fact}(Q^2;\mu_\text{F}^2,\mu_\text{R}^2)\right]_\text{NaiAn}
   = {\mathcal A}_{1}(\mu_\text{R}^2)\,
      {\cal F}_{\pi}^\text{LO}(Q^2;\mu_\text{F}^2)
   + \frac{\left[{\mathcal A}_{1}(\mu_\text{R}^2)\right]^2}{\pi}\,
      {\cal F}_{\pi}^{\text{NLO}}(Q^2;\mu_\text{F}^2,\mu_\text{R}^2)\,.~
 \label{eq:pffNaivAn}
\end{eqnarray}
Заметим, что в этом подходе квадрат аналитической константы связи
не имеет дисперсионного представления,
а значит и пионный формфактор тоже.

(ii) Следуя принципу аналитизации пионного формфактора ``как целого''~\cite{KS01,Ste02},
в работе~\cite{BPSS04} была предложена схема ``максимальной аналитизации'',
когда $\alpha_s^2(\mu^2)\to{\mathcal A}_{2}(\mu^2)$:
\begin{eqnarray}
 \left[F_{\pi}^\text{Fact}(Q^2;\mu_\text{F}^2,\mu_\text{R}^2)\right]_\text{MaxAn}
  = {\mathcal A}_{1}(\mu_\text{R}^2)\,
     {\cal F}_{\pi}^\text{LO}(Q^2;\mu_\text{F}^2)
  + \frac{{\mathcal A}_{2}(\mu_\text{R}^2)}{\pi}\,
     {\cal F}_{\pi}^\text{NLO}(Q^2;\mu_\text{F}^2,\mu_\text{R}^2)\,.~
 \label{eq:pffMaxAn}
\end{eqnarray}
\end{subequations}
В обоих случаях в качестве аналитических зарядов ${\mathcal A}_{1}(\mu_\text{R}^2)$
и ${\mathcal A}_{2}(\mu_\text{R}^2)$
использовались двухпетлевые глобальные заряды
${\mathcal A}_{1}^{(2);\text{\tiny glob}}(\mu_\text{R}^2)$
и ${\mathcal A}_{2}^{(2);\text{\tiny glob}}(\mu_\text{R}^2)$.

\begin{figure}[bht]
 \vspace*{0mm}
  \centerline{\includegraphics[width=0.32\textwidth]{
   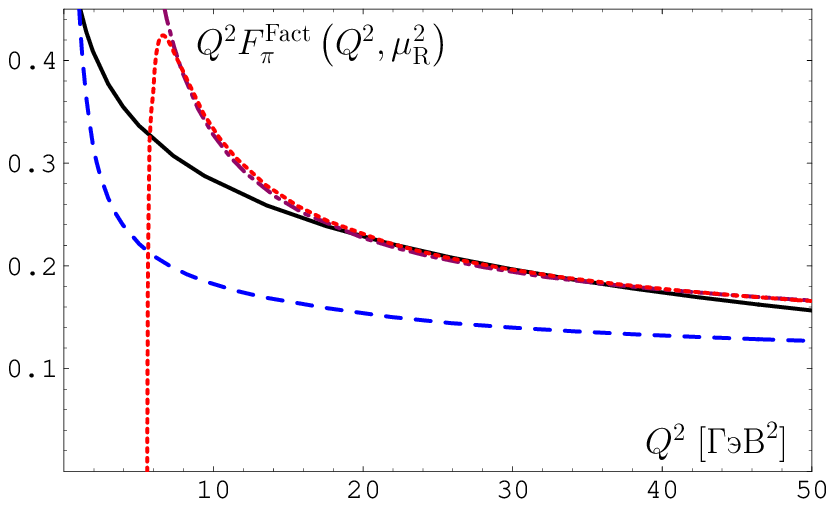}~
              \includegraphics[width=0.32\textwidth]{
   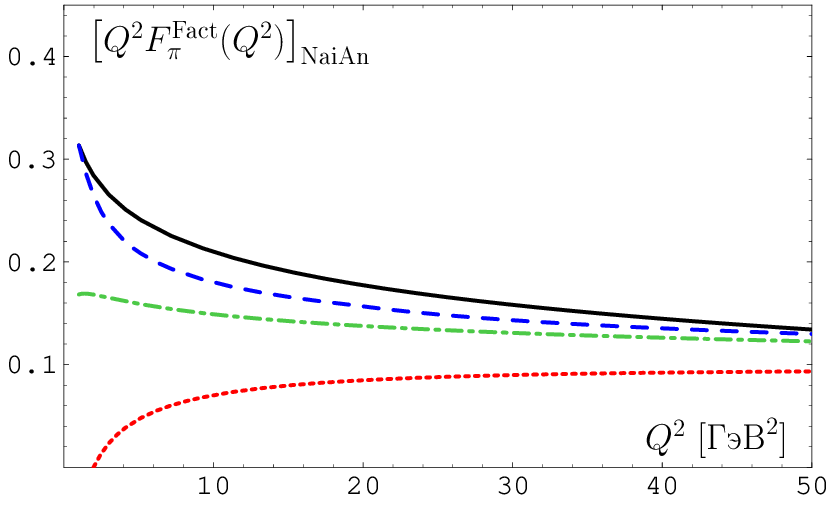}~
              \includegraphics[width=0.32\textwidth]{
   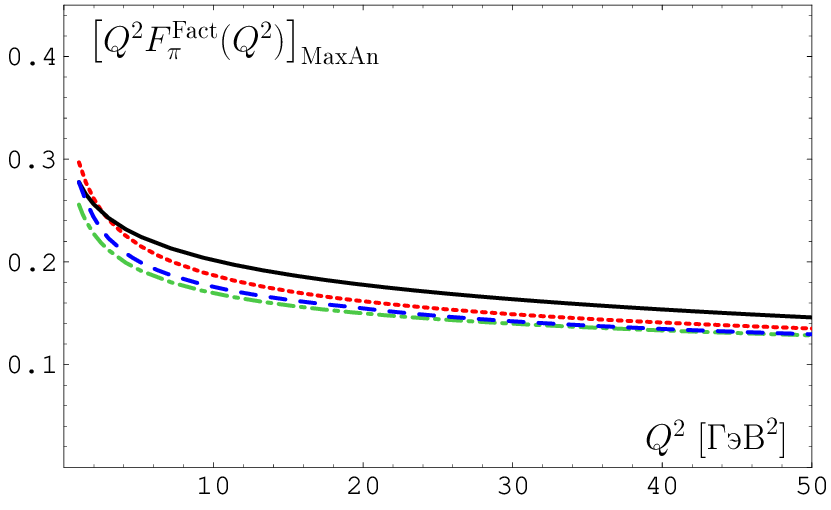}}
\vspace*{0mm}
\caption{\footnotesize
  Результаты для $Q^2 F_{\pi}^{\text{Fact}}$,
  полученные в стандартной ТВ (слева),
  АТВ с ``наивной аналитизацией'' (в центре) и
  с ``максимальной аналитизацией'' (справа).
  Обозначения кривых см. в тексте. Все расчеты проведены для
  реалистической АР~\cite{BMS01,BMS01c}.
  \label{fig:PFF.PT.ATP-Nai.ATP-Max}}
\end{figure}

На рис.~\ref{fig:PFF.PT.ATP-Nai.ATP-Max}
мы показываем полученные результаты
для факторизуемой части пионного формфактора
в различных подходах:
в стандартной теории возмущений КХД (слева),
в схемах АТВ с ``наивной аналитизацией'' (в центре) и
с ``максимальной аналитизацией'' (справа).
В случае стандартной теории возмущений
мы видим большое расхождение предсказаний в области
$Q^2=1-50$~ГэВ$^2$.
На левом рисунке показаны предсказания,
отвечающие следующим масштабам перенормировки:
стандартному $\mu_\text{R}^2=Q^2$ (штрихованная линия),
$\overline{\strut \text{BLM}}$ с $\mu_\text{min}^2=1$~ГэВ$^2$
(сплошная линия),
FAC (штрих-пунктирная линия)
и PMS (пунктирная линия).
Простой BLM-выбор приводит к результатам,
просто не помещающимся в видимой части рисунка.

В случае АТВ с ``наивной аналитизацией'' (на рисунке в центре)
мы видим не столь большое расхождение предсказаний в области
$Q^2=1-50$~ГэВ$^2$ по сравнению со стандартной ТВ КХД,
особенно если учесть, что теперь показаны предсказания,
полученные с BLM-предписанием (пунктирная линия) и
в $\alpha_\text{V}$-схеме (штрих-пунктирная линия).
Но все же это расхождение заметно
и оставляет вопрос о выборе масштаба открытым.
А вот для АТВ с ``максимальной аналитизацией'' (на рисунке справа)
вопрос о выборе масштаба практически не встает вообще:
все схемы
(обозначения кривых те же, что и на центральном рисунке)
дают практически одни и те же результаты!
Этот нетривиальный результат применения АТВ,
с нашей точки зрения, свидетельствует о мощи
принципа аналитичности в квантовой теории поля.

 \subsection{ДАТВ: Зависимость от масштаба факторизации}
  \label{subsec:pionFF.FAPT}
В этом подразделе мы будем обсуждать, что дает применение ДАТВ
при анализе пионного ФФ.
ДАТВ может применяться в двух разных подходах:
\begin{itemize}
  \item Можно, как это было сделано в разделе~\ref{subsec:pionFF.APT},
   зафиксировать $\mu_\text{F}^2=Q^2$ и использовать ДАТВ для корректного учета
   эволюционных факторов $E_n^{\text{LO}}(Q^2,\mu_0^2)\sim \alpha_s^{\nu_n}(Q^2)$
   в гегенбауэровских коэффициентах $a_2(Q^2)$ и $a_4(Q^2)$;
  \item Можно же, напротив, зафиксировать $\mu_\text{F}^2=\textsl{const}$
   и использовать ДАТВ для корректного учета появляющегося эволюционного
   логарифма $\ln(Q^2/\mu_\text{F}^2)$ в (\ref{eq:Fcal.(1)F}).
\end{itemize}
Исторически, в работе~\cite{BKS05} мы проанализировали только второй
вариант применения ДАТВ.
Первый вариант до сих пор не анализировался и мы восполним этот пробел
в разделе~\ref{subsec:pionFF.APT.Global}.
В этом же разделе мы займемся подходом работы~\cite{BKS05} и сначала обсудим,
какие изменения по сравнению с разделом~\ref{subsec:pionFF.APT}
необходимо сделать при использовании ДАТВ
для учета эволюционного логарифма
в (\ref{eq:Fcal.(1)F}).

\begin{figure}[thb]
 \vspace*{0mm}
  \centerline{\includegraphics[width=0.32\textwidth]{
   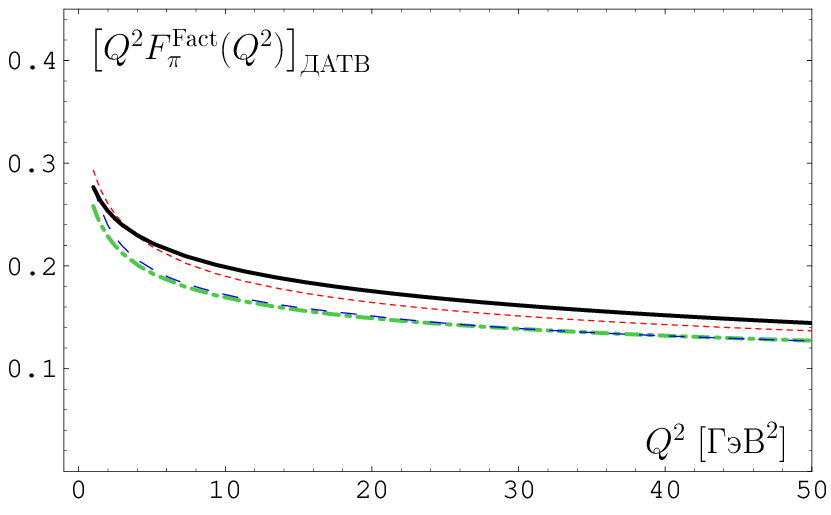}~
              \includegraphics[width=0.32\textwidth]{
   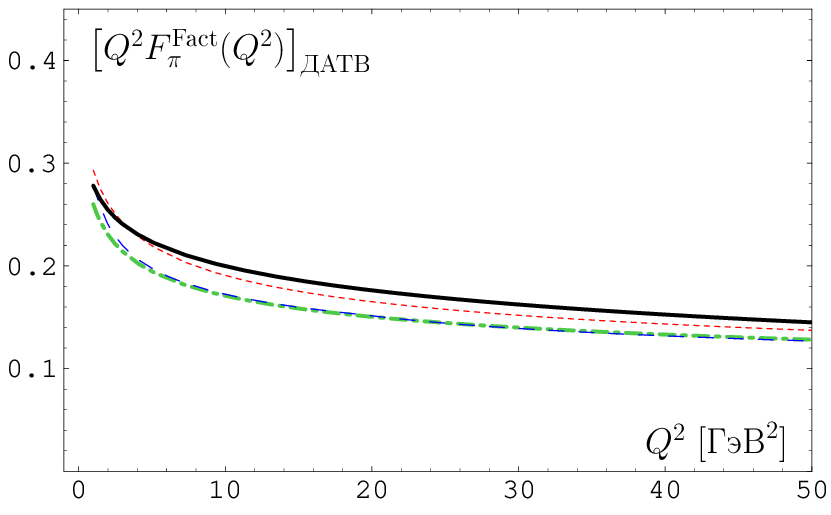}~
              \includegraphics[width=0.32\textwidth]{
   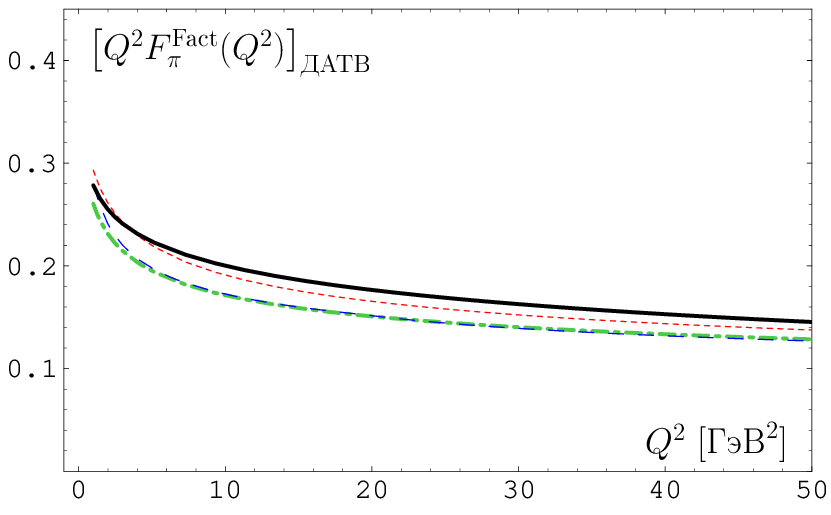}}
\vspace*{0mm}
\caption{\footnotesize
  Результаты для $Q^2 F_{\pi}^{\text{Fact}}$,
  полученные в подходе ДАТВ с различными масштабами
  факторизации: $\mu_\text{F}^2=1$~ГэВ$^2$ (слева),
  $\mu_\text{F}^2=5.76$~ГэВ$^2$  (в центре) и
  $\mu_\text{F}^2=10$~ГэВ$^2$  (справа).
  Обозначения кривых см. в тексте.
  Все расчеты проведены для реалистической АР~\cite{BMS01,BMS01c}.
  \label{fig:ffact-FAPT-BMS-muF}}
\end{figure}

Итак, мы фиксируем масштаб факторизации $\mu_\text{F}^2=\textsl{const}$,
так что гегенбауэровские коэффициенты $a_2(\mu_\text{F}^2)$ и $a_4(\mu_\text{F}^2)$
также становятся константами,
а масштаб перенормировки выбираем пропорциональным $Q^2$:
$\mu_\text{R}^2=\lambda_R Q^2$.
Взамен у нас возникает добавок
$(\alpha_s^2(\lambda_R Q^2)/\pi)\,C_\text{F}\,{\cal F}_\pi^{(1,\text{F})}(Q^2;\mu_\text{F}^2)$,
где ${\cal F}_\pi^{(1,\text{F})}(Q^2;\mu_\text{F}^2)$ содержит $\ln(Q^2/\mu_\text{F}^2)$,
см.~(\ref{eq:Fcal.(1)F}).
Дальше мы преобразуем этот логарифм очевидным образом:
\begin{eqnarray}
 \label{eq:Log.Transform}
  \ln\left(\frac{Q^2}{\mu_\text{F}^2}\right)
   &=& \ln\left(\frac{\lambda_R Q^2}{\Lambda_3^2}\right)
     + \ln\left(\frac{\Lambda_3^2}{\lambda_R\mu_\text{F}^2}\right)
    = L(\lambda_R\,Q^2)\,
    + \ln\left(\frac{\Lambda_3^2}{\lambda_R\mu_\text{F}^2}\right)\,,
\end{eqnarray}
и тогда
\begin{eqnarray}
  \left[\alpha_s^2(\lambda_R Q^2)\,
        \ln\left(\frac{Q^2}{\mu_\text{F}^2}\right)
  \right]_\text{ДАТВ}
  &\rightarrow&
      {\cal L}_{2;1}^{(2);\text{\tiny glob}}(\lambda_R\,Q^2)\,
    +
      {\mathcal A}_2^{(2);\text{\tiny glob}}(\lambda_R\,Q^2)\,
       \ln\left(\frac{\Lambda_3^2}{\lambda_R\mu_\text{F}^2}\right)~~~
 \label{eq:alpha_s^2.Log.1}\\
  &=& {\cal L}_{2;1}^{(2);\text{\tiny glob}}(\lambda_R\,Q^2)\,
    - {\mathcal A}_2^{(2);\text{\tiny glob}}(\lambda_R\,Q^2)\,
       L(\lambda_R\,Q^2)\nonumber\\
  &+& {\mathcal A}_2^{(2);\text{\tiny glob}}(\lambda_R\,Q^2)\,
       \ln\left(\frac{Q^2}{\mu_\text{F}^2}\right)\,.~~~~~
 \label{eq:alpha_s^2.Log.2}
\end{eqnarray}
Последнее слагаемое в этом выражении,
${\mathcal A}_2^{(2);\text{\tiny glob}}(\lambda_R\,Q^2)\,
  \ln\left(\frac{Q^2}{\mu_\text{F}^2}\right)$,
дает аналитизацию всего вклада
$\alpha_s^2(\lambda_R Q^2)\,\ln\left(\frac{Q^2}{\mu_\text{F}^2}\right)$
в подходе ``максимальной аналитизации'' АТВ.
Следовательно, разность
${\cal L}_{2;1}^{(2);\text{\tiny glob}}(\lambda_R\,Q^2)
-{\mathcal A}_2^{(2);\text{\tiny glob}}(\lambda_R\,Q^2)\,L(\lambda_R\,Q^2)$
и дает эффект аналитизации по ДАТВ.
Получаемые при учете этого вклада ФФ мы будем обозначать
$\left[F_{\pi}^{\text{Fact}}(Q^2)\right]_{\text{ДАТВ}}$.
На рисунке~\ref{fig:ffact-FAPT-BMS-muF} мы показываем результаты расчетов
таких ФФ при различных значениях масштаба факторизации
(см.~подпись под рисунком).
Различные кривые на рисунке отвечают различным схемам
и выборам масштаба перенормировки
(обозначения согласованы с обозначениями
 рис.~\ref{fig:PFF.PT.ATP-Nai.ATP-Max}):
сплошная (черная) линия соответствует $\overline{\strut \text{BLM}}$-предписанию
с $\mu_\text{min}^2=1$~ГэВ$^2$,
штрихованная (синяя) линия --- стандартному выбору $\lambda_R=1$,
штрих-пунктирная (зеленая, почти сливается со штрихованной) --- $\alpha_V$-схеме,
пунктирная (красная) --- BLM-предписанию.
Заметим, что при построении этих графиков мы пользовались
точным аналитизированным выражением
${\cal L}_{2;1}^{(2);\text{\tiny glob}}(\lambda_R\,Q^2)$,
которое очень близко к приближенному,
рассчитываемому по формуле (\ref{eq:Glo.alpha^2.Log.2-loop}),
и на 10--20\% отличается от приближенного
выражения (\ref{eq:Glo.alpha^2.Log.nai}),
использовавшегося в работе~\cite{BKS05}.

Мы видим опять, что все кривые очень близки друг к другу:
ширина полосы относительно центральной линии имеет порядок $7.5\%$,
а если не учитывать достаточно экзотическое
$\overline{\strut \text{BLM}}$-предписание --- то и вовсе  $5\%$.
Зависимость самих результатов от выбора масштаба факторизации
также оказывается очень малой:
относительная разность между результатами,
отвечающими $\mu_\text{F}^2=1$~ГэВ$^2$ и $\mu_\text{F}^2=10$~ГэВ$^2$,
составляет $1\%$ для случая $\alpha_V$-схемы
(для других выборов масштаба перенормировки --- еще меньше),
а если сравнивать случай $\mu_\text{F}^2=1$~ГэВ$^2$
со случаем $\mu_\text{F}^2=50$~ГэВ$^2$,
тогда относительная разница результатов доходит до $2\%$.

Таким образом, использование принципа аналитичности
также снижает зависимость от выбора масштаба факторизации.

 \subsection{АТВ и ДАТВ: Изящное решение проблемы порогов тяжелых кварков}
  \label{subsec:pionFF.APT.Global}
 Итак, мы хотим теперь зафиксировать $\mu_\text{F}^2=Q^2$
и использовать ДАТВ для корректного учета эволюционных факторов
$E_n^{\text{LO}}(Q^2,\mu_0^2)\sim \alpha_s^{\nu_n}(Q^2)$
в гегенбауэровских коэффициентах $a_2(Q^2)$ и $a_4(Q^2)$.
 Нам интересно сравнить полученный при таком образе действий
результат с тем,
что мы получили в предыдущем разделе.

Но прежде чем приступить к расчетам и получающимся результатам,
давайте обсудим чрезвычайно красивое (по мнению автора)
решение проблемы порогов тяжелых кварков,
предоставляемое АТВ.
Напомним сами себе, в чем состоит проблема.
Формулы (\ref{eq:Q2pff})--(\ref{eq:Q2.pff.NLO.Dia})
показывают,
что неведущая поправка ${\cal F}_{\pi}^{(1,\beta)}$ пропорциональна
коэффициенту $b_0$ и, таким образом, зависит явно от $N_f$.
Это и есть проблема: при выборе $\mu_\text{R}$ в
формулах (\ref{eq:Q2pff})--(\ref{eq:Q2.pff.NLO.Dia})
для ФФ по стандартному рецепту $\mu_\text{R}^2=Q^2$
и сдвигов $\mu_\text{R}^2$ на порогах тяжелых кварков так,
чтобы обеспечить непрерывность самого ФФ,
мы получим ФФ пиона как непрерывную функцию,
но уж никак не аналитическую.

Решение этой проблемы в АТВ полностью аналогично
рецепту глобализации эффективного заряда в АТВ:
зависящие от $N_f$ пертурбативные результаты
дают нам зависящие от $N_f$ спектральные плотности,
разрывные на порогах,
однако генерируемые ими через дисперсионные интегральные представления
эффективные заряды в евклидовой области
являются аналитическими функциями!

Подробнее это выглядит так.
В формулах (\ref{eq:Q2pff})--(\ref{eq:Q2.pff.NLO.Dia})
у нас есть три типа величин,
которые зависят явно от $N_f$:
$\alpha_s(Q^2)$,
$\alpha_s(Q^2)^2$ и
$b_0(N_f)\alpha_s(Q^2)^2$,
причем первые две величины непрерывны на порогах,
а последняя --- разрывна.
Очевидно, что эта последняя величина генерирует следующую спектральную
плотность:
\begin{eqnarray}
 \label{eq:rho.b0.alpha2}
  \frac{1}{\pi}\,\textbf{Im}{}\left[b_0(N_f)\alpha_s[L-i\pi]^2\right]
   &\equiv& \bar{\rho}_{2;b_0}[L;N_f]
   \ =\  \frac{4\,\pi}{\beta_f}\,
          \rho_{2}[L]\,.
\end{eqnarray}
Таким образом,
у нас кроме ${\mathcal A}_1^\text{\tiny glob}[L]$ и
${\mathcal A}_2^\text{\tiny glob}[L]$
появится еще одна глобализованная аналитическая функция
${\mathcal A}_{2;b_0}^\text{\tiny glob}[L]$,
определяемая глобальной спектральной плотностью
\begin{eqnarray}
 \rho_{2;b_0}^{\text{\tiny glob}}[L]
  &=& \bar{\rho}_{2;b_0}\left[L;3\right]
       \theta\left(L<L_{4}\right)
    + \bar{\rho}_{2;b_0}\left[L+\lambda_4;4\right]\,
       \theta\left(L_{4}\leq L<L_{5}\right)
  \nonumber\\
  &+& \bar{\rho}_{2;b_0}\left[L+\lambda_5;5\right]\,
       \theta\left(L_{5}\leq L<L_{6}\right)
    + \bar{\rho}_{2;b_0}\left[L+\lambda_6;6\right]\,
       \theta\left(L_{6}\leq L\right)\,.~~~
 \label{eq:Rho_[2.b0].Glo}
\end{eqnarray}
На каждом из этих трех базовых аналитических эффективных зарядов
еще ``вырастут веточки'' за счет наличия эволюционных факторов.
Разберем их появление на примере ${\mathcal A}_1^\text{\tiny glob}[L]$.
У нас есть следующие эволюционные факторы:
единица и пять факторов типа $(\alpha_s[L])^{\nu_i}$
с
$\nu_1=\gamma_2^{(0)}/(2b_0)$,
$\nu_2=\gamma_4^{(0)}/(2b_0)$,
$\nu_3=2\gamma_2^{(0)}/(2b_0)$,
$\nu_4=(\gamma_2^{(0)}+\gamma_4^{(0)})/(2b_0)$
и
$\nu_5=2\gamma_4^{(0)}/(2b_0)$,
т.~е. кроме ${\mathcal A}_1^\text{\tiny glob}[L]$
у нас появятся еще 5 дробно-индексных аналитических заряда
типа ${\mathcal A}_{1+\nu_i}^\text{\tiny glob}[L]$.

\begin{figure}[thb]
 \vspace*{0mm}
  \centerline{\includegraphics[width=0.47\textwidth]{
   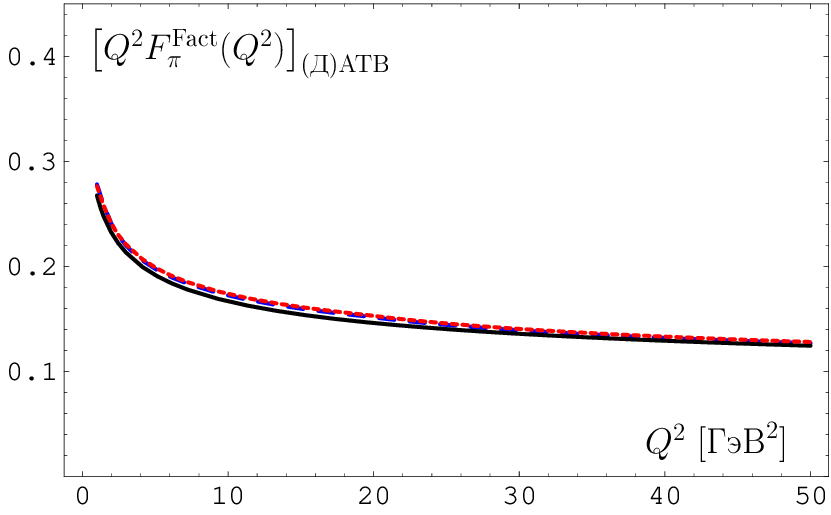}~~~
              \includegraphics[width=0.47\textwidth]{
   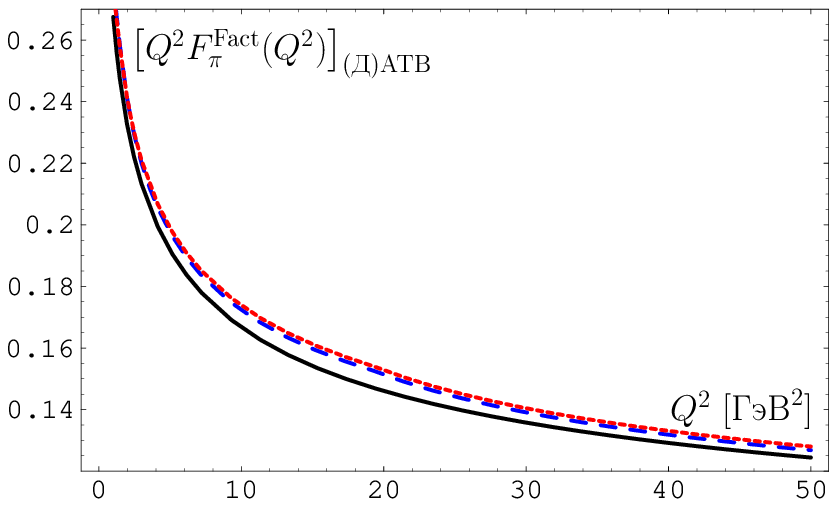}}
\vspace*{0mm}
\caption{\footnotesize
  Результаты для $Q^2 F_{\pi}^{\text{Fact}}$,
  полученные в подходах АТВ$_\text{MaxAn}$,
  ДАТВ c фиксированным масштабом факторизации
  и ДАТВ с $\mu_\text{F}^2=Q^2$.
  Обозначения кривых см. в тексте.
  На правом рисунке просто увеличен масштаб оси ординат,
  чтобы лучше можно было увидеть все три кривые.
  \label{fig:ffact-APTMax-FAPTmuF-FAPT}}
\end{figure}

Теперь мы можем взглянуть на результаты расчетов
пионного ФФ в разных подходах:
на рисунке~\ref{fig:ffact-APTMax-FAPTmuF-FAPT}
пунктирная (красная) линия отвечает подходу АТВ с ``максимальной аналитизацией'',
штрихованная (синяя) линия --- ДАТВ c $\mu_\text{F}^2=5.76$~ГэВ$^2$,
в то время как сплошная (черная) линия показывает результат ДАТВ с $\mu_\text{F}^2=Q^2$
(везде выбран масштаб перенормировки $\mu_\text{R}^2=Q^2$).

Полученное согласие
(на уровне 1.5\%, что можно увидеть на правой части
рисунка~\ref{fig:ffact-APTMax-FAPTmuF-FAPT})
впечатляет!

 \subsection{Переход в область Минковского:
             роль дисперсионного представления}
  \label{subsec:pionFF.FAPT.Minkovsky}
Из дисперсионного представления (\ref{eq:An.AU_nu.Glo.L})
для аналитического эффективного заряда КХД
${\mathcal A}_{\nu}^{\text{\tiny glob}}[L]$,
мы имеем следующее представление для него
в области $Q^2=-s$ с $s>0$
\begin{eqnarray}
 \label{eq:An_A_SS_TL}
  {\mathcal A}_{\nu}^{\text{\tiny glob}}\left(-s\right)
  &=& {\cal P.V.}\int_0^{\infty}\!
       \frac{\rho_{\nu}^{\text{\tiny glob}}\left(\sigma\right)}
            {\sigma-s}\,
        d\sigma
      +i\,\pi\,\rho_{\nu}^{\text{\tiny glob}}\left(s\right)\,.
\end{eqnarray}
Заметим здесь, что из дисперсионного представления ФФ пиона следует,
что именно эти выражения
должны входить в выражение для факторизуемой части пионного ФФ
(с $\nu=1$ и $\nu=2$)
во временноподобной области передач импульса:
\begin{eqnarray}
 s\,F_{\pi}^\text{Fact}(-s)
  &\sim & {\mathcal A}_{1}^{\text{\tiny glob}}(-\lambda_\text{R}s)\,
               {\cal F}_{\pi}^\text{LO}
        + \frac{{\mathcal A}_{2}^{\text{\tiny glob}}(-\lambda_\text{R}s)}{\pi}\,
               {\cal F}_{\pi}^\text{NLO}
 \label{eq:SL_pffKS}
\end{eqnarray}
где ${\cal F}_{\pi}^\text{LO}$ и ${\cal F}_{\pi}^\text{NLO}$
--- числовые множители, зависящие от параметров
$a_2(\mu_\text{F}^2)$ и $a_4(\mu_\text{F}^2)$ пионной АР,
а $\lambda_\text{R}$ --- числовой параметр,
характеризующий выбор масштаба перенормировки
$\mu^2_\text{R}=\lambda_\text{R}Q^2$.
Появление мнимых частей в ФФ выражениях совершенно естественно
и связано с рождением реальных физических частиц.
Конечно, в нашей модели эффективного заряда безмассовой КХД,
в которой нет учета адронизации свободных кварков
в реальные физические частицы и пороги тяжелых кварков
учитываются достаточно грубо,
эта мнимая часть отлична от 0 даже и в области малых
$s\sim m_\pi^2$,
где никаких физических частиц не рождается,
но этот недостаток может быть исправлен,
как это предложено в работах~\cite{NP04,ANP05,NP05}.

Тем не менее, в этой области вклад факторизованной части
пионного ФФ в суммарный ФФ (\ref{eq:pff}) мал
(детали см. в~\cite{BPSS04}),
так что этот недостаток не столь важен.
В области же $s\geq m_\rho^2$ у нашего ФФ всюду имеется
мнимая часть и это хорошо объясняется эффектами рождения
реальных адронов.
Как раз отсутствие мнимой части у ФФ пиона в этой области
должно было бы вызывать недоумение.
А именно такое отсутствие гарантируется использованием
при расчете ФФ аналитического эффективного заряда
${\mathfrak A}_{\nu}(s)$,
предложенного в~\cite{Rad82,KP82} для описания сечения
$e^{+}e^{-}$-аннигиляции в адроны.
Для сечения, которое должно быть всегда вещественным,
такое предложение совершенно оправдано и дает,
как мы говорили во введении,
вполне разумные результаты.
А вот для расчетов ФФ такое использование оказывается
физически неоправданным --- здесь нужно
использовать заряд (\ref{eq:An_A_SS_TL}),
т.~е. ${\mathcal A}_{\nu}^{\text{\tiny glob}}\left(-s\right)$.

\begin{figure}[b]
 \centerline{\includegraphics[width=0.47\textwidth]{
  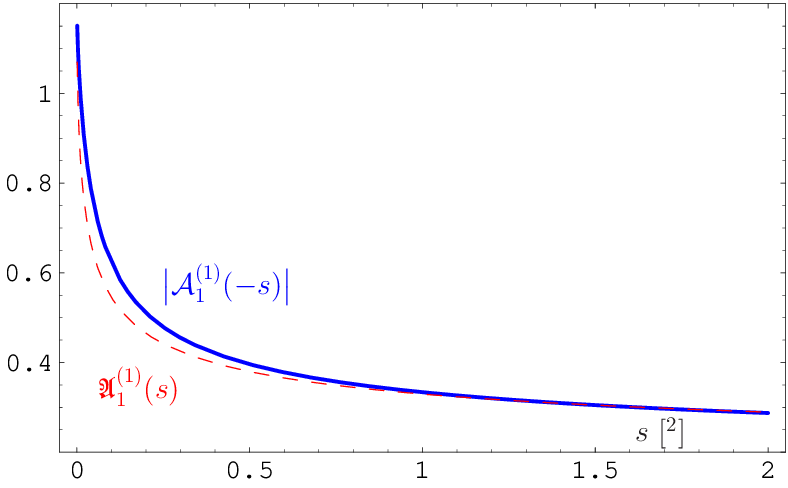}~~~%
             \includegraphics[width=0.47\textwidth]{
  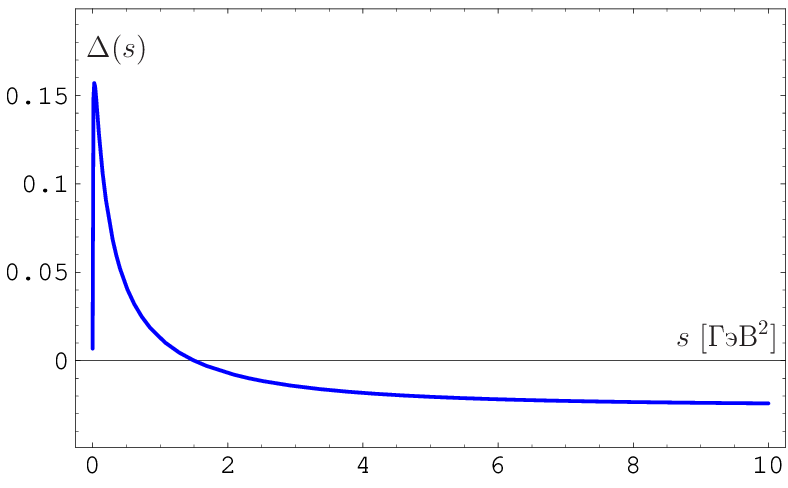}}
   \caption{\footnotesize
    Слева показаны зависимости $\big|{\mathcal A}_{1}^{(1)}[L_s-i\pi]\big|$
    (сплошная линия) и ${\mathfrak A}_{1}^{(1)}[L_s]$ (штрихованная линия),
    справа --- $\Delta(s)$.\label{fig:Comp.Im.Re.Abs}}
\end{figure}

Насколько важно такое уточнение численно
для абсолютной величины ФФ?
На левой панели рис.~\ref{fig:Comp.Im.Re.Abs}
мы сравниваем зависимости
$\big|{\mathcal A}_{1}^{(1)}[L_s-i\pi]\big|$ (сплошная линия)
и
${\mathfrak A}_{1}^{(1)}[L_s]$ (штрихованная линия),
а на правой панели этого рисунка показана относительная разность
этих двух функций,
$\Delta(s)=1-{\mathfrak A}_{1}^{(1)}(s)/\big|{\mathcal A}_{1}^{(1)}(-s)\big|$.
Хорошо видно, что максимальное отличие имеет порядок
$+15\%$ и достигается при  $s\approx0.03$~ГэВ$^2$.

Учитывая, что в реальных приложениях анализируются
достаточно большие значения $s\sim6-10$~ГэВ$^2$,
получается, что численно заряд
${\mathfrak A}_{1}(s)$
оказывается достаточно близок (на уровне 3\%)
к $\big|{\mathcal A}_{1}(-s)\big|$.
Тем не менее,
нам представляется важным сделанное уточнение,
поскольку оно показывает не абсолютный характер
аналитического зарядов ${\mathfrak A}_{\nu}(s)$
в области Минковского:
они пригодны только для описания величин типа сечений
$R(s)=\sigma_{e^+e^-\to\text{hadrons}}(s)/\sigma_{e^+e^-\to\mu^+\mu^-}(s)$,
но не ФФ,
для которых аналитические свойства,
т.~е. дисперсионные представления типа (\ref{eq:A.E}),
диктуют использование
зарядов ${\mathcal A}_{\nu}^{\text{\tiny glob}}\left(-s\right)$.

  Интересным следствием использования формулы (\ref{eq:An_A_SS_TL})
для ФФ пиона в минковской области
является фиксация масштаба перенормировки
на значении $\mu^2_\text{R}=Q^2/4$.
Действительно,
мнимая часть
${\mathcal A}_{\nu}^{\text{\tiny glob}}\left(-\lambda_\text{R}s\right)$
пропорциональна спектральной плотности
$\pi\,\rho_{\nu}^{\text{\tiny glob}}\left(\lambda_\text{R}s\right)$,
которая скачками меняется при пересечении порогов тяжелых кварков,
т.~е. при $\lambda_\text{R}s=m_Q^2$ (где $Q=c, b, t$),
см. обсуждение в разделе~\ref{subsec:APT.Global}.
И только при выборе $\lambda_\text{R}=1/4$,
т.~е. $\mu^2_\text{R}=Q^2/4$,
положения этих нерегулярностей в мнимой части ФФ пиона
совпадают с кинематическими порогами образования
пары тяжелых кварков, $s_{Q\bar{Q}}=4 m_Q^2$.

\section{РАСЧЕТ ШИРИНЫ РАСПАДА $H^0\to\bar{b}b$ В ДАТВ}
 \label{sec:Higgs.width}
  Займемся теперь приложением ДАТВ в задаче расчета
ширина распада бозона Хиггса $H^0$ на кварк-антикварковую пару $b\bar{b}$.
В стандартной теории возмущений из-за аналитического продолжения
из глубоко евклидовой в глубоко минковскую область
возникают $\pi^2$-добавки в коэффициенты разложения:
$r_k=d_k+O(\pi^2)$, см., например, \cite{BCK05}.
Эти вклады могут быть велики,
особенно для высших коэффициентов разложения.
Следовательно, даже при высоких энергиях,
существенных для анализа распада хиггсовского бозона,
учет этих вкладов во всех порядках разложения представляется важным.
Но именно такой учет и доставляется применением ДАТВ\footnote{%
Интересно заметить, что в работе~\cite{GLK84} были фактически записаны
основные формулы 1-петлевой ДАТВ в пространстве Минковского
и даже обсуждена функциональная схема перехода из евклидовой области к минковской.
Но затем авторы проводили разложение по $1/L$, возвращаясь таким образом
к стандартной теории возмущений КХД.}:
аналитические эффективные заряды содержат все $\pi^2$-вклады
внутри себя
явно по построению!

Расчет  $\Gamma(\text{H} \to b\bar{b})$ с использованием функции $R_\text{S}$
проводится в $\overline{\text{MS}\vphantom{^1}}$-схеме с учетом эволюции как эффективного заряда и его степеней,
так и массы $b$-кварка, следуя работе~\cite{BMS06}.
Мы сравним полученные результаты  с полученными также в $\overline{\text{MS}\vphantom{^1}}$-схеме
в работах~\cite{Che96,ChKS97,BKM01,BCK05},
где широко использовался заряд $\displaystyle a_\text{s}=\alpha_\text{s}/\pi$,
так что и мы будем его использовать.

\subsection{Стандартная теория возмущений для $\mathbf{R}_\text{S}$}
\label{subsec:stand-Higgs}

Распад бозона Хиггса в кварк-антикварковую пару $b\bar{b}$
в КХД описывается с помощью коррелятора
двух скалярных (S) $b$-кварковых токов
$J^\text{S}_b=\bar{\Psi}_b\Psi_b$
\begin{eqnarray}
\label{eq:Scalar.Corr}
  \Pi(Q^2)
  &=& (4\pi)^2 i\int dx e^{iqx}\langle 0|\;T[\;J^\text{S}_b(x)
                                               J^\text{S}_{b}(0)\,]\;
                                      |0\rangle\,,
\end{eqnarray}
с $Q^2=-q^2$. Его мнимая часть
$R_\text{S}(s)=\textbf{Im}\,\Pi(-s-i\epsilon)/{(2\pi s)}$
определяет полную ширину распада:
\begin{eqnarray}
\Gamma(\text{H} \to b\bar{b})
=\frac{G_F}{4\sqrt{2}\pi}M_\text{H}
m_{b}^{2}(M_\text{H}^2) R_\text{S}(M_\text{H}^2)
\label{decay_rate_for_b}\,,
\end{eqnarray}
где $M_\text{H}$ --- масса хиггсовского бозона,
а $m_b(Q^2)$ --- эволюционирующая масса $b$-кварка.
Явные многопетлевые расчеты обычно проводятся в евклидовой области
для соответствующей коррелятору (\ref{eq:Scalar.Corr})
функции Адлера $D_\text{S}$
\cite{GLK84,KK94,Che96,BCK05,BKM01},
где хорошо работает теория возмущений КХД:
\begin{eqnarray}
 \widetilde{D}_{\text{S}}(Q^2;\mu^2)
  &=& 3\,m_b^2(Q^2)
       \left[1+\sum_{n \geq 1} d_n(Q^2/\mu^2)~a_\text{s}^n(\mu^2)
        \right]\,.
\label{eq:D-s}
\end{eqnarray}
Здесь мы использовали обозначение $\displaystyle a_\text{s}={\alpha_\text{s}/ \pi}$,
а тильда над $D_{\text{S}}(Q^2;\mu^2)$ напоминает,
что эта функция включает в себя также фактор $m_b^2(Q^2)$.
В согласии с обсуждением в разделе~\ref{sec:2} после уравнения~(\ref{eq:R})
мы можем сразу же записать для $\widetilde{R}_\text{S}$ соответствующее
разложение (см., например,~\cite{ChKS97}):
\begin{eqnarray}
  \widetilde{R}_\text{S}(s)
\equiv
  \widetilde{R}_\text{S}(s,s)
=
  3m^{2}_{b}(s)\left[ 1 + \sum_{n\geq 1}^{} r_{n}~a_\text{s}^n(s)
               \right]\, .
  \label{eq:R-s}
\end{eqnarray}
Коэффициенты $r_n$ здесь содержат характерные `$\pi^2$-вклады',
связанные с интегральным преобразованием $\hat{R}$
(аналитическим продолжением) степеней логарифмов,
появляющихся в $\widetilde{D}_{\text{S}}$.
Эти логарифмы возникают из двух различных источников:
одни --- из коэффициентов $d_n(Q^2/\mu^2)$ в $\widetilde{D}_\text{S}$,
появление логарифмов в которых связано с зависимостью эффективного заряда
$\alpha_\text{s}(\mu^2)$ от $\mu^2$,
другие связаны с наличием фактора $m_{b}^{2}(Q^2)$ и
определяются комбинацией аномальных размерностей массы кварка $\gamma_i$
(см. Приложение~\ref{app:D.Rep})
и коэффициентов $\beta$-функции $b_j$, умноженных на степени $\pi^2$
\cite{KK94,BKM01,Che96,ChKS97}.
Как показали недавние расчеты, выполненные в~\cite{BCK05},
учет таких $\pi^2$-вкладов может быть существенным:
\begin{eqnarray}
 \left[{3m_b^2}\right]^{-1}\widetilde{R}_\text{S}
 &=& 1
   + 5.667\,a_s
   + a_s^2\left[51.57
             - \underline{15.63}
             - N_f\left({1.907} - \underline{0.548}\right)
          \right]
 \nonumber\\
 && ~\,
   + a_s^3\left[648.7 - \underline{484.6}
             - N_f\left(63.74 - \underline{37.97}\right)
             + N_f^2\left(0.929 - \underline{0.67}\right)
          \right]
 \nonumber\\
 && ~\,
   + a_s^4\left[9470.8 - \underline{9431.4}
             - N_f\left(1454.3 - \underline{1233.4}\right)
          \right.
 \nonumber\\
 && ~~~~~~~~~~
          \left.
             + N_f^2\left(54.78 - \underline{45.10}\right)
             - N_f^3\left(0.454 -\underline{0.433}\right)
          \right]\,.
 \label{eq:RSnumx}
\end{eqnarray}
В этом выражении мы подчеркнули те $\pi^2$-вклады,
которые возникают за счет аналитического продолжения
из евклидовой области в минковскую.
Легко видно, что полный вклад таких членов по порядку величины сравним
со значением исходного коэффициента $d_n$,
особенно в случае коэффициента $d_4$.
Таким образом,
необходимо провести учет таких вкладов во всех порядках теории возмущений.
Именно это и позволяет сделать подход ДАТВ,
причем такой учет производится автоматически
без использования дополнительных процедур оптимизации.
Действительно, в ДАТВ нам не нужно разлагать перенормировочные множители
в ряд по логарифмам --- вместо этого мы можем преобразовать их
в минковскую область `как целое' посредством $\textbf{A}_\text{M}$-операции.

Завершим обсуждение стандартного подхода
численным результатом в $O(a_s^4)$-порядке,
взятым из~\cite{BCK05}:
\begin{eqnarray}
 \left[{3m_b^2}\right]^{-1}\widetilde{R}_\text{S}
 &=& 1 + 5.6668\,a_s
       + 29.147\,a_s^2
       + 41.758\,a_s^3
       - 825.7\,a_s^4
       \label{eq:R-Ch}      \\
 &=& 1 + 0.2075
       + 0.0391
       + 0.0020
       - 0.00148\,.
 \label{eq:R-Ch-numer}
\end{eqnarray}
В последнем уравнении выбрано $a_\text{s}=a_\text{s}(M_\text{H}^2)= 0.0366$,
что отвечает массе бозона Хиггса $M_\text{H} = 120$~ГэВ.

\subsection{Анализ $\widetilde{R}_\text{S}$ в ДАТВ}
\label{subsec:FAPT-R}
Сначала мы построим аналитизацию выражения
для $\widetilde{R}_\text{S}$,
считая, что коэффициенты $d_n$ в (\ref{eq:D-s}) не зависят от $N_f$,
а именно, зафиксируем их на значениях, отвечающих $N_f=5$,
см. Приложение~\ref{app:D.Rep}, формулы (\ref{eq:d_1-d_4}).
Начнем с учета эффектов, связанных с перенормировкой массы $b$-кварка.
Эволюция $m_{(l)}(Q^2)$ в $l$-петлевом приближении,
как показано в Приложении~\ref{app:D.Rep},
может быть записана в виде (\ref{eq:m2-hat-run}).
Таким образом, задача аналитизации выражения
для $\widetilde{D}_\text{S}$ свелась
к задаче аналитизации произведений вида
\begin{eqnarray}
 \label{eq:An.Mass.nu0.a^n}
   m_{(l)}^2[L]\left(\alpha_\text{s}[L]\right)^{n}
  &=& \hat{m}_{(l)}^2
       \left(\alpha_\text{s}[L]\right)^{n+\nu_0}
        f_{(l)}\left(\alpha_s[L]\right)\,,
\end{eqnarray}
где $\hat{m}_{(l)}^2$ есть просто числовая константа.
В принципе, можно действовать так как было предложено в~\cite{BMS06}:
разложить $f_{(l)}(\alpha_s[L])$ в ряд по степеням $(\alpha_s[L])^{m}$
и затем применять ДАТВ к этому разложению,
которое,
благодаря аналитизации,
очень быстро сходится
(достаточно учета четырех порядков для достижения точности, лучшей $0.01\%$).

Однако, мы предлагаем здесь использовать всю мощь ДАТВ и
получить точные формулы для аналитизации выражений (\ref{eq:An.Mass.nu0.a^n})
для случаев, когда $f_{(l)}(\alpha_s[L])$ представима в виде
(\ref{eq:f_2}) или (\ref{eq:f_3L.Pade}).
Разберем подробно случай двухпетлевой эволюции (\ref{eq:f_2}),
$$f_{(2)}(\alpha_s)=\left[1 + \delta_1\,\alpha_s\right]^{\nu_1}\,,$$
после чего обобщение на случай (\ref{eq:f_3L.Pade})
станет очевидным.
Итак, мы хотим построить глобальный аналитический образ
выражения
\begin{eqnarray}
 \label{eq:An.Evo.nu.delta.mu}
  B_{\nu}^{(2)}[L;\left\{\delta,\mu\right\}]
   \equiv
    \left(\alpha_\text{s}^{(2)}[L]\right)^{\nu}
     \left[1 + \delta\,\alpha^{(2)}_{s}[L]\right]^{\mu}\,,
\end{eqnarray}
т.~е. нам нужно построить спектральные плотности
$\bar{\rho}^{(2)}_{B_{\nu}\left\{\delta,\mu\right\}}[L;N_f]$,
отвечающие пертурбативному вкладу $B_{\nu}^{(2)}[L;\left\{\delta,\mu\right\}]$
при заданных числах флейворов $N_f$.
Спектральная плотность у нас определяется
мнимой частью $B_{\nu}^{(2)}[L-i\pi;\left\{\delta,\mu\right\}]$,
причем мы знаем из (\ref{eq:a_(2).1.Lamb.R.phi}),
что
\begin{eqnarray*}
  \alpha_\text{s}^{(2)}[L-i\pi;N_f]
   = \frac{\displaystyle e^{i\varphi_{(2)}[L]}}{\beta_f\,R_{(2)}[L]}\,.
\end{eqnarray*}
Тогда мы можем записать
\begin{subequations}
\begin{eqnarray}
  1+\delta\,\alpha_\text{s}^{(2)}[L-i\pi;N_f]
   = \frac{\displaystyle e^{i\varphi_{\Delta}[L,\delta,N_f]}}{R_{\Delta}[L,\delta,N_f]}\,,
 \label{eq:1.delta.alpha}
\end{eqnarray}
где
\begin{eqnarray}
 R_{\Delta}[L,\delta,N_f]
  &=& \frac{\beta_f\,R_{(2)}[L]}
           {\sqrt{\beta_f^2\,R_{(2)}^2[L]
                + 2\,\delta\,\beta_f\,R_{(2)}[L]\cos\varphi_{(2)}[L]
                + \delta^2
                 }
           }\,,\\
 \varphi_{\Delta}[L,\delta,N_f]
  &=& \varepsilon(\delta)\,
       \arccos
        \left[\frac{\beta_f\,R_{(2)}[L]+\delta\,\cos\varphi_{(2)}[L]}
                   {\sqrt{\beta_f^2\,R_{(2)}^2[L]
                         + 2\,\delta\,\beta_f\,R_{(2)}[L]\cos\varphi_{(2)}[L]
                         + \delta^2
                         }
                   }
        \right]\,,
\end{eqnarray}
\end{subequations}
где $\varepsilon(x)$ --- функция знака переменной $x$
(т.~е. $\varepsilon(x)=+1$ для $x>0$ и $\varepsilon(-x)=-\varepsilon(x)$).
После этого мнимая часть (\ref{eq:An.Evo.nu.delta.mu}) определяется
элементарно,
так что искомые спектральные плотности есть
\begin{eqnarray}
 \label{eq:rho-bar.Evo.nu.mu}
  \bar{\rho}^{(2)}_{B_{\nu}\left\{\delta,\mu\right\}}[L;N_f]
   = \frac{\sin\left[\nu\,\varphi_{(2)}[L]
                   + \mu\,\varphi_{\Delta}[L,\delta,N_f]
               \right]}
          {\left(\beta_f\,R_{(2)}[L]\right)^\nu
           \left(R_{\Delta}[L,\delta,N_f]\right)^\mu}
   \,,
\end{eqnarray}
а глобальная спектральная плотность
$\rho^{(2);\text{\tiny glob}}_{B_{\nu}\left\{\delta,\mu\right\}}[L]$
определяется по ним по формулам
типа (\ref{eq:Rho[L].Glo.nu}).
В результате аналитический образ выражения (\ref{eq:An.Mass.nu0.a^n}),
который мы будем обозначать
${\mathfrak B}_{n+\nu_{0}}^{(2);\text{\tiny glob}}[L]$,
определяется в двухпетлевом приближении
с помощью спектральной плотности
$\rho^{(2);\text{\tiny glob}}_{B_{n+\nu_0}\left\{\delta_1,\nu_1\right\}}[L]$
(выражения для $\delta_1$ и $\nu_1$ см. в Приложении~\ref{app:Anom.Dim.Quark.Mass}
 --- формулы (\ref{eq:f_2})).

В трехпетлевом Паде-приближении необходимые  спектральные плотности
имеют следующий вид
\begin{eqnarray}
 \label{eq:rho-bar.Evo.nu.mu1.mu2}
  \bar{\rho}^{(3)}_{B_{\nu}\left\{\delta_1,\mu_1;\delta_2,\mu_2\right\}}[L;N_f]
   = \frac{\sin\left[\nu\,\varphi_{(3\text{-P})}[L]
                   + \mu_1\,\varphi_{\Delta}[L,\delta_1,N_f]
                   + \mu_2\,\varphi_{\Delta}[L,\delta_2,N_f]
               \right]}
          {\left(\beta_f\,R_{(3\text{-P})}[L]\right)^\nu
           \left(R_{\Delta}[L,\delta_1,N_f]\right)^{\mu_1}
           \left(R_{\Delta}[L,\delta_2,N_f]\right)^{\mu_2}}
   \,,
\end{eqnarray}
глобальная спектральная плотность
$\rho^{(3);\text{\tiny glob}}_{B_{\nu}\left\{\delta_1,\mu_1;\delta_2,\mu_2\right\}}[L]$
определяется по ним по формулам типа (\ref{eq:Rho[L].Glo.nu}),
и аналитический образ выражения (\ref{eq:An.Mass.nu0.a^n}),
обозначаемый
${\mathfrak B}_{n+\nu_{0}}^{(3);\text{\tiny glob}}[L]$,
определяется с помощью спектральной плотности
$\rho^{(3);\text{\tiny glob}}_{B_{n+\nu_0}\left\{\delta_{22},\nu_{22};-\delta_{23},\nu_{23}\right\}}[L]$
(выражения для $\delta_{22}$, $\nu_{22}$, $\delta_{23}$ и $\nu_{23}$
 см. в Приложении~\ref{app:Anom.Dim.Quark.Mass}
 --- формулы (\ref{eq:f_3L.Pade.Gen})).

Для получения $\widetilde{R}_\text{S}^{\text{ДАТВ;5}}$,
мы применим операцию $\textbf{A}_\text{M}$,
описанную в разделе~\ref{sec:2}\footnote{%
Верхний индекс ``5'' в $\widetilde{R}_\text{S}^{(l)\text{ДАТВ;5}}$
напоминает нам,
что мы зафиксировали значения коэффициентов $d_n$ при $N_f=5$.}:
\begin{eqnarray}
 \widetilde{R}_\text{S}^{(l)\text{ДАТВ;5}}(s)
   =  \textbf{A}_\text{M}[\widetilde{D}^{(l);5}_{\text{S}}]
   =  3\,\hat{m}_{(l)}^2\,
      \left[{\mathfrak B}_{\nu_{0}}^{(l);\text{\tiny glob}}(s)
          + \sum_{n\geq1}^{l} d_{n}(5)
             \frac{{\mathfrak B}_{n+\nu_{0}}^{(l);\text{\tiny glob}}(s)}
                  {\pi^{n}}
      \right]\,,~~~
 \label{eq:R-MFAPT}
\end{eqnarray}
где верхний индекс $(l)$ обозначает петлевой порядок эволюции и
в то же самое время порядок пертурбативного разложения $D_S$-функции.
Подчеркнем, что это выражение строится из тех же самых
коэффициентов $d_{n}$,
что и евклидовая $\widetilde{D}^{(l);5}_{\text{S}}$-функция,
и что глобальные аналитические заряды
${\mathfrak B}_{n+\nu_{0}}^{(l);\text{\tiny glob}}$
вобрали в себя все эффекты эволюции массы,
а также все $\pi^2$-вклады.

В то же время, мы можем построить полную аналитизацию
$\widetilde{R}_\text{S}^{\text{ДАТВ}}$,
явно учитывая зависимость коэффициентов разложения
от $N_f$
в полной аналогии с тем,
как мы сделали это для ФФ пиона в разделе~\ref{subsec:pionFF.APT.Global}:
там пертурбативные формулы тоже содержали явную $N_f$-зависимость
и это привело просто к определению новых спектральных плотностей.
Точно также мы можем действовать и здесь:
кроме спектральных плотностей $\rho^{(l);\text{\tiny glob}}_{B_{n+\nu_0}\left\{\ldots\right\}}$,
отвечающих глобальным аналитическим зарядам
${\mathfrak B}_{n+\nu_{0}}^{(l);\text{\tiny glob}}$,
появятся также дополнительные глобальные спектральные плотности,
построенные на основе выражений типа
$d_n(N_f)\,\bar{\rho}_{B_{n+\nu_0}\left\{\ldots\right\}}^{(l)}$,
которые дадут нам новые аналитические заряды
${\mathfrak B}_{n+\nu_{0};d_n}^{(l);\text{\tiny glob}}$.
Соответствующая полная ДАТВ формула будет такой:
\begin{eqnarray}
 \widetilde{R}_\text{S}^{(l)\text{ДАТВ}}(s)
  =  \textbf{A}_\text{M}[\widetilde{D}^{(l)}_{\text{S}}]
   =  3\,\hat{m}_{(l)}^2\,
      \left[{\mathfrak B}_{\nu_{0}}^{(l);\text{\tiny glob}}(s)
          + \sum_{n\geq1}^{l}
             \frac{{\mathfrak B}_{n+\nu_{0};d_n}^{(l);\text{\tiny glob}}(s)}
                  {\pi^{n}}
      \right]\,.~~~
 \label{eq:R-MFAPT-Total}
\end{eqnarray}
С точки зрения принципа аналитизации всего выражения ``как целого''
такой подход представляется нам наиболее последовательным.
В минковской области он приводит к результатам,
отличным от частичной аналитизации, когда коэффициенты разложения
``замораживаются'' на каком-либо значении $N_f$
(в этой задаче естественной представляется обычно заморозка
 на значении $N_f=5$).
В самом деле,
когда мы вставляем $N_f$-зависящие части в спектральную плотность,
в минковской области мы получаем в некотором роде усреднение
коэффициентов $d_n(N_f)$ по нескольким значениям $N_f$.
Так, в области $s\sim (100~\text{ГэВ})^2$
коэффициенты будут эффективно усредняться по значениям
$N_f=5$ (интегрирование по $L_s\lesssim12$)
и $N_f=6$ (интегрирование по $L_s\gtrsim12$),
что приведет к понижению результата для
$\widetilde{R}_\text{S}^{(l)\text{ДАТВ}}(s)$
в сравнении с
$\widetilde{R}_\text{S}^{(l)\text{ДАТВ;5}}(s)$.

\subsection{Сравнение различных подходов к расчету $\widetilde{R}_\text{S}$}
\label{subsec:various-Rs}
\begin{figure}[h]
 \centerline{~\includegraphics[width=0.49\textwidth]{
  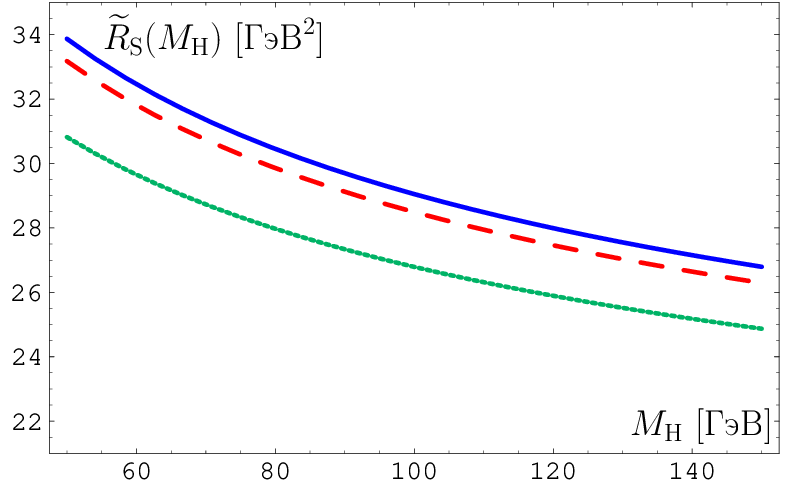}~
              \includegraphics[width=0.49\textwidth]{
  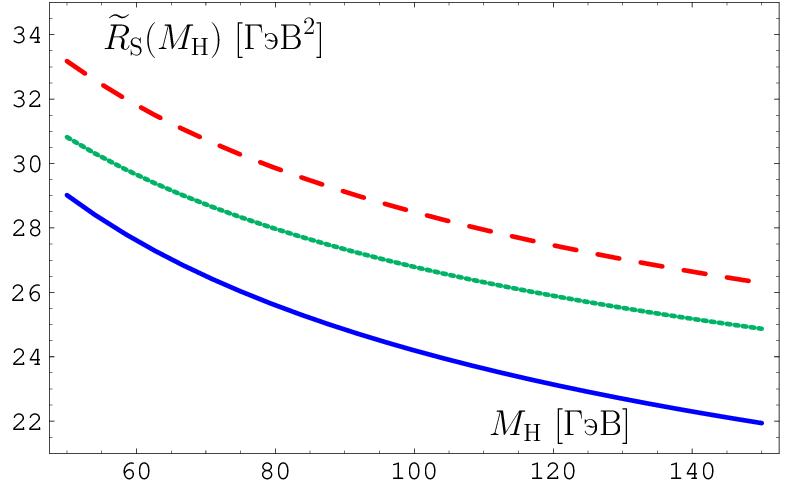}~}%
   \vspace{-0.5cm}
   \caption{Результаты расчета величины $\widetilde{R}_\text{S}(M^2_\text{H})$
   в различных подходах (пояснения см. в тексте).\label{fig:R_S}}
\end{figure}

В этом разделе мы сравниваем результаты расчетов $\widetilde{R}_\text{S}$
в различных подходах.
\begin{itemize}
\item Бродхерст, Катаев и Максвелл (БКМ) \cite{BKM01}
в подходе так называемой ``наивной неабелинизации'' (ННА)
использовали для расчета $\widetilde{R}_{\text{S}}$
оптимизацию степенного разложения,
основанную на методе ``контурного интегрирования''.
Их результаты очень близки к однопетлевому подходу ДАТВ
(см. более подробное обсуждение в \cite{BMS06}):
\begin{eqnarray}
 \widetilde{R}_{\text{S}}^{(l=1)\text{ДАТВ}}(s)
  = 3\,\hat{m}_{(l=1)}^2\,
      \left[{\mathfrak A}_{\nu_{0}}^{(1);\text{\tiny glob}}(s)
          + \sum_{n\geq1}^{4} d_{n}(5)
             \frac{{\mathfrak A}_{n+\nu_{0}}^{(1);\text{\tiny glob}}(s)}
                  {\pi^{n}}
      \right]\,.~~~
 \label{eq:R.MFAPT.1L}
 \end{eqnarray}
 Имеет смысл уточнить используемые значения $\Lambda^{(1)}$.
 В работе~\cite{BMS06} применялось значение $\Lambda_{N_f=3}^{(1);\text{KPS}}=312$~МэВ,
 предложенное в~\cite{KPS02} для $N_f=3$,
 а по сути должно было бы использоваться отвечающее этому выбору
 значение $\Lambda_{N_f=5}^{(1);\text{KPS}}=221$~МэВ.
 Здесь же мы используем $\Lambda_{N_f=5}^{(1);Z}=111$~МэВ,
 получаемое из требования ${\mathfrak A}_{1}^{(1);\text{\tiny glob}}(m_Z^2)=0.120$.
 В результате соответствующая этому выбору пунктирная (зеленая) кривая
 на рис.~\ref{fig:R_S}
 оказывается ниже чисто пертурбативного результата
 (штрихованная красная кривая) на 8\%
 (в~\cite{BMS06} соответствующая кривая шла выше на 16\%,
  а при выборе $\Lambda_{N_f=5}^{(1);\text{KPS}}=221$~МэВ ---
  она оказывается выше на 8\%).

\item Байков, Четыркин и Кюн (БЧК) \cite{BCK05}
использовали стандартную теорию возмущений
в ${\cal O}(a_\text{s}^{4})$-порядке, см. (\ref{eq:RSnumx}),
с $\Lambda_{N_f=5}^{(4)}=231$~МэВ:
\begin{eqnarray}
 \widetilde{R}_{\text{S}}^{(l=4)\text{БЧК}}(s)
  &=& 3 m_{(l=4)}^2(s)
        \left[1 + \sum_{n\geq 1}^{4} r_{n}(5)
                   \left(\frac{\alpha_\text{s}^{(l=4)}}{\pi}\right)^{n}
        \right]\,.~~~
 \label{eq:R-Che}
\end{eqnarray}

\item В подходе ДАТВ с ``заморозкой'' коэффициентов $d_n(N_f)$ на значении
$d_n(5)$ (также как и в предыдущих двух подходах)
мы получаем $\widetilde{R}_{\text{S}}^{(3)\text{ДАТВ};5}(s)$,
см.~(\ref{eq:R-MFAPT}),
и результат для $\Lambda_{N_f=5}^{(4)}=261$~МэВ
(нормировка на ${\mathfrak A}_{1}^{(3);\text{\tiny glob}}(m_Z^2)=0.120$)
представлен на левой части
рис.~\ref{fig:R_S} сплошной (синей) линией.
Видно, что он очень близок к результату БЧК
(отличие имеет порядок 2\%, а если сравнивать с трехпетлевым результатом БЧК ---
 1.5\%).

\item В подходе ДАТВ с полной аналитизацией $N_f$-зависимостей
мы получаем $\widetilde{R}_{\text{S}}^{(3);\text{ДАТВ}}(s)$,
см.~(\ref{eq:R-MFAPT-Total}).
Результат показан сплошной (синей) кривой на правой части
рис.~\ref{fig:R_S}
и, как мы и ожидали,
оказался меньше ``замороженного'' результата.
Это отличие меняется от 12.5\% (при $M_\text{H}=50$~ГэВ)
до 16.5\% (при $M_\text{H}=150$~ГэВ).

\end{itemize}

Таким образом, мы можем сделать вывод,
что ряд стандартной теории возмущений КХД и
нестепенное разложение ДАТВ
в области больших логарифмов $L$
оказываются близки в сценарии с заморозкой коэффициентов $d_n(N_f)$
на значениях, отвечающих $N_f=5$.
Полная ДАТВ с аналитизацией всей зависимости пертурбативных результатов
от $N_f$ естественно дает другой,
в нашем случае меньший, результат.
Удивляться тут нечему,
поскольку, как мы уже говорили, такая аналитизация
эффективно усредняет коэффициенты $d_n(N_f)$,
что приводит к их уменьшению и,
как результат, к уменьшению всей суммы нестепенного ряда.
Чтобы проанализировать этот эффект,
определим эффективные коэффициенты отвечающие полностью аналитизированному
ряду $\widetilde{R}_\text{S}^{(3);\text{ДАТВ}}[L]$
следующим образом:
\begin{eqnarray}
 \label{eq:d.n.eff}
  d_{n}^\text{eff}[L]
   &=& \frac{{\mathfrak B}_{n+\nu_{0};d_n}^{(l);\text{\tiny glob}}[L]}
            {{\mathfrak B}_{n+\nu_{0}}^{(l);\text{\tiny glob}}[L]}\,.
\end{eqnarray}
Мы видим из таблицы~\ref{tab:d.n.eff},
что эффективные коэффициенты действительно уменьшились,
в среднем на 16--18\%.
\begin{table}[h]\vspace*{-3mm}
\caption{Эффективные значения коэффициентов $d_{n}^\text{eff}[L]$,
 см.~(\ref{eq:d.n.eff}), при $L=11-13$. В двух последних колонках приведены
 значения соответствующих сумм ряда для $\widetilde{R}_\text{S}^{(3)}[L]$
 в случае ДАТВ с ``заморозкой'' коэффициентов на $N_f=5$ и в случае полной
 ДАТВ. \label{tab:d.n.eff}\vspace*{+1mm}}
\begin{ruledtabular}
\begin{tabular}{ccccc||cc}
      &~~$n=0$~~&~~$n=1$~~&~~$n=2$~~&~~$n=3$~~&~~$\widetilde{R}_\text{S}^{(3);\text{ДАТВ;5}}[L]$
                                                        &~~$\widetilde{R}_\text{S}^{(3);\text{ДАТВ}}[L]$
\\ \hline \hline
 $d_{n}(N_f=5)$
      & 1.00    & 5.67    & 42.0    & 353     &  --     &  --
\\
 $d_{n}^\text{eff}[L=11]$
      & 0.84    & 4.90    & 36.1    & 292     & 32.01   & 27.16
\\
 $d_{n}^\text{eff}[L=12]$
      & 0.83    & 4.77    & 34.6    & 271     & 28.75   & 23.90
\\
 $d_{n}^\text{eff}[L=13]$
      & 0.82    & 4.63    & 32.8    & 245     & 26.07   & 21.21
\end{tabular}
\end{ruledtabular}
\end{table}

Физически эффект полной аналитизации ряда для $\widetilde{R}_\text{S}(s)$
отвечает учету вкладов петель с $t$-кварками
даже в области $\sqrt{s}\leq 175$~ГэВ,
где в обычной теории возмущений дают вклад только пять кварков
$u$, $d$, $s$, $c$ и $b$.
И мы видим, что эффект этот не мал!

\section{СУММИРОВАНИЕ РЯДОВ ТЕОРИИ ВОЗМУЩЕНИЙ В АТВ И ДАТВ}
 \label{sec:summing}
 \subsection{Однопетлевые АТВ и ДАТВ ($N_f=3$)}
 \label{sec:Sum.ATP}
 Рассмотрим следующий ряд
\begin{subequations}
\begin{eqnarray}
 \label{eq:Ini.Series}
  D[L] = d_0+\sum_{n=1}^{\infty} d_n\,a_s^n[L]
\end{eqnarray}
с коэффициентами
\begin{eqnarray}
 \label{eq:Coeff.d.n}
  d_n = d_1 \int_{0}^\infty\!\!P(t)\,t^{n-1}dt\,,
 \quad\text{причем}\quad
  \int_{0}^\infty\!\!\!P(t)\,dt = 1\,.
\end{eqnarray}
Для облегчения восприятия последующих формул мы вводим
следующее обозначение:
\begin{equation}
 \langle\langle{f(t)}\rangle\rangle_{P}
  \equiv
   \int_{0}^\infty\!\!f(t)\,P(t)\,dt\,,
\label{eq:average}
\end{equation}
\end{subequations}
так что
$d_{n+1} = d_1\,\langle\langle{t^{n}}\rangle\rangle_{P}$\,.
Рассмотрим пару примеров явного вида функции $P(t)$:
\begin{subequations}
\begin{itemize}
  \item функция $P(t;\tau_0)=\delta(t-\tau_0)$ приводит к коэффициентам,
   зависящим от порядка $n$ степенным образом
   \begin{eqnarray}
    \label{eq:P.delta}
     d_{n}&=& d_1 \tau_0^{n-1}\,.
   \end{eqnarray}

  \item функция
   \begin{eqnarray}
    \label{eq:P.exponent}
      P(t;c,\delta)&=& e^{-t/c}\,
                       \frac{(t/c)^{\delta}}
                            {c\,\Gamma(1+\delta)}
   \end{eqnarray}
   генерирует факториально растущие коэффициенты
   \begin{eqnarray}
    \label{eq:d.factorial}
     d_{n}&=& d_1\,c^{n-1}\,
               \frac{\Gamma (n+\delta)}{\Gamma(1+\delta)}\,,
   \end{eqnarray}
  которые имитируют ``липатовское'' поведение при больших $n\gg1$
  в квантовой теории поля, см.~\cite{Lip76,KS80}.
\end{itemize}
\end{subequations}

Благодаря однопетлевому рекуррентному соотношению
\begin{eqnarray}
 \label{eq:Rec.Rel.n+1}
  \frac{1}{\Gamma(n+1)}\left(-\frac{d}{dL}\right)^n{\mathcal A}_{1}[L]
   = {\mathcal A}_{n+1}[L]
\end{eqnarray}
аналитический образ этого ряда, как показано Михайловым~\cite{MS04},
представляется следующей функцией:
\begin{subequations}
\label{eq:Sum.Integ.Repr.APT}
\begin{eqnarray}
 {\cal D}[L]
   &=& d_0 + \sum_{n=0}^{\infty}
              d_{n+1}\,{\mathcal A}_{n+1}[L]
   = d_0 + d_1\left\langle\!\!\!\left\langle{\sum_{n=0}^{\infty}\frac{(-t)^{n}}{n!}\,
                            \frac{d^n}{dL^n}\,{\mathcal A}_{1}[L]
                     }\right\rangle\!\!\!\right\rangle_{\!\!P}  \nonumber\\
   &=& d_0 + d_1 \langle\langle{{\mathcal A}_{1}[L-t]}\rangle\rangle_{P}\,.~
 \label{eq:Sum.Integ.Repr.APT.Eucl}
\end{eqnarray}
Эта формула и является формулой суммирования рядов теории возмущений в АТВ,
причем она справедлива и для суммирования в минковской области:
\begin{eqnarray}
 {\cal R}[L]
   &=& d_0 + \sum_{n=0}^{\infty}
              d_{n+1}\,{\mathfrak A}_{n+1}[L]
   = d_0 + d_1 \langle\langle{{\mathfrak A}_{1}[L-t]}\rangle\rangle_{P}\,.~
 \label{eq:Sum.Integ.Repr.APT.Mink}
\end{eqnarray}
\end{subequations}

Если теперь рассмотреть ряд, стартующий не с 1,
а с дробной степени эффективного заряда,
что имеет место, например,
для ширины распада $H^0\to\bar{b}b$, см.~\cite{BMS06},
\begin{eqnarray}
 {\cal R}_{\nu}
  &=& d_0\,{\mathfrak A}_{\nu}
   + \sum_{n=0}^{\infty}d_{n+1}\,{\mathfrak A}_{n+1+\nu}\,,
  \label{eq:R-MFAPT.sum}
\end{eqnarray}
то благодаря рекуррентному соотношению
\begin{eqnarray}
\label{eq:Rec.Rel.n+1+nu}
 {\mathfrak A}_{n+\nu}[L]
 = \frac{\Gamma(\nu)}{\Gamma(n+\nu)}
    \left(-\frac{d}{dL}\right)^{n}{\mathfrak A}_{\nu}[L]\,,
 \end{eqnarray}
которое есть прямое следствие (\ref{eq:AU.nu.Laplace}),
мы получим
\begin{eqnarray}
{\cal R}_{\nu}
  &=& d_0\,{\mathfrak A}_{\nu}
   + d_1 \langle\langle{X(t;1+\nu)}\rangle\rangle_{P}\,, \label{eq:Q1}
\end{eqnarray}
где
\begin{eqnarray}
 \label{eq:A1}
 X(t;1+\nu) &\equiv &
  \sum_{n=0}^{\infty}
   \frac{(-\hat{x})^n\Gamma(1+\nu)}{\Gamma(n+1+\nu)}\,
    {\mathfrak A}_{1+\nu}[L]\Big|_{\hat{x}\to t\,d/dL}\,.
\end{eqnarray}
Имеется красивое интегральное представление
для интересующего нас ряда (\ref{eq:A1}),
см. формулу (5.2.7.20) в~\cite{PBM-EF81-rus}:
\begin{eqnarray}
 \label{eq:Sum.Gamma.ratio}
  \sum_{n=0}^{\infty}
   \frac{(-\hat{x})^n\Gamma(1+\nu)}{\Gamma(n+1+\nu)}
   = \int_0^{1}\!
      \exp\left(\hat{x}\cdot u^{1/\nu}-\hat{x}\right)du\,.
\end{eqnarray}
Вспоминая, что $\hat{x}=t\cdot{d}/{dL}$,
так что оператор $e^{z\,\hat{x}}$
при действии на функцию $A[L]$ просто сдвигает ее аргумент:
$e^{z\,\hat{x}}A[L]=A[L+zt]$,
мы имеем
\begin{eqnarray}
 X(t;1+\nu)
  = \int_0^{1}\!
     {\mathfrak A}_{1+\nu}
      \left[L+t\left(u^{1/\nu}-1\right)\right]
       du\,.
 \label{eq:X.A}
\end{eqnarray}
Подставляя (\ref{eq:X.A}) в (\ref{eq:Q1}),
получаем окончательно:
\begin{subequations}
\label{eq:Sum.Integ.Repr.FAPT}
\begin{eqnarray}
{\cal R}_{\nu}[L]
  &=& d_0\,{\mathfrak A}_{\nu}[L]
    + d_1 \left\langle\!\!\!\left\langle{\int_0^{1}\!\!
              {\mathfrak A}_{1+\nu}\left[L-t\left(1-u^{1/\nu}\right)\right]\,du
              }\right\rangle\!\!\!\right\rangle_{\!\!P}\nonumber\\
  &=& d_0\,{\mathfrak A}_{\nu}[L]
    + d_1 \langle\langle{{\mathfrak A}_{1+\nu}[L-t]}\rangle\rangle_{P_{\nu}}\,,~~~
 \label{eq:Sum.Integ.Repr.FAPT.Mink}
\end{eqnarray}
где
\begin{eqnarray}
 P_{\nu}(t)
  &\equiv&
   \int_0^{1}\!\!P\left(\frac{t}{1-u^{1/\nu}}\right)
    \frac{du}{1-u^{1/\nu}}
  \ =\ \int_0^{1}\!\!P\left(\frac{t}{1-t'}\right)
        \Phi_{\nu}(t')\frac{dt'}{1-t'}\,;
 \label{eq:P.nu}\\
 \Phi_{\nu}(t)
  &\equiv& \nu\,t^{\nu-1}
   \mathop{\longrightarrow}\limits_{\nu\to0+}
    \delta(t)\,.
 \label{eq:Phi.nu.0}
\end{eqnarray}
Опять же, точно такая же формула справедлива
и для суммирования в евклидовой области:
\begin{eqnarray}
 {\cal D}_{\nu}[L]
   \equiv d_0\,
       {\mathcal A}_{\nu}[L]
   + \sum_{n=0}^{\infty}
      d_{n+1}\,
       {\mathcal A}_{n+1+\nu}[L]
   = d_0\,
       {\mathcal A}_{\nu}[L]
   + d_1\langle\langle{{\mathcal A}_{1+\nu}[L-t]}\rangle\rangle_{P_{\nu}}\,.~~~
 \label{eq:Sum.Integ.Repr.FAPT.Eucl}
\end{eqnarray}
\end{subequations}

Интересно отметить, что в случае функции $P(t;c,\delta)$ из (\ref{eq:P.exponent})
интеграл для функции $P_{\nu}(t)$ в (\ref{eq:P.nu}) вычисляется точно,
так что мы имеем
\begin{eqnarray}
 \label{eq:P.nu.factorial.delta}
  P_{\nu}(t;\delta)
  = \frac{\nu\,_1F_1\left(1-\nu;1-\delta;-t/c\right)}
         {c\,\delta}
  - \frac{\Gamma(1+\nu)\Gamma(-\delta)\,
          _1F_1\left(1+\delta-\nu;1+\delta;-t/c\right)}
         {c\,\Gamma(1+\delta)\,\Gamma(\nu-\delta)\,(c/t)^{\delta}}\,,~~~
\end{eqnarray}
где $_1F_1$ --- конфлюентная гипергеометрическая функция Куммера,
определяемая как
\begin{eqnarray}
 \label{eq:eq:1_F_1.Kummer}
  _1F_1(a;b;z)
   = \sum\limits_{n=0}^{\infty}
      \frac{\Gamma(a+n)\,\Gamma(b)\,z^n}{\Gamma(a)\,\Gamma(b+n)\,n!}\,.
\end{eqnarray}
Для целых значений $\delta=m\geq0$ формулы упрощаются к виду
\begin{eqnarray}
 \label{eq:P.nu.factorial.m}
  P_{\nu}(t;m)
  = \frac{\Gamma (1+\nu)}
         {c\,\Gamma(1+m)}\,
     G_{1,2}^{2,0}
      \left(\frac{t}{c}\left|
       \begin{array}{c}\nu\\0,m\end{array}
             \right.
      \right),
\end{eqnarray}
где $G_{1,2}^{2,0}(z|...)$ --- $G$-функция Мейера,
определяемая как
\begin{eqnarray}
 \label{eq:G_1.2^2.0.Meijer}
 G_{1,2}^{2,0}
      \left(z
            \left|
             \begin{array}{c} a \\
                             {b_1,b_2}
             \end{array}
            \right.
      \right)
 = \frac{1}{2\pi\,i}
    \oint_{C}
     \frac{\Gamma(b_1+s)\Gamma(b_2+s)}{\Gamma(a+s)}\,
      \frac{ds}{z^s}\,,
\end{eqnarray}
причем контур интегрирования $C$ выбирается так,
что все полюса функций $\Gamma(b_1+s)$ и $\Gamma(b_2+s)$
находятся по одну сторону от него (внутри или снаружи).

\subsection{Глобальная однопетлевая АТВ в области Минковского}
 \label{sec:Sum.Glob.ATP.Minkow}
Теперь мы обсудим ситуацию с суммированием рядов в глобальной АТВ
с одним порогом тяжелого кварка,
флейвору которого мы будем приписывать значение 4.
Для простоты начнем с области Минковского,
где формулы оказываются более компактными.
В этом случае однопетлевая спектральная плотность
$\rho^\text{\tiny glob}_n[L]$,
где $L=\ln\left(s/\Lambda_3^2\right)$,
выражается через спектральные плотности с 3 и 4 флейворами,
$\bar{\rho}_n[L;3]$ и $\bar{\rho}_n[L+\lambda_4;4]$:
\begin{eqnarray}
 \label{eq:global_PT_Rho_4}
  \rho_n^\text{\tiny glob}[L]
  &=& \bar{\rho}_n\left[L;3\right]
       \theta\left(L<L_{4}\right)
    + \bar{\rho}_n\left[L+\lambda_4;4\right]\,
       \theta\left(L_{4}\leq L\right)\,,~~~
\end{eqnarray}
с $\lambda_4\equiv\ln\left(\Lambda_3^2/\Lambda_4^2\right)$,
$L_{4}\equiv\ln\left(M_{4}^2/\Lambda_3^2\right)$,
и
\begin{eqnarray}
 \bar{\rho}_{n}\left[L;N_f\right]
  = \frac{\rho_{n}[L]}{\beta_f^n}
  \equiv
    \frac{\sin\left[n~\arccos\left(L_\sigma/\sqrt{L_\sigma^2+\pi^2}\right)\right]}
         {\pi\,\left[\beta_f\sqrt{L^2_\sigma+\pi^2}\right]^{n}}\,.
\label{eq:bar.SD.n}
\end{eqnarray}
Аналитизированные степени эффективного заряда
в области Минковского есть при этом
\begin{eqnarray}
 \label{eq:An_U_n_Glo}
  {\mathfrak A}_n^\text{\tiny glob}[L]
  = \int\limits_{L}^{\infty}\!
       \rho_n^\text{\tiny glob}\left[L_\sigma\right]\,
        dL_\sigma
  = \theta\left(L<L_{4}\right)
       \int\limits_{L}^{L_{4}}\!
        \bar{\rho}_{n}\left[L_{\sigma};3\right]\,
         dL_\sigma
  +\!\! \int\limits_{\text{max}(L,L_{4})}^{\infty}\!\!\!\!\!\!\!\!
        \bar{\rho}_{n}\left[L_{\sigma}+\lambda_{4};4\right]\,
         dL_\sigma\,.~~~~~
\end{eqnarray}

Для однопетлевых спектральных плотностей,
отвечающих фиксированному числу флейворов,
$\rho_{n}\left[L\right]$
и $\bar{\rho}_{n}\left[L;N_f\right]$,
имеются рекуррентные соотношения:
\begin{subequations}
\label{eq:Reccur.SD.n+1.1}
\begin{eqnarray}
 \label{eq:Reccur.Nf.SD.n+1.n}
  \rho_{n+1}[L]
   &=& \frac{1}{n}
        \left(-\frac{d}{dL}\right)
         \rho_{n}[L]\
    =\ \frac{1}{\Gamma(n+1)}
        \left(-\frac{d}{dL}\right)^n
         \rho_{1}[L]\,;
\\
  \bar{\rho}_{n+1}[L;N_f]
   &=& \frac{1}{n\,\beta_f}
        \left(-\frac{d}{dL}\right)
         \bar{\rho}_{n}[L;N_f]
    = \frac{1}{\Gamma(n+1)\,\beta_f^{n+1}}
        \left(-\frac{d}{dL}\right)^n
         \rho_1[L]\,,
 \label{eq:Reccur.Nf.bar.SD.n+1.1}
\end{eqnarray}
\end{subequations}
которые справедливы (за исключением промежуточных средних равенств)
для  $n\geq0$
и позволяют нам немедленно переписать (\ref{eq:An_U_n_Glo})
в более полезном виде:
\begin{eqnarray}
  {\mathfrak A}_{n+1}^\text{\tiny glob}[L]
  &=& \theta\left(L<L_{4}\right)
       \left[\frac{\bar{\rho}_{n}[L;3]-\bar{\rho}_{n}[L_{4};3]}
                  {n\,\beta_3}
           + \frac{\bar{\rho}_{n}[L_{4}+\lambda_{4};4]}
                  {n\,\beta_4}
       \right] \nonumber\\
\label{eq:Reccur.Glo.U.n+1.n}
  &+& \theta\left(L\geq L_{4}\right)
       \frac{\bar{\rho}_{n}[L+\lambda_{4};4]}
            {n\,\beta_4}\,.
\end{eqnarray}
Используя (\ref{eq:Reccur.SD.n+1.1})
и соотношение
\begin{eqnarray}
 \label{eq:rho.1}
  \rho_{1}[L]
   &=& \left(\frac{-d}{dL}\right){\mathfrak A}_1[L]
\end{eqnarray}
мы можем свести (\ref{eq:Reccur.Glo.U.n+1.n}) к представлению,
в котором вся $n$-зависимость становится явной:
\begin{eqnarray}
  {\mathfrak A}_{n+1}^\text{\tiny glob}[L]
  &=& \frac{\theta\left(L<L_{4}\right)}
           {\Gamma(n+1)}
       \Big\{\left[\left(\frac{-1}{\beta_3}\,\frac{d}{dL}\right)^{n}
                    \bar{\mathfrak A}_{1}[L;3]
                 - \left(\frac{-1}{\beta_3}\,\frac{d}{dL_{4}}\right)^{n}
                    \bar{\mathfrak A}_{1}[L_{4};3]
             \right] \nonumber\\
  & &~~~~~~~~~~~~~~~+
                   \left(\frac{-1}{\beta_4}\,\frac{d}{dL_{4}}\right)^{n}
                    \bar{\mathfrak A}_{1}[L_{4}+\lambda_{4};4]
       \Big\}\nonumber\\
 \label{eq:Reccur.Glo.U.n+1.1}
  &+& \frac{\theta\left(L\geq L_{4}\right)}
           {\Gamma(n+1)}
        \left(\frac{-1}{\beta_4}\,\frac{d}{dL}\right)^{n}
         \bar{\mathfrak A}_{1}[L+\lambda_{4};4]\,.
\end{eqnarray}
Таким образом,
общая структура $n$-зависимости в (\ref{eq:Reccur.Glo.U.n+1.1})
есть просто $\hat{x}_f^n/\Gamma(n+1)$ с $\hat{x}_f=[-1/\beta_f](d/dL)$.
Но мы знаем,
как такие зависимости суммируются, см. (\ref{eq:Sum.Integ.Repr.APT}):
\begin{eqnarray}
 {\cal R}^\text{\tiny glob}[L]
   &\equiv&
           d_0 + \sum_{n=0}^{\infty}
                  d_{n+1}\,{\mathfrak A}^\text{\tiny glob}_{n+1}[L]\
   \equiv\ d_0 + d_1\,\sum_{i}\,\theta_i[L]\,{\mathfrak S}_{f;i}[L+\lambda_f]\,;\\
 {\mathfrak S}_{f;i}[L]
   &\sim&
          \sum_{n=0}^{\infty}
            \frac{\langle\langle{t^n}\rangle\rangle_{P}}{\beta_f^n\,\Gamma(n+1)}
             \left(-\frac{d}{dL}\right)^n
              \bar{\mathfrak A}_{1}[L;N_f]
     = \langle\langle{\bar{\mathfrak A}_{1}\!\left[L-t/\beta_f;N_f\right]}\rangle\rangle_{P}\,.~
 \label{eq:sum.Sigma}
\end{eqnarray}
Собирая вместе вклады от различных $\theta$-структур в (\ref{eq:Reccur.Glo.U.n+1.1})
и вставляя их в (\ref{eq:sum.Sigma})
мы получаем ответ:
\begin{eqnarray}
 {\cal R}^\text{\tiny glob}[L]
  = d_0
  &+& d_1\,\theta\left(L\!<\!L_{4}\right)
        \left\langle\!\!\!\left\langle{\bar{\mathfrak A}_{1}\!\Big[L\!-\!\frac{t}{\beta_3};3\Big]\!
           + \Delta_{4}\bar{\mathfrak A}_{1}[t]
            }\right\rangle\!\!\!\right\rangle_{\!\!P}\nonumber\\
  &+& d_1\,\theta\left(L\!\geq\!L_{4}\right)
          \left\langle\!\!\!\left\langle{\bar{\mathfrak A}_{1}\!\Big[L+\lambda_4\!-\!\frac{t}{\beta_4};4\Big]
              }\right\rangle\!\!\!\right\rangle_{\!\!P}\,,~~~
 \label{eq:sum.R.Glo.4}
\end{eqnarray}
где обозначено (с $\lambda_3\equiv0$)
$$\Delta_{f}\bar{\mathfrak A}_{1}[t] \equiv
  \bar{\mathfrak A}_{1}\!\Big[L_{f}+\lambda_{f}-\frac{t}{\beta_f};f\Big]
- \bar{\mathfrak A}_{1}\!\Big[L_{f}+\lambda_{f-1}-\frac{t}{\beta_{f-1}};f-1\Big]\,.$$

Если мы примем во внимание все пороги,
т.~е. в дополнение к порогу $L_4$ учтем также пороги $L_5$ и $L_6$,
тогда ответ запишется в следующем виде
\begin{eqnarray}
 {\cal R}^\text{\tiny glob}[L]
  =  d_0
 &+& d_1\sum\limits_{f=3}^{6}
      \theta\left(L_{f}\!\leq\!L\!<\!L_{f+1}\right)
       \left\langle\!\!\!\left\langle{%
        \bar{\mathfrak A}_{1}\!\Big[L\!+\!\lambda_f\!-\!\frac{t}{\beta_f};f\Big]
        }\right\rangle\!\!\!\right\rangle_{\!\!P}\nonumber\\
 &+& d_1\sum\limits_{f=3}^{5}
      \theta\left(L_{f}\!\leq\!L\!<\!L_{f+1}\right)\!\!
       \sum\limits_{k=f+1}^{6}\!
        \langle\langle{\Delta_{k}\bar{\mathfrak A}_{1}[t]
                      }\rangle\rangle_{P},~
 \label{eq:sum.R.Glo.456}
\end{eqnarray}
где мы определили $L_3=-\infty$ и $L_7=+\infty$.

\subsection{Глобальная однопетлевая АТВ в евклидовой области}
 \label{sec:Sum.Glob.ATP.Euclid}
Теперь мы готовы рассмотреть вопрос о суммировании рядов
в глобальной АТВ в евклидовой области
\begin{eqnarray}
 \label{eq:D.Glo.EAPT}
  {\cal D}^\text{\tiny glob}[L]
   &\equiv&
           d_0 + \sum_{n=0}^{\infty}
                  d_{n+1}\,{\mathcal A}^\text{\tiny glob}_{n+1}[L]\,,
\end{eqnarray}
 также сперва с учетом порога только одного тяжелого кварка.
Аналитизированные степени эффективного заряда здесь есть
\begin{eqnarray}
 \label{eq:An_A_n_Glo}
  {\mathcal A}_n^\text{\tiny glob}[L]
  = \int\limits_{-\infty}^{L_{4}}
     \frac{\bar{\rho}_{n}\left[L_{\sigma};3\right]\,dL_\sigma}
          {1+e^{L-L_{\sigma}}}\,
        dL_\sigma
  + \int\limits_{L_{4}}^{\infty}
     \frac{\bar{\rho}_{n}\left[L_{\sigma}+\lambda_4;4\right]\,dL_\sigma}
          {1+e^{L-L_{\sigma}}}\,
        dL_\sigma\,.
\end{eqnarray}
Для спектральных плотностей,
отвечающих фиксированному числу флейворов,
имеется рекуррентное соотношение (\ref{eq:Reccur.Nf.SD.n+1.n}),
из которого и следует рекуррентное соотношение (\ref{eq:Rec.Rel.n+1})
для ${\mathcal A}_{n+1}[L]$ и ${\mathcal A}_{1}[L]$,
а также соотношения (\ref{eq:Reccur.Nf.bar.SD.n+1.1}),
что позволяет нам записать
\begin{eqnarray}
  {\mathcal A}_{n+1}^\text{\tiny glob}[L]
  &=& \frac{1}{n}\,
       \left[\frac{\bar{\rho}_n\left[L_{4}+\lambda_{4};4\right]}
                  {\beta_4\left(1+e^{L-L_{4}}\right)}
           - \frac{\bar{\rho}_n\left[L_{4};3\right]}
                  {\beta_3\left(1+e^{L-L_{4}}\right)}
       \right]
 \nonumber\\
 \label{eq:Reccur.Glo.A.n+1.n}
  &-& \frac{1}{n}\,
       \frac{d}{d L}
        \left[ \int\limits_{-\infty}^{L_{4}}\!
                \frac{\bar{\rho}_n\left[L_{\sigma};3\right]\,dL_\sigma}
                     {\beta_3\left(1+e^{L-L_{\sigma}}\right)}
            +  \int\limits_{L_{4}}^{\infty}\!
                \frac{\bar{\rho}_n\left[L_{\sigma}+\lambda_4;4\right]\,dL_\sigma}
                     {\beta_4\left(1+e^{L-L_{\sigma}}\right)}
         \right]\,.~~~~~
\end{eqnarray}
Продолжая спускаться по лестнице $n+1\to n\to n-1\to\ldots\to1$,
получим для $n\geq1$
\begin{eqnarray}
  {\mathcal A}_{n+1}^\text{\tiny glob}[L]
  &=& \frac{1}{n!}\,
       \left[-\frac{d}{dL}\right]^{n}
        \left\{\int\limits_{-\infty}^{L_{4}}\!
               \frac{\bar{\rho}_{1}\left[L_{\sigma};3\right]\,dL_\sigma}
                    {\beta_3^n\left(1+e^{L-L_{\sigma}}\right)}
            + \int\limits_{L_{4}}^{\infty}\!
               \frac{\bar{\rho}_{1}\left[L_{\sigma}+\lambda_4;4\right]\,dL_\sigma}
                    {\beta_4^n\left(1+e^{L-L_{\sigma}}\right)}
        \right\}
 \nonumber\\
  &+& \sum\limits_{k=0}^{n-1}\,
       \frac{\Gamma(n-k)}{\Gamma(n+1)}
       \left[\frac{\bar{\rho}_{n-k}\left[L_{4}+\lambda_{4};4\right]}
                  {\beta_4^{k+1}}
           - \frac{\bar{\rho}_{n-k}\left[L_{4};3\right]}
                  {\beta_3^{k+1}}
       \right]
        \left[-\frac{d}{dL}\right]^k
         \frac{1}{1+e^{L-L_{4}}}\,.~~~
\label{eq:Reccur.Glo.A.n+1.1}
\end{eqnarray}
Воспользуемся теперь соотношением (\ref{eq:Reccur.Nf.bar.SD.n+1.1})
и свойством $d/dL=-d/dL_4$ для функций,
зависящих только от разности $L-L_4$,
чтобы записать для $n\geq1$:
\begin{eqnarray}
 \label{eq:Reccur.Glo.A.n+1.1.imp}
  {\mathcal A}^\text{\tiny glob}_{n+1}[L]
  &=& \frac{1}{\beta_3^{n}\,n!}
       \left\{\left[-\frac{d}{dL}\right]^{n}
               \int\limits_{-\infty}^{L_{4}}\!
                \frac{\bar{\rho}_{1}\left[L_{\sigma};3\right]\,dL_\sigma}
                     {1+e^{L-L_{\sigma}}}
            - \Phi_{n-1}(L,L_4;3)
       \right\}\nonumber\\
  &+& \frac{1}{\beta_4^{n}\,n!}
       \left\{\left[-\frac{d}{dL}\right]^{n}
               \int\limits_{L_{4}}^{\infty}\!
                \frac{\bar{\rho}_{1}\left[L_{\sigma}+\lambda_4;4\right]\,dL_\sigma}
                     {1+e^{L-L_{\sigma}}}
            + \Phi_{n-1}(L,L_4;4)
        \right\}\,,~
\end{eqnarray}
где ($n\geq0$ и $\lambda_3=0$)
\begin{eqnarray}
 \Phi_{n}(L,\lambda;f)
  &\equiv& \sum\limits_{k=0}^{n}\,
            \hat{x}^{n-k}\,\hat{y}^k\,
             \frac{\bar{\rho}_{1}\left[\lambda+\lambda_f;f\right]}
               {1+e^{L-L_4}}
 \label{eq:Phi.n.x.y}
   = \frac{\hat{y}^{n+1}-\hat{x}^{n+1}}{\hat{y}-\hat{x}}\,
       \frac{\bar{\rho}_{1}[\lambda+\lambda_f;f]}{1+e^{L-L_4}}
        \Big|_{\footnotesize\begin{array}{c}
               \hat{x}\to-d/d\lambda\\
               \hat{y}\to d/dL_4
               \end{array}}\,.~~~~~~
\end{eqnarray}
Тогда
\begin{eqnarray}
 {\cal D}^\text{\tiny glob}[L]
   = d_0
  &+& d_1\left\langle\!\!\!\left\langle{\int\limits_{-\infty}^{L_{4}}\!
                \frac{\bar{\rho}_{1}\left[L_{\sigma};3\right]\,dL_\sigma}
                     {1+e^{L-L_{\sigma}-t/\beta_3}}
             + \int\limits_{L_{4}}^{\infty}\!
                \frac{\bar{\rho}_{1}\left[L_{\sigma}+\lambda_4;4\right]\,dL_\sigma}
                     {1+e^{L-L_{\sigma}-t/\beta_4}}
              }\right\rangle\!\!\!\right\rangle_{\!\!P}
 \nonumber\\
  &+& d_1\left\langle\!\!\!\left\langle{\sum_{n=0}^{\infty}
               \frac{t^{n+1}}{(n+1)!}\,
                  \left[\frac{\Phi_{n}(L,L_4;4)}{\beta_4^{n+1}}
                      - \frac{\Phi_{n}(L,L_4;3)}{\beta_3^{n+1}}
                  \right]
              }\right\rangle\!\!\!\right\rangle_{\!\!P}\,.~~~
 \label{eq:sum.a.n+1.Phi.n}
\end{eqnarray}
При этом, благодаря (\ref{eq:Phi.n.x.y}),
можно просуммировать по $n$ в (\ref{eq:sum.a.n+1.Phi.n}) явно.
Действительно, положим $\alpha=t/\beta_f$:
\begin{eqnarray}\!\!\!\!\!
 \sum_{n=0}^{\infty}
  \frac{\alpha^{n+1}}{(n+1)!}\,
   \Phi_{n}(L,L_4;f)
  &=& \frac{\exp(\alpha\hat{y})-\exp(\alpha\hat{x})}{\hat{y}-\hat{x}}\,
      \frac{\bar{\rho}_{1}(\lambda+\lambda_f)}{1+e^{L-L_4}}\Big|_{\lambda\to L_4}
 \nonumber\\
  &=& \int_{0}^{\infty}\!\!db
   \left[e^{(\alpha-b)\hat{y}+b\hat{x}}-e^{-b\hat{y}+(\alpha+b)\hat{x}}\right]\,
          \frac{\bar{\rho}_{1}(\lambda+\lambda_f)}{1+e^{L-L_4}}\Big|_{\lambda\to L_4}
 \nonumber\\
  &=& \int_{0}^{\infty}\!\!db
   \left[\frac{\bar{\rho}_{1}(L_4+\lambda_f-b)}{1+e^{L-L_4-\alpha+b}}
       - \frac{\bar{\rho}_{1}(L_4+\lambda_f-\alpha-b)}{1+e^{L-L_4+b}}
   \right]
 \nonumber\\
  &=& \alpha\int_{0}^{1}\!
       \frac{\bar{\rho}_{1}(L_4+\lambda_f-\alpha x)\,dx}{1+e^{L-L_4-\alpha\bar{x}}}
        \,.
 \label{eq:sum.a.Phi}
\end{eqnarray}
И теперь мы получаем
конечную формулу суммирования ряда (\ref{eq:Ini.Series}):
\begin{eqnarray}
 {\cal D}^\text{\tiny glob}[L]
   = d_0
   + d_1\,\left\langle\!\!\!\left\langle{\int\limits_{-\infty}^{L_{4}}
                \frac{\bar{\rho}_{1}\left[L_{\sigma};3\right]\,dL_\sigma}
                     {1+e^{L-L_{\sigma}-t/\beta_3}}
              + \int\limits_{L_{4}}^{\infty}
                 \frac{\bar{\rho}_{1}\left[L_{\sigma}+\lambda_4;4\right]\,dL_\sigma}
                      {1+e^{L-L_{\sigma}-t/\beta_4}}
              + \Delta_{4}[L,t]
               }\right\rangle\!\!\!\right\rangle_{\!\!P}\,,~
\label{eq:sum.D.Glo.4}
\end{eqnarray}
где
\begin{eqnarray}
 \Delta_{f}[L,t]
  \equiv \int\limits_{0}^{1}\!
         \frac{\bar{\rho}_{1}\left[L_f+\lambda_f-tx/\beta_f;N_f\right]\,t}
              {\beta_f\,\left[1+e^{L-L_f-t\bar{x}/\beta_f}\right]}\,dx
      - \int\limits_{0}^{1}\!
         \frac{\bar{\rho}_{1}\left[L_f+\lambda_{f-1}-tx/\beta_{f-1};N_{f-1}\right]\,t}
              {\beta_{f-1}\,\left[1+e^{L-L_f-t\bar{x}/\beta_{f-1}}\right]}\,dx\,.~~~
 \label{eq:Delta.f.Eucl}
\end{eqnarray}
Заметим, что если учесть все пороги,
т.~е. в дополнение к $L_4$
еще $L_5$ и $L_6$,
то конечная формула изменится очевидным образом
\begin{eqnarray}
 {\cal D}^\text{\tiny glob}[L]
   = d_0
   + d_1\,\sum\limits_{f=3}^{6}
            \left\langle\!\!\!\left\langle{\int\limits_{L_{f}}^{L_{f+1}}\!
                  \frac{\bar{\rho}_{1}\left[L_{\sigma}+\lambda_f;N_f\right]\,dL_\sigma}
                       {1+e^{L-L_{\sigma}-t/\beta_f}}
                }\right\rangle\!\!\!\right\rangle_{\!\!P}
   + d_1\,\sum\limits_{f=4}^{6}
            \langle\langle{\Delta_{f}[L,t]}\rangle\rangle_{P}\,,~~~
 \label{eq:sum.D.Glo.456}
\end{eqnarray}
где, как и ранее в разделе~\ref{sec:Sum.Glob.ATP.Minkow},
мы пользуемся соглашением $L_3=-\infty$ и $L_7=+\infty$.

\subsection{Глобальная однопетлевая ДАТВ}
В этом разделе мы обсудим обобщение метода суммирования рядов
в однопетлевой ДАТВ
на случай глобальной ДАТВ.
Мы будем рассматривать здесь
следующие ряды
\begin{subequations}
 \label{eq:R-D.Glo.M-E.FAPT}
\begin{eqnarray}
 \label{eq:R.Glo.MFAPT}
  {\cal R}_{\nu}^\text{\tiny glob}[L]
   &\equiv&
       d_0\,{\mathfrak A}^\text{\tiny glob}_{\nu}[L]
    + \sum_{n=0}^{\infty}
       d_{n+1}\,{\mathfrak A}^\text{\tiny glob}_{n+1+\nu}[L]\,;\\
 \label{eq:D.Glo.EFAPT}
  {\cal D}_{\nu}^\text{\tiny glob}[L]
   &\equiv&
       d_0\,{\mathcal A}^\text{\tiny glob}_{\nu}[L]
    + \sum_{n=0}^{\infty}
       d_{n+1}\,{\mathcal A}^\text{\tiny glob}_{n+1+\nu}[L]
\end{eqnarray}
\end{subequations}
и начнем со случая одного порога
тяжелого кварка ($f=4$).
Фактически, нам надо объединить результаты
разделов~\ref{sec:Sum.ATP}, \ref{sec:Sum.Glob.ATP.Minkow}
и \ref{sec:Sum.Glob.ATP.Euclid}.
Аналитизированные $n+1+\nu$-степени эффективного заряда есть
\begin{eqnarray}
 \label{eq:Glo-MFAPT.An_U_nu_Glo}
  {\mathfrak A}_{n+1+\nu}^\text{\tiny glob}[L]
  = \theta\left(L<L_{4}\right)
       \int\limits_{L}^{L_{4}}\!
        \bar{\rho}_{n+1+\nu}\left[L_{\sigma};3\right]\,
         dL_\sigma
  +\!\! \int\limits_{\text{max}(L,L_{4})}^{\infty}\!\!\!\!\!\!\!\!
        \bar{\rho}_{n+1+\nu}\left[L_{\sigma}+\lambda_{4};4\right]\,
         dL_\sigma\,;~~~~~\\
 \label{eq:Glo-EFAPT.An_U_nu_Glo}
 {\mathcal A}_{n+1+\nu}^\text{\tiny glob}[L]
  = \int\limits_{-\infty}^{L_{4}}
     \frac{\bar{\rho}_{n+1+\nu}\left[L_{\sigma};3\right]\,dL_\sigma}
          {1+e^{L-L_{\sigma}}}\,
        dL_\sigma
  + \int\limits_{L_{4}}^{\infty}
     \frac{\bar{\rho}_{n+1+\nu}\left[L_{\sigma}+\lambda_4;4\right]\,dL_\sigma}
          {1+e^{L-L_{\sigma}}}\,
        dL_\sigma\,.~~~~~
\end{eqnarray}
Для однопетлевых спектральных плотностей,
отвечающих фиксированному числу флейворов,
$\rho_{n+1+\nu}\left[L\right]$
и $\bar{\rho}_{n+1+\nu}\left[L;N_f\right]$,
имеются рекуррентные соотношения:
\begin{subequations}
 \label{eq:Glo-MFAPT.Reccur.SD.n+nu.nu}
 \begin{eqnarray}
 \label{eq:Glo-MFAPT.Reccur.Nf.SD.n+nu.nu}
  \rho_{n+1+\nu}[L]
   &=& \frac{1}{n+\nu}
        \left(-\frac{d}{dL}\right)
         \rho_{n+\nu}[L]\
    =\ \frac{\Gamma(1+\nu)}{\Gamma(n+1+\nu)}
        \left(-\frac{d}{dL}\right)^n
         \rho_{1+\nu}[L]\,;
\\
  \bar{\rho}_{n+1+\nu}[L;N_f]
   &=& \frac{1}{(n+\nu)\,\beta_f}
        \left(-\frac{d}{dL}\right)
         \bar{\rho}_{n+\nu}[L;N_f]
   \nonumber\\
   &=& \frac{\Gamma(1+\nu)}{\Gamma(n+1+\nu)\,\beta_f^{n+1+\nu}}
        \left(-\frac{d}{dL}\right)^n
         \rho_{1+\nu}[L]\,,
 \label{eq:Glo-MFAPT.Reccur.Nf.bar.SD.n+nu.nu}
\end{eqnarray}
\end{subequations}
которые справедливы для  $n\geq0$.
Пользуясь этими соотношениями и аккуратно повторяя все шаги
раздела~\ref{sec:Sum.Glob.ATP.Minkow},
мы получим для ряда (\ref{eq:R.Glo.MFAPT})
следующий ответ:
\begin{eqnarray}
 {\cal R}^\text{\tiny glob}_{\nu}[L]
  = d_0\,{\mathfrak A}^\text{\tiny glob}_{\nu}[L]
  &+& d_1\,\theta\left(L\!<\!L_{4}\right)
        \left\langle\!\!\!\left\langle{\bar{\mathfrak A}_{1+\nu}\!\Big[L\!-\!\frac{t}{\beta_3};3\Big]\!
           + \Delta_{4}\bar{\mathfrak A}_{1+\nu}[t]
            }\right\rangle\!\!\!\right\rangle_{\!\!P_{\nu}}\nonumber\\
  &+& d_1\,\theta\left(L\!\geq\!L_{4}\right)
          \left\langle\!\!\!\left\langle{\bar{\mathfrak A}_{1+\nu}\!\Big[L+\lambda_4\!-\!\frac{t}{\beta_4};4\Big]
              }\right\rangle\!\!\!\right\rangle_{\!\!P_{\nu}}\,,~~~
 \label{eq:Glo-MFAPT.sum.R.Glo.4}
\end{eqnarray}
где обозначено (с $\lambda_3\equiv0$)
\begin{eqnarray}
 \Delta_{f}\bar{\mathfrak A}_{1+\nu}[t] \equiv
  \bar{\mathfrak A}_{1+\nu}\!\Big[L_{f}+\lambda_{f}-\frac{t}{\beta_f};f\Big]
- \bar{\mathfrak A}_{1+\nu}\!\Big[L_{f}+\lambda_{f-1}-\frac{t}{\beta_{f-1}};f-1\Big]\,.
 \label{eq:Glo-MFAPT.Delta.1+nu}
\end{eqnarray}

Аналогичная процедура,
повторяющая основные шаги раздела~\ref{sec:Sum.Glob.ATP.Euclid}
и выполняемая для ряда (\ref{eq:D.Glo.EFAPT}),
приводит к ответу
\begin{eqnarray}
 {\cal D}^\text{\tiny glob}_{\nu}[L]
   = d_0
   + d_1\,\left\langle\!\!\!\left\langle{\int\limits_{-\infty}^{L_{4}}
                \frac{\bar{\rho}_{1+\nu}\left[L_{\sigma};3\right]\,dL_\sigma}
                     {1+e^{L-L_{\sigma}-t/\beta_3}}
              + \int\limits_{L_{4}}^{\infty}
                 \frac{\bar{\rho}_{1+\nu}\left[L_{\sigma}+\lambda_4;4\right]\,dL_\sigma}
                      {1+e^{L-L_{\sigma}-t/\beta_4}}
              + \Delta_{4;\nu}[L,t]
               }\right\rangle\!\!\!\right\rangle_{\!\!P_{\nu}}\,,~~~
\label{eq:EAPT.sum.D.Glo.4}
\end{eqnarray}
где
\begin{eqnarray}
 \Delta_{f;\nu}[L,t]
  \equiv
   \int\limits_{0}^{1}\!
    \frac{\bar{\rho}_{1+\nu}\!\left[L_f+\lambda_f-\frac{tx}{\beta_f};N_f\right]t}
         {\beta_f\,\left[1+e^{L-L_f-t\bar{x}/\beta_f}\right]}\,dx
 - \int\limits_{0}^{1}\!
    \frac{\bar{\rho}_{1+\nu}\!\left[L_f+\lambda_{f-1}-\frac{tx}{\beta_{f-1}};N_{f-1}\right]t}
         {\beta_{f-1}\,\left[1+e^{L-L_f-t\bar{x}/\beta_{f-1}}\right]}\,dx\,.~~~
 \label{eq:EAPT.Delta.f.Eucl}
\end{eqnarray}

Если мы примем во внимание все пороги,
то тогда ответы запишутся в следующем виде
\begin{subequations}
\begin{eqnarray}
 {\cal R}^\text{\tiny glob}_{\nu}[L]\!
  = d_0\,{\mathfrak A}^\text{\tiny glob}_{\nu}[L]
 &+& d_1\sum\limits_{f=3}^{6}
      \theta\left(L_{f}\!\leq\!L\!<\!L_{f+1}\right)
       \left\langle\!\!\!\left\langle{\bar{\mathfrak A}_{1+\nu}\!\Big[L\!+\!\lambda_f\!
                   -\!\frac{t}{\beta_f};f\Big]
                      }\right\rangle\!\!\!\right\rangle_{\!\!P_{\nu}}
 \nonumber\\
  &+& d_1\sum\limits_{f=3}^{5}
       \theta\left(L_{f}\!\leq\!L\!<\!L_{f+1}\right)\!\!
        \sum\limits_{k=f+1}^{6}\!
         \langle\langle{\Delta_{k}\bar{\mathfrak A}_{1+\nu}[t]}\rangle\rangle_{P_{\nu}},~~~
 \label{eq:Glo-MFAPT.sum.R.Glo.456}\\
 {\cal D}^\text{\tiny glob}_{\nu}[L]
   = d_0\,{\mathcal A}^\text{\tiny glob}_{\nu}[L]
  &+& d_1\sum\limits_{f=3}^{6}
       \left\langle\!\!\!\left\langle{\int\limits_{L_{f}}^{L_{f+1}}\!
        \frac{\bar{\rho}_{1+\nu}\left[L_{\sigma}+\lambda_f;N_f\right]\,dL_\sigma}
             {1+e^{L-L_{\sigma}-t/\beta_f}}
                     }\right\rangle\!\!\!\right\rangle_{\!\!P_{\nu}}
 \nonumber\\
  &+& d_1\sum\limits_{f=4}^{6}
       \langle\langle{\Delta_{f;\nu}[L,t]}\rangle\rangle_{P_{\nu}}\,,~~~
 \label{eq:EAPT.sum.D.Glo.456}
\end{eqnarray}
\end{subequations}
где, как и ранее,
мы пользуемся соглашением $L_3=-\infty$ и $L_7=+\infty$.
Мы видим, что эффект ДАТВ в глобальном подходе свелся к замене исходной
производящей функции $P(t)$ ее дробным аналогом $P_{\nu}(t)$,
см.~(\ref{eq:P.nu}),
а также тривиальным изменением индексов $1\to1+\nu$ у функций,
входящих под знак операции усреднения
$\langle\langle{...}\rangle\rangle_{P_{\nu}}$,
а именно ${\mathfrak A}^\text{\tiny glob}_{1}[L]$,
${\mathcal A}^\text{\tiny glob}_{1}[L]$
и $\bar{\rho}_{1}[L]$.

\subsection{Приложения техники суммирования в однопетлевой (Д)АТВ}

Для иллюстрации возможностей развитой техники суммирования рядов теории
возмущений в однопетлевой АТВ и ДАТВ сошлемся
на нашу с С.~Михайловым недавнюю работу~\cite{BM08},
доложенную на семинаре, посвященном памяти Игоря Соловцова,
который проходил в лаборатории теоретической физики
им.~Н.~Н.~Боголюбова ОИЯИ (Дубна) 17--18 января 2008 г.
В этой работе мы проанализировали пертурбативные коэффициенты
$d_n$ для распада хиггсовского бозона $H^0\to\bar{b}b$,
см. раздел~\ref{sec:Higgs.width},
построили для них достаточно аккуратную модель
(с параметрами $c=2.4$ и $\beta=-0.52$):
\begin{subequations}
\label{eq:Higgs.Model}
\begin{eqnarray}
 \label{eq:Higgs.d_n.Model}
  \tilde{d}_n^\text{H}
   = c^{n-1}\,
     \frac{\Gamma (n+1)+\beta\,\Gamma (n)}{1+\beta}\,,
\end{eqnarray}
основанную на производящей функции
\begin{eqnarray}
 \label{eq:Higgs.P(t).Model}
  P_\text{H}(t)
  &=& \frac{\beta+t/c}{c\,(\beta+1)}\,e^{-t/c}\,,
\end{eqnarray}
\end{subequations}
и применили затем описанную здесь
технику однопетлевого суммирования к ряду (\ref{eq:R-MFAPT}).
В результате мы показали,
что для расчета ширины распада бозона Хиггса с точностью 1\%
в области значений его массы $m_\text{H}=60-180$~ГэВ$^2$
вполне достаточно учета вкладов с коэффициентами
$d_0$, $d_1$, $d_2$ и $d_3$,
а учет вклада с $d_4$ приводит к улучшению
точности до 0.5\%.

\section{ЗАКЛЮЧЕНИЕ}
 \label{sec:Conclusion}

В этой работе мы дали обзор основных элементов глобальной
версии ДАТВ,
причем старались сделать это в таком виде,
чтобы читатель имел возможность применять ее на практике
для расчетов реальных процессов.

Мы рассмотрели основы аналитической теории возмущений
на примере расчета $D$-функции Адлера
в пространственноподобной области
и связанного с ним $R$-отношения
для $e^+e^-$-аннигиляции в адроны
во временноподобной области.
Мы кратко обсудили формализм АТВ в однопетлевом приближении
для случая фиксированного числа ароматов
и также рассмотрели усложнения,
связанные с учетом порогов тяжелых кварков
при построении глобальной версии АТВ.
Мы перечислили недостатки АТВ,
связанные с необходимостью аналитизации более сложных выражений,
возникающих в реальной КХД при использовании методов ренормгруппы
и факторизации,
и как способ их исправления
рассмотрели дробно-аналитическую теорию возмущений.
Мы кратко обсудили случай однопетлевой ДАТВ,
объяснили, как можно действовать в случае учета высших петель,
и обсудили глобальный вариант ДАТВ,
в котором учитываются пороги тяжелых кварков.

В качестве одного из приложений ДАТВ в евклидовой области
мы рассмотрели
расчет факторизуемой части формфактора пиона
и продемонстрировали,
что использование АТВ и ДАТВ приводит к существенному снижению зависимости
результатов от выбора схемы и масштабов перенормировки и факторизации.
Кроме того, мы показали,
что в АТВ и ДАТВ проблема учета порогов
в пертурбативных расчетах решается естественным путем,
а переход в область Минковского с помощью дисперсионного представления
для формфактора пиона
говорит о необходимости использовать в этой области
эффективные заряды ${\mathcal A}_{\nu}(-s)$,
обладающие мнимыми частями,
а не эффективные заряды ${\mathfrak A}_{\nu}(s)$,
пригодные для расчетов поправок к сечениям реакций.
Кроме того, при таком переходе возникает естественное
предписание для масштаба перенормировки
$\mu^2_\text{R}=Q^2/4$,
при котором скачки мнимой части формфактора пиона
в минковской области совпадают с порогами рождения пар тяжелых кварков,
$s_\text{thr}=4m_Q^2$.

В качестве приложения ДАТВ в минковской области
мы рассмотрели расчет полной ширины распада бозона Хиггса
в кварк-антикварковую $\bar{b}b$-пару.
Область значений энергии в системе центра масс,
интересная для эксперимента,
здесь очень велика,
$\sqrt{s}\gtrsim100$~ГэВ.
Поэтому результаты ДАТВ с $N_f=5$,
отвечающие  переносу $\pi^2$-вкладов
из коэффициентов теории возмущений
в аналитические эффективные заряды ${\mathfrak A}_{n+\nu}$,
прекрасно согласуются с результатами стандартной теории возмущений
уже на уровне двухпетлевого приближения.
В то же время,
результат глобальной версии ДАТВ отличается
от них на уровне 14\%,
что связано с учетом эффектов виртуальных $t$-кварков
в петлевых поправках.

Наконец, мы обсудили суммирование нестепенных рядов типа
$\sum_{n}d_n{\mathfrak A}_{n+\nu}[L]$
и
$\sum_{n}d_n{\mathcal A}_{n+\nu}[L]$
в АТВ (с $\nu=0$) и ДАТВ (с $\nu\neq0$).
Мы продемонстрировали в однопетлевой АТВ,
как такое суммирование можно провести точно~\cite{MS04}
и выразить ответ в виде интеграла от ${\mathcal A}_{1}[L-t]$
по $t$ с весом $P(t)$,
определяемыми коэффициентами пертурбативного ряда $d_n$.
Мы показали, что аналогичное суммирование можно провести
и в случае однопетлевой ДАТВ: при этом сумма ряда
$\sum_{n}d_n{\mathcal A}_{n+\nu}[L]$
выражается тоже интегралом по $t$,
но уже от ${\mathcal A}_{1+\nu}[L-t]$
и с модифицированным весом $P_{\nu}(t)$.
Мы получили также основные формулы глобализации
(учета порогов тяжелых кварков)
для этих методов суммирования в АТВ и ДАТВ.

В качестве возможных направлений развития этого подхода
хочется отметить:
\begin{itemize}
  \item Применение ДАТВ в анализе данных глубоко неупругого рассеяния;
  \item Применение АТВ и ДАТВ для анализа суммирования пертурбативных
   поправок к различным процессам.
  \item Обобщение техники суммирования рядов в АТВ и ДАТВ
   на двухпетлевой случай.
\end{itemize}

\acknowledgments
 Автор благодарен своим друзьям и соавторам С.~Михайлову
и Н.~Стефанису за понимание и поддержку,
без которых эта работа никогда не была бы завершена.
Я также признателен И.~Аникину, О.~Теряеву и Д.~В.~Ширкову
за плодотворные обсуждения и ценные советы,
К.~Гёке и Н.~Стефанису за теплый прием в Рурском университете Бохума,
где эта работа была начата и частично реализована.
Работа выполнена при поддержке грантов РФФИ
№~06-02-16215, 07-02-91557 и 08-01-00686,
программы сотрудничества БРФФИ--ОИЯИ (контракт №~F06D-002),
грантов 2007--2008~гг. программы Гейзенберг--Ландау
и гранта DFG (проект DFG 436 RUS 113/881/0).

\begin{appendix}
\appendix
\section{Двухпетлевые ренормгрупповые решения для эффективного заряда в КХД}
 \label{App:RG-solution-2L}

\textbf{1.} Разложение $\beta$-функции дается правой частью следующего
уравнения
\begin{eqnarray}
 \frac{d}{dL}\left(\frac{\alpha_\text{s}}{4 \pi}\right)
=
  \beta\left( \frac{\alpha_\text{s}}{4 \pi}\right)
= - b_0\left(\frac{\alpha_\text{s}}{4 \pi}\right)^2
  - b_1\left(\frac{\alpha_\text{s}}{4 \pi}\right)^3
  - b_2\left(\frac{\alpha_\text{s}}{4 \pi}\right)^4\,- \ldots \, ,
\label{eq:betaf}
\end{eqnarray}
где $L=\ln(\mu^2/\Lambda^2)$ и
\begin{eqnarray}
    b_0 &=& \frac{11}{3}\,C_\text{A} - \frac{4}{3}\,T_\text{R} N_f
    \,;\qquad
    b_1 = \frac{34}{3}\,C_{\text{A}}^{2}
        - \left(4C_\text{F}
        + \frac{20}{3}\,C_\text{A}\right)T_\text{R} N_f ;\nonumber \\
    b_2 &=&   \frac{2857}{54} C_A^3
 +2 C_F^2 T_\text{R} N_f - \frac{205}{9} C_F C_A T_\text{R} N_f
 - \frac{1415}{27} C_A^2 T_\text{R} N_f
 + \frac{44}{9} C_F (T_\text{R} N_f)^2 \nonumber \\
 &&
 + \frac{158}{27} C_A (T_\text{R} N_f)^2\,,
 \label{eq:beta0&1&2}
\end{eqnarray}
причем $C_\text{F}=\left(N_\text{c}^{2}-1\right)/2N_\text{c}=4/3$,
$C_\text{A}=N_\text{c}=3$, $T_\text{R}=1/2$, и $N_f$ обозначает
число активных флейворов кварков.
Соответствующее двухпетлевое уравнение РГ для эффективного заряда
$a=b_0\,\alpha_\text{s}/(4\pi)$
\begin{eqnarray}
 \frac{d a_{(2)}}{dL}
  = - a_{(2)}^2\left[1 + c_1\,a_{(2)}\right]
 \text{,~~где~}
  c_1\equiv \frac{b_1}{b_0^2}\,.
\label{eq:beta.new.2L}
\end{eqnarray}
Интегрируя по $L$, можно показать,
что $a_{(2)}[L]$ удовлетворяет следующему нелинейному уравнению
\begin{eqnarray}
 \label{eq:App-RGExact}
 \frac{1}{a_{(2)}} +
 c_1
     \ln\left[\frac{a_{(2)}}{1+c_1 a_{(2)}}
     \right] = L\,,
\end{eqnarray}
точное решение которого известно~\cite{Mag99,GGK98}
\begin{eqnarray}
 \label{eq:App-Exactsolution.2L}
 a_{(2)}[L] =
 -\frac{1}{c_1} \frac{1}{1+W_{-1}(z_W(L))}\, ,
 \end{eqnarray}
где
$z_W(L)=\left(1/c_1\right) \exp\left(-1+i\pi-L/c_1\right)$,
а $W_{k}$, $k=0,\pm 1,\ldots $,
обозначают ветви многозначной функции Ламберта
$W(z)$,
определяемой как решение уравнения
\begin{eqnarray}
z=W(z)\, e^{W(z)}\,.
\end{eqnarray}
Обзор свойств этой специальной функции
может быть найден в~\cite{CGHJK96,Mag99,Mag00}.
Заметим здесь также, что эта функция определена
в известных программах символьных расчетов
Mathematica\footnote{В версиях 3, 4 и 5 пакета Mathematica
функция $W_{k}(z)$ обозначается именем ProductLog$[k,z]$.}
и Maple.

\textbf{2.}~Разложение решения $a_{(2)}[L]$ уравнения (\ref{eq:beta.new.2L})
по однопетлевому решению $a_{}=1/L$
в $O(a_{}^{4})$-порядке строится следующим образом.
Сначала мы переписываем уравнение (\ref{eq:App-RGExact})
в виде
\begin{subequations}
\begin{eqnarray}
 \label{eq:A7a}
  a_{(2)}
  = \Phi_{(2)}\left(a_{},a_{(2)}\right)
  \equiv
    \frac{a_{}}
         {1+c_1a_{}\left[\ln(1+c_1 a_{(2)})
                       - \ln a_{(2)}
                   \right]}\,,
\end{eqnarray}
а затем получаем его решения методом итераций:
\begin{eqnarray}
 \label{eq:123.Iter}
  a_{(2)}^\text{1-iter}
   = \Phi_{(2)}\left(a_{},a_{}\right)\,,\quad
  a_{(2)}^\text{2-iter}
   = \Phi_{(2)}\left(a_{},a_{(2)}^\text{1-iter}\right)\,,\quad
  a_{(2)}^\text{3-iter}
   = \Phi_{(2)}\left(a_{},a_{(2)}^\text{2-iter}\right)\,.
\end{eqnarray}
\end{subequations}
Третьей итерации достаточно,
чтобы получить затем правильные коэффициенты разложения
$a_{(2)}$ в ряд по $a_{}$
в $O(a_{}^{3})$-порядке из выражения (\ref{eq:123.Iter}):
\begin{eqnarray}
  a_{(2)} = a_{}
    &+& c_1\,a_{}^2\,\ln a_{}
     +  c_1^2\,a_{}^3\,\left(\ln^2 a_{} + \ln a_{} -1\right)
  \nonumber \\
    &+& c_1^3\,a_{}^4\,\left(\ln^3 a_{}+\frac5{2}\ln^2 a_{}
                           - 2~\ln a_{} - \frac1{2}
                       \right)
     + O\left(c_{1}^4\right)\,.
 \label{eq:a.PT.2L}
\end{eqnarray}

\begin{figure}[h]
 \centerline{\includegraphics[width=0.45\textwidth]{
  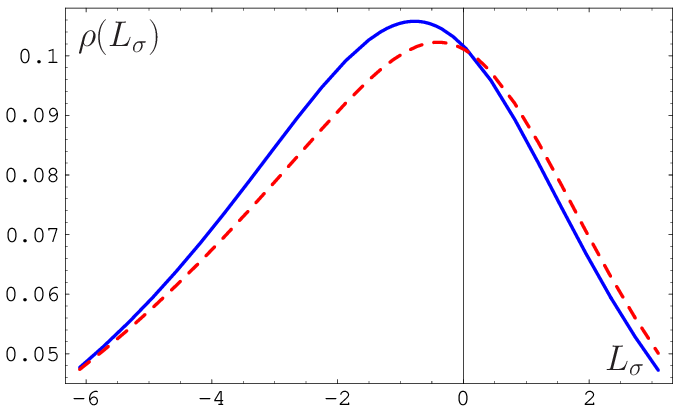}~~~~%
             \includegraphics[width=0.45\textwidth]{
  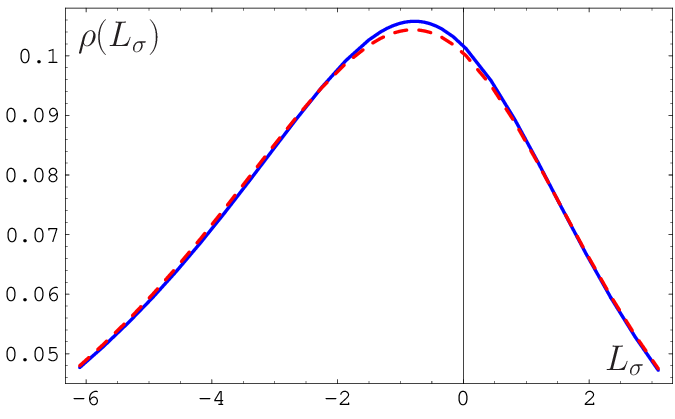}}%
   \caption{\label{fig:exact-2it}\footnotesize
   Сравнение спектральных плотностей, отвечающих 1-й (слева) и 2-й (справа)
   итерациям (\ref{eq:123.Iter}), с точной спектральной двухпетлевой
   плотностью.
   Штрихованная красная линия отвечает на левом графике
   $\rho^\text{(2)it-1}_{1}[L_\sigma]$, см.~(\ref{eq:Sp.Den.iteration-1}),
   а на правом графике --- $\rho^\text{(2)it-2}_{1}[L_\sigma]$,
   см.~(\ref{eq:Sp.Den.iteration-2}),
   в то время как сплошная синяя линия на обоих графиках представляет
   $\rho^{(2)}_1[L_\sigma]$, см.~(\ref{eq:SpDen.nu.(2)}).
   Для лучшего сравнения показана та область значений $L_\sigma$,
   где отличия сравниваемых плотностей максимальны.}
\end{figure}

\textbf{3.}~Рассмотрим теперь вопрос,
насколько хорошо спектральные плотности
$\rho^\text{(2)it-1}_{1}[L_\sigma]$ и $\rho^\text{(2)it-2}_{1}[L_\sigma]$,
отвечающие  итерационным решениям (\ref{eq:123.Iter}),
приближают точную спектральную плотность
$ \rho_{1}^{(2)}[L_\sigma]$.
Строим сначала спектральную плотность 1-й итерации:
\begin{subequations}
 \label{eq:Sp.Den.iteration-1}
\begin{eqnarray}
 \label{eq:Sp.Den.it-1}
  \rho_{\nu=1}^{(2)\text{it-1}}[L_\sigma]
  &=& \frac{1}{\pi}\,
     \frac{\sin[\varphi_{(2)}^\text{it-1}[L_\sigma]]}
          {R_{(2)}^\text{it-1}[L_\sigma]}\,,\\
 R_{(2)}^\text{it-1}[L]
  &=& \sqrt{\left[L + c_1\ln r[L]\right]^2
          + \left[\pi+c_1\phi[L]\right]^2
            }\,,
\label{eq:R2.it1}\\
 \varphi_{(2)}^\text{it-1}[L]
  &=& \arccos\left[\frac{L + c_1\ln r[L]}
                        {R_{(2)}^\text{it-1}[L]}
             \right]\,,
\label{eq:varphi2.it1}
\end{eqnarray}
где
\begin{eqnarray}
 r[L]
  = \sqrt{\left[L+c_1\right]^2+ \pi^2};~~~
 \phi[L]
  = \arccos\left[\frac{L+c_1}{r[L]}
             \right]\,.
\label{eq:phi.it1}
\end{eqnarray}
\end{subequations}
Эта спектральная плотность оказывается достаточно близкой к точной
плотности $\rho^\text{(2)}_{1}(\sigma)$:
максимальное отличие имеет порядок 5\%,
см. левый график на рис.\ \ref{fig:exact-2it}.
Теперь обратимся к приближенному решению,
генерируемому второй итерацией (\ref{eq:123.Iter}).
Оно дает нам такие плотность, модуль и фазу:
\begin{subequations}
 \label{eq:Sp.Den.iteration-2}
\begin{eqnarray}
 \label{eq:Sp.Den.it-2}
  \rho_{\nu=1}^{(2)\text{it-2}}[L_\sigma]
  &=& \frac{1}{\pi}\,
     \frac{\sin[\varphi_{(2)}^\text{it-2}[L_\sigma]]}
          {R_{(2)}^\text{it-2}[L_\sigma]}\,,\\
 R_{(2)}^\text{it-2}[L]
  &=& \sqrt{\left[L + c_1\ln R[L]\right]^2
          + \left[\pi + c_1\Phi[L]\right]^2
          }\,,
\label{eq:R2}\\
 \varphi_{(2)}^\text{it-2}[L]
  &=&\arccos\left(\frac{L + c_1\ln R[L]}
                       {R_{(2)}^\text{it-2}[L]}
            \right)\,,
\label{eq:varphi2}
 \end{eqnarray}
где
\begin{eqnarray}
\label{eq:R[L]}
 R[L]
  &=& \sqrt{\left[L + c_1 + c_1\ln r[L]\right]^2
         + \left[\pi+c_1 \phi[L]\right]^2
         }\,,\\
 \Phi[L]
  &=& \arccos\left(\frac{L+c_1+c_1\ln r[L]}{R[L]}\right)\,.
\label{eq:Phi[L]}
\end{eqnarray}
\end{subequations}
Эта спектральная плотность
оказывается гораздо ближе к
точной, $\rho^\text{(2)}_{1}(\sigma)$:
отличие в области максимума имеет порядок 1\%,
см. правый график на рис.\ \ref{fig:exact-2it}.

\section{Трехпетлевые ренормгрупповые решения для эффективного заряда в КХД}
 \label{App:RG-solution.3-loop}
\textbf{1.} В трехпетлевом приближении $\beta$-функция
задается следующим выражением
\begin{eqnarray}
  \beta_{(3)}\left( \frac{\alpha_\text{s}}{4 \pi}\right)
= - b_0\left(\frac{\alpha_\text{s}}{4 \pi}\right)^2
  - b_1\left(\frac{\alpha_\text{s}}{4 \pi}\right)^3
  - b_2\left(\frac{\alpha_\text{s}}{4 \pi}\right)^4\,,
\label{eq:betaf.3L}
\end{eqnarray}
Соответствующее трехпетлевое уравнение РГ для эффективного заряда
таково
\begin{eqnarray}
 \frac{d a_{(3)}}{dL}
  = - a_{(3)}^2\left[1 + c_1\,a_{(3)}+ c_2\,a_{(3)}^2\right]
 \text{,~~где~}
  c_2\equiv \frac{b_2}{b_0^3}\,.
\label{eq:beta.new.2L.3-loop}
\end{eqnarray}
Интегрируя по $L$, можно показать,
что $a_{(3)}[L]$ удовлетворяет следующему нелинейному уравнению\footnote{%
Заметим, что для $N_f=3$, $c_1\simeq0.624$, $c_2\simeq0.883$,
так что $c_1^2-4 c_2\simeq-2.908<0$.
Тем не менее, правая часть (\ref{eq:RG.Exact.3L})
записана в таком виде, что она остается чисто вещественной и в этом случае.}
\begin{eqnarray}
 \label{eq:RG.Exact.3L}
  L\!=\!\frac{1}{a_{(3)}}
    + c_1 \ln\!\left[\frac{a_{(3)}}{\sqrt{1+c_1a_{(3)}+c_2a_{(3)}^2}}
             \right]
    + \frac{c_1^2-2c_2}{\sqrt{c_1^2-4c_2}}
       \ln\left[\frac{2+a_{(3)}\left(c_1-\sqrt{c_1^2-4 c_2}\right)}
                     {2+a_{(3)}\left(c_1+\sqrt{c_1^2-4 c_2}\right)}
          \right].~~
\end{eqnarray}
Точное решение этого уравнения неизвестно,
поэтому для его применения
обычно используют численные методы поиска решений
или разложение по $a=1/L$ в области больших $L$:
\begin{eqnarray}
 \label{eq:a.PT.3L}
  a_{(3)}
   &=& a_{}
     + a_{}^2\, c_1\,\ln a_{(1)}
     + a_{}^3\,
        \left[c_1^2\, \left(\ln^2 a_{}
            + \ln a_{}
             -1 \right) +c_2\right]  \nonumber \\
&& ~~+a_{}^4\,
        \left[ c_1^3\, \left(\ln^3 a_{}
            + \frac5{2}\ln^2 a_{}
            -2~\ln a_{}
            -\frac1{2} \right) +3 c_2 c_1\ln a_{}\right]\nonumber \\
 &&            ~~+ O\left(a_{}^{5}\ln^{4} a_{}\right)\,.
\end{eqnarray}

\textbf{2.} Рассмотрим здесь решение РГ уравнения
для эффективного заряда в модифицированном по Паде
трехпетлевом приближении в КХД~\cite{KM01,KM03},
где $\beta$-функция (\ref{eq:betaf}) задается так:
\begin{eqnarray}
 \beta_{(3\text{-P})}\left(\frac{\alpha_\text{s}}{4\pi}\right)
  = - b_0\left(\frac{\alpha_\text{s}}{4 \pi}\right)^2
      \left[1
         +  \frac{b_1\,\alpha_\text{s}/(4\pi)}
                 {b_0\left(1-b_2\,\alpha_\text{s}/(4b_1\pi)\right)}
     \right]\,.
\label{eq:betaf.3L.Pade}
\end{eqnarray}
Легко видеть, что первые три члена в разложениях (\ref{eq:betaf.3L})
и (\ref{eq:betaf.3L.Pade}) совпадают,
зато их асимптотики при $a\to\infty$ совершенно различны:
$\beta_{(3)}(a)\sim -a^4$ и $\beta_{(3\text{-P})}(a)\sim -a^2$.
Соответствующее РГ уравнение (\ref{eq:beta.new.2L.3-loop}) модифицируется к виду
\begin{eqnarray}
 \frac{d a_{(3\text{-P})}}{dL}
  = - a_{(3\text{-P})}^2
      \left[1
      + \frac{c_1\,a_{(3\text{-P})}}
             {1-(c_2/c_1)\,a_{(3\text{-P})}}
      \right]\,.
\label{eq:beta.new.2L.3L.Pade}
\end{eqnarray}
Его решение, как нетрудно убедиться,
имеет вид
\begin{eqnarray}
 \label{eq:App-RGExact.3L}
  \frac{1}{a_{(3\text{-P})}}
    + c_1 \ln\left[\frac{a_{(3\text{-P})}}{1
    + \left(c_1-c_2/c_1\right)a_{(3\text{-P})}}\right]
  = L\,,
\end{eqnarray}
которое очень похоже по форме на (\ref{eq:App-RGExact}).
Неудивительно поэтому, что точное решение (\ref{eq:App-RGExact.3L})
также может быть выражено через функцию Ламберта $W(z)$,
а именно:
\begin{eqnarray}
 \label{eq:App-Exactsolution.2L.3L}
 a_{(3\text{-P})}[L] =
 -\frac{1}{c_1}\,
   \frac{1}{1-c_2/c_1^2+W_{-1}\big(z_{W}^{(3\text{-P})}(L)\big)}\, ,
 \end{eqnarray}
где
$z_{W}^{(3\text{-P})}(L)
=\left(1/c_1\right) \exp\left[-1+i\pi+c_2/c_1^2-L/c_1\right]$.
Относительная точность этого решения
в сравнении с численным решением стандартного трехпетлевого
уравнения  (\ref{eq:beta.new.2L.3-loop})
лучше 1\% для $L\geq 7$
(и лучше $0.5\%$ для $L\geq9$).

\textbf{3.} Для модифицированного по Паде трехпетлевого приближения
соответствующие спектральные плотности определяются
с помощью (\ref{eq:App-Exactsolution.2L.3L})
\begin{subequations}
\begin{eqnarray}
 \label{eq:Spec.Den.Lamb.3L.Pade}
  \rho^{(3\text{-P})}_{\nu}[L_\sigma]
   &=& \frac{1}{\pi}\,
        \frac{\sin[\nu~\varphi_{(3\text{-P})}[L_\sigma]]}
             {\left(R_{(3\text{-P})}[L_\sigma]\right)^{\nu}}\,,
  \\ \label{eq:Lamb.R.3L.Pade}
  R_{(3\text{-P})}[L]
   &=& c_1\,\Big|1-\frac{c_2}{c_1^2}
                  +W_{\pm1}\big[z_{W}^{(3\text{-P})}(L\mp i\pi)\big]\Big|\,,~
  \\ \label{eq:Lamb.phi.3L.Pade}
  \varphi_{(3\text{-P})}[L]
   &=& \arccos\left[
        \textbf{Re}\left(
         \frac{-R_{(3\text{-P})}[L]}
              {c_1\,\left(1-c_2/c_1^2+\big[z_{W}^{(3\text{-P})}(L-i\pi)\big]\right)}
                   \right)
              \right]\,.
\end{eqnarray}
\end{subequations}
Следующие явные выражения для аналитических образов
эффективного заряда для области Минковского
\begin{eqnarray}
 \label{eq:U_1.3L}
  {\mathfrak A}_{1}^{(3\text{-P})}\left[L_s\right]
   = \frac{1}{\pi}
     \left\{\pi
   - \frac{c_1^2}{c_1^2-c_2}\,
     \textbf{Im}\left[\ln{W_{1}\left(z_s\right)
     \vphantom{\frac{c_2}{c_1^2}}}
                \right]
   + \frac{c_2}{c_1^2-c_2}\,
     \textbf{Im}\left[\ln\left(1-\frac{c_2}{c_1^2}+W_1\left(z_s\right)
                         \right)
                \right]
      \right\}~~~~
\end{eqnarray}
с $z_s=z_{W}^{(3\text{-P})}[L_s]$
были получены Маградзе~\cite{Mag00}.
Относительная точность этих решений по сравнению с результатами
численного интегрирования обычной,
немодифицированной по Паде,
спектральной плотности  $\rho^{(3)}_1[L_\sigma]$
оказывается лучше $0.25\%$ при $L_s\geq 2$.

\section{Эволюция пионной АР}
\label{app:DA.Evo}
АР пиона эволюционирует с изменением масштаба нормировки $\mu^2$
в соответствии с уравнением эволюции ЕРБЛ~\cite{ER80,ER80tmf,LB79,LB80}
\begin{eqnarray}
 \frac{d \varphi_\pi(x,\mu_\text{F}^2)}{d \ln\mu_\text{F}^2}
  = V(x,u,\alpha_s(\mu_\text{F}^2))\mathop{\otimes}\limits_{u}
    \varphi_\pi(u,\mu_\text{F}^2)\, ,
 \label{eq:ERBL.Evo}
\end{eqnarray}
где $V(x,u,\alpha_s)$ --- пертурбативно рассчитываемое
ядро эволюции,
которое в ведущем $O(\alpha_s)$-порядке имеет вид
\begin{eqnarray}
  V(x,u,\alpha_s)
        =  \frac{\alpha_s}{4 \pi} \, V_0(x,u)
        \,.
\label{eq:kernel}
\end{eqnarray}
Решение этого уравнения в ведущем порядке есть
(\ref{eq:AR.2-Geg})--(\ref{eq:anLO}).
Удобно представлять АР
$\varphi_\pi(x,\mu^2)$ в виде
разложения по полиномам Гегенбауэра
$C_{k}^{3/2}(2 x-1)$,
которые являются собственными функциями
ядра эволюции ЕРБЛ в ведущем порядке, $V_0$,
т.~е.
\begin{eqnarray}
 \varphi_\pi(x,\mu^2)
  = 6 x (1-x)
     \left[ 1
          + \sum_{m=1}^{\infty} a_{2m}(\mu^2) \, C_{2m}^{3/2}(2x -1)
     \right]\,.
\label{eq:phi0-m}
\end{eqnarray}
При этом вся зависимость от $\mu^2$ переходит в коэффициенты разложения
$a_m(\mu^2)$.
Имеется разложение аномальных размерностей,
являющихся собственными значениями ядра $V_0$,
по степеням $\alpha_s$:
\begin{subequations}
\begin{eqnarray}
  \gamma_n(\alpha_\text{s})
   =  \frac{\alpha_\text{s}}{4 \pi}\gamma_n^{(0)}
   + \ldots\,,
 \label{eq:gamman}
\end{eqnarray}
где аномальные размерности ведущего порядка есть
\begin{eqnarray}
 \gamma_n^{(0)}
   = 2 C_\text{F}\left[4 S_1(n+1) 
                 - 3 - \frac{2}{(n+1) (n+2)}
           \right]
 \label{eq:gamma0}
\end{eqnarray}
\end{subequations}
с $S_1(n+1)=\sum_{i=1}^{n+1} 1/i =\psi(n+2)-\psi(1)$,
а функция $\psi(z)$ определена как $\psi(z)= d\ln\Gamma(z)/d z$.

\section{Разложение скалярной $D$-функции}
 \label{app:D.Rep}

Первые три коэффициента $d_1$, $d_2$, $d_3$  разложения
$D_\text{S}$-функции двух скалярных кварковых токов,
\begin{eqnarray}
 D_\text{S}(Q^2)
  &=& 3\,m_b^2(Q^2)
       \left[1+\sum_{n>0} d_n
        \left({\alpha_s(Q^2)\over\pi}\right)^n
        \right]\, ,
  \label{dn}
\end{eqnarray}
были рассчитаны в~\cite{Che96} и равны
\begin{subequations}
 \label{eq:d_1-d_4}
\begin{eqnarray}
 d_1 &=& C_\text{F}
           \left[\frac{17}{4}\right]\,,
 \label{eq:d_1}\\
 d_2 &=& C_\text{F}^2
          \left[\frac{691}{64}
                  - \frac{9}{4}\,\zeta(3)
          \right]
      +\,C_\text{F}\,C_\text{A}
          \left[\frac{893}{64}
                  - \frac{31}{8}\,\zeta(3)
          \right]
      +\,T_\text{R}\,N_f\,C_\text{F}
          \left[-\frac{65}{16}
                  + \zeta(3)
          \right]\,,~~
  \label{eq:d_2}
\end{eqnarray}
\begin{eqnarray}
 d_3 &=& C_\text{F}^3
          \left[\frac{23443}{768}
                 - \frac{239}{16}\,\zeta(3)
                 + \frac{45}{8}\,\zeta(5)
          \right]
      +\,C_\text{F}^2\,C_\text{A}
          \left[\frac{13153}{192}
                 - \frac{1089}{32}\,\zeta(3)
                 + \frac{145}{16}\,\zeta(5)
          \right]
\nonumber\\
     &+& C_\text{F}\,C_\text{A}^2
          \left[\frac{3894493}{62208}
                 - \frac{2329}{96}\,\zeta(3)
                 + \frac{25}{48}\,\zeta(5)
          \right]
\nonumber\\
     &+& T_\text{R}\,N_f\,C_\text{F}^2
          \left[-\frac{88}{3}
                 + \frac{65}{4}\,\zeta(3)
                 + \frac{3}{4}\,\zeta(4)
                 - 5\,\zeta(5)
          \right]
\nonumber\\
     &+& T_\text{R}\,N_f\,C_\text{F}\,C_\text{A}
          \left[-\frac{33475}{972}
                 + \frac{22}{3}\,\zeta(3)
                 - \frac{3}{4}\,\zeta(4)
                 + \frac{5}{6}\,\zeta(5)
          \right]
\nonumber\\
     &+& T_\text{R}^2\,N_f^2\,C_\text{F}
          \left[\frac{15511}{3888}
                 - \zeta(3)
          \right]\,.
 \label{eq:d_3}
\end{eqnarray}
Коэффициент $d_4$ был получено совсем недавно в~\cite{BCK05}:
\begin{eqnarray}
 d_4 &=& N_f^3
            \left[-\frac{520771}{559872}
                  + \frac{65}{432}\,\zeta(3)
                  + \frac{1}{144}\,\zeta(4)
                  + \frac{5}{18}\,\zeta(5)
            \right]
\nonumber\\
     &+& N_f^2
            \left[\frac{220313525}{2239488}
                  - \frac{11875}{432}\,\zeta(3)
                  + \frac{5}{6}\,\zeta^2(3)
                  + \frac{25}{96}\,\zeta(4)
                  - \frac{5015}{432}\,\zeta(5)
            \right]
\nonumber\\
     &+& N_f\left[-\frac{1045811915}{373248}
                  + \frac{5747185}{5184}\,\zeta(3)
                  - \frac{955}{16}\,\zeta^2(3)
                  - \frac{9131}{576}\,\zeta(4)
                  + \frac{41215}{432}\,\zeta(5)
                   \right. \nonumber \\
& & \phantom{N_f^0}
               \left.~
                  + \frac{2875}{288}\,\zeta(6)
                  + \frac{665}{72}\,\zeta(7)
            \right] \nonumber\\
     &+& N_f^0 \left[\frac{10811054729}{497664}
                  - \frac{3887351}{324}\,\zeta(3)
                  + \frac{458425}{432}\,\zeta^2(3)
                  + \frac{265}{18}\,\zeta(4)
                  + \frac{373975}{432}\,\zeta(5)
               \right. \nonumber \\
     & & \phantom{N_f^0}
               \left.~
                  - \frac{1375}{32}\,\zeta(6)
                  - \frac{178045}{768}\,\zeta(7)
               \right]\,.
\label{eq:d_4}
\end{eqnarray}
\end{subequations}

\section{Аномальные размерности и эволюция кварковых масс}
 \label{app:Anom.Dim.Quark.Mass}
\textbf{1.} Коэффициенты $\gamma_i$ определяют разложение аномальной размерности
кварковой массы аналогично тому,
как это сделано в (\ref{eq:betaf}) по отношению к разложению
$\beta$-функции, а именно,
\begin{eqnarray}
 \frac{d}{dL}\ln\left(m[L]\right)
  \equiv \gamma_\text{mass}\left(\frac{\alpha_\text{s}[L]}{4 \pi}\right)
  = - \sum_{i \geq 0}
  \gamma_i\left(\frac{\alpha_\text{s}[L]}{4 \pi}\right)^{i+1}\,.
\label{eq:gammaf}
\end{eqnarray}
Их явные выражения таковы, см.~\cite{Che97},
\begin{eqnarray}
 \gamma_0
  &=& 3 C_\text{F}\,;
   \label{g0}\\
 \gamma_1
  &=& \bigg[\frac{202}{3}-\frac{20}{9}\,N_f
       \bigg]\,;
   \label{g1}\\
 \gamma_2
  &=& \bigg[1249
            -\bigg(\frac{2216}{27}+\frac{160}{3}\zeta(3)\bigg)\,N_f
            -\frac{140}{81}\,N_f^2
       \bigg]\,;
   \label{g2}
\end{eqnarray}
\begin{eqnarray}
 \gamma_3
  &=& \bigg[\frac{4603055}{162}
            +\frac{135680}{27}\zeta(3)
            -8800\zeta(5)
       \nonumber\\
  & &{}     -\bigg(\frac{91723}{27}
                  +\frac{34192}{9}\zeta(3)
                  -880\zeta(4)
                  -\frac{18400}{9}\zeta(5)
             \bigg)\,N_f
        \nonumber\\
  & &{}     +\bigg(\frac{5242}{243}
                  +\frac{800}{9}\zeta(5)
                  -\frac{160}{3}\zeta(4)
             \bigg)\,N_f^2
            -\bigg(\frac{332}{243}
                  -\frac{64}{27}\zeta(3)
             \bigg)\,N_f^3
       \bigg]\,,
  \label{g3}
\end{eqnarray}
где $\zeta(\nu)$ обозначает $\zeta$-функцию Римана.
Эволюция $m_{(l)}(Q^2)$ в $l$-петлевом приближении
описывается следующим общим решением уравнений РГ
\begin{subequations}
\begin{eqnarray}
\label{eq:m2-def}
  m_{(l)}^2(Q^2)
  &=& m_{(l)}^2(\mu^2)\,
       \exp\left[2\int_{\alpha_\text{s}(\mu^2)/(4\pi)}^{\alpha_\text{s}(Q^2)/(4\pi)}
                 \frac{\gamma_\text{mass}(x)}{\beta(x)}\, dx
                            \right] \\
  &=& m_{(l)}^2(\mu^2)\,
       \frac{\left[\alpha_\text{s}(Q^2)\right]^{\nu_0}
              f_{(l)}(\alpha_s(Q^2))}
            {\left[\alpha_\text{s}(\mu^2)\right]^{\nu_0}
              f_{(l)}(\alpha_s(\mu^2))}\,,
\label{eq:m-run}
\end{eqnarray}
где
\begin{eqnarray}
 \nu_0 = 2\frac{\,\gamma_0}{b_0}
 \label{eq:nu.0.nu.1}
\end{eqnarray}
и функция $f_{(l)}(\alpha_s)$,
задаваемая
\begin{eqnarray}
 f_{(l)}(\alpha_s)
 = \exp\left[2\int_{0}^{\alpha_\text{s}/(4\pi)}
             \left(\frac{\gamma_{m}^{(l)}(x)}{\beta^{(l)}(x)}
                  -\frac{\gamma_0\,x
                         }{b_0\,x^2
                         }\right)\, dx
                   \right]\,,
\label{eq:phi}
\end{eqnarray}
\end{subequations}
аккумулирует эффекты многопетлевой (начиная с двухпетлевой) эволюции
$m_{(l)}^2(Q^2)$ с $Q^2$.
В однопетлевом приближении ($l=1$), $f_{(l)}(\alpha_\text{s})$
по определению задается равной единице.
С другой стороны, при $l=2$ и $l=3$ мы имеем
\begin{eqnarray}
 f_{(2)}(\alpha_s)
 = \left[1 + \delta_1\,\alpha_s\right]^{\nu_1}\,,
 \quad\text{где~~~}
 \delta_1 = \frac{b_1}{4\pi b_0} = \frac{c_1 b_0}{4\pi}\,,~
 \nu_1 = 2 \left(\frac{\gamma_1}{b_1}
                 -\frac{\gamma_0}{b_0}
           \right)\,,
\label{eq:f_2}
\end{eqnarray}
и
\begin{subequations}
\begin{eqnarray}
 f_{(3)}(\alpha_s)
 = \left[1 + \delta_{1}\,\alpha_s+ \delta_{2}\,\alpha_s^2\right]^{\nu_{20}}
    \exp\left[\nu_{21}\,
              \arccos\left(
                   \frac{1+\delta_{1}\,\alpha_s/2}
                        {\sqrt{1 + \delta_{1}\,\alpha_s+ \delta_{2}\,\alpha_s^2}}
                      \right)
        \right]\,,
\label{eq:f_3}
\end{eqnarray}
где
\begin{eqnarray}
 \label{eq:delta_21}
  \delta_2 = \frac{b_2}{16\pi^2 b_0}\,,~
  \nu_{20} = \left(\frac{\gamma_2}{b_2}
                  -\frac{\gamma_0}{b_0}
             \right)\,,~
  \nu_{21} = \frac{-2\,b_1}
                  {\sqrt{4b_2b_0-b_1^2}}
              \left(\frac{\gamma_2}{b_2}
                   -2\frac{\gamma_1}{b_1}
                   +\frac{\gamma_0}{b_0}
              \right)\,.
\end{eqnarray}
\end{subequations}
Для нас важен случай РГ уравнения
для эффективного заряда в модифицированном по Паде
трехпетлевом приближении в КХД~\cite{KM01,KM03},
где $\beta$-функция задается уравнением (\ref{eq:betaf.3L.Pade}).
В этом подходе
аналогичным образом задается и аномальная размерность:
\begin{eqnarray}
 \gamma_\text{mass}^{(3\text{-P})}\left(\frac{\alpha_\text{s}}{4\pi}\right)
  = -\left(\frac{\alpha_\text{s}}{4 \pi}\right)
      \left[\gamma_0
         +  \frac{\gamma_1\,\alpha_\text{s}/(4\pi)}
                 {1-\gamma_2\,\alpha_\text{s}/(4\gamma_1\pi)}
     \right]\,.
\label{eq:gammaf.3L.Pade}
\end{eqnarray}
Тогда функция $f_{(3\text{-P})}(a_s)$
очень напоминает $f_{(2)}(a_s)$:
\begin{subequations}
\label{eq:f_3L.Pade.Gen}
\begin{eqnarray}
 f_{(3\text{-P})}(a_s)
 = \left[1 + \delta_{22}\,\alpha_s\right]^{\nu_{22}}\,
   \left[1 - \delta_{23}\,\alpha_s\right]^{\nu_{23}}\,,
\label{eq:f_3L.Pade}
\end{eqnarray}
где (напомним: $c_1=b_1/b_0^2$ и $c_2=b_2/b_0^3$)
\begin{eqnarray}
 \delta_{22}
  &=& b_0\,\frac{c_1^2-c_2}{4\,\pi c_1}\,,\quad
  \nu_{22}
  \ =\ \frac{c_1}
            {2\,\pi\,\delta_{22}}\,
       \left(\frac{\gamma _1}
                  {4\,\pi(\delta_{22}+\delta_{23})}
           - \gamma_0
       \right)\,,
\label{eq:nu22.delta22_3L.Pade}\\
 \delta_{23}
  &=& \frac{\gamma_2}
           {4\,\pi\,\gamma_1}\,,\quad
  \nu_{23}
  \ =\ \frac{\gamma_1}
            {2\,\pi\,\delta_{23}}\,
        \left(\frac{c_1}
                   {4\,\pi(\delta_{22}+\delta_{23})}
             -\frac{1}{b_0}
        \right)\,.
\label{eq:nu23.delta23_3L.Pade}
\end{eqnarray}
\end{subequations}

Вводя ренормгрупповой инвариант $\hat{m}_{(l)}$,
см., например,~\cite{BKM01,KKP00},
\begin{eqnarray}
 \label{eq:m2-hat}
  \hat{m}_{(l)}
   &=& m_{(l)}(\mu^2)
        \left\{\left[\alpha_\text{s}(\mu^2)\right]^{\nu_0}
         f_{(l)}(\alpha_s(\mu^2))\right\}^{-1/2}\,,
\end{eqnarray}
можно переписать (\ref{eq:m-run}) в виде
\begin{eqnarray}
 \label{eq:m2-hat-run}
  m_{(l)}^2(Q^2)
   &=& \hat{m}_{(l)}^2
        \left[\alpha_\text{s}(Q^2)\right]^{\nu_0}
         f_{(l)}(\alpha_s(Q^2))\,.
\end{eqnarray}
Отметим, что для полюсной массы $b$-кварка, равной $4.07$~ГэВ,
ренормгрупповые инварианты $\hat{m}_{(2)}=8.012$~ГэВ и
$\hat{m}_{(3-P)}=7.995$~ГэВ.
Разложение $f_{(l)}(x)$ в трехпетлевом приближении
есть
(оно одно и то же как в обычной трехпетлевой схеме,
 так и в ее Паде-модификации)
\begin{eqnarray}
  f_{(3)}(\alpha_s)
   &=& 1
    +  \frac{\alpha_s}{\pi}\,
        \frac{b_1}{2\,b_0}
         \left(\frac{\gamma_1}{b_1}
             - \frac{\gamma_0}{b_0}
         \right)\nonumber\\
   &+& \frac{\alpha_s^2}{\pi^2}\,
        \frac{b_1^2}{16\,b_0^2}
         \left[\frac{\gamma_0}{b_0}-\frac{\gamma_1}{b_1}
                 + 2\,\left(\frac{\gamma_0}{b_0}
                           -\frac{\gamma_1}{b_1}
                      \right)^2
                 + \frac{b_0 b_2}{b_1^2}
                    \left(\frac{\gamma_2}{b_2}
                        - \frac{\gamma_0}{b_0}
                    \right)
         \right]
    + O\left(\alpha_s^3\right)\,,
 \label{varphi}
\end{eqnarray}
что находится во взаимнооднозначном соответствии с формулой (15),
полученной Четыркиным в~\cite{Che97}.\footnote{%
Заметим, однако, что Четыркин разлагает выражение
$\sqrt{f_{(l)}(x)}\equiv c(x)$
и использует отличную от нашей нормировку коэффициентов,
а именно, $\bar{\beta}_n=b_n/(4^nb_0)$ и $\bar{\gamma}_n=\gamma_n/(4^nb_0)$.}

\end{appendix}


\end{document}